\documentclass{article}

\usepackage{latexsym}
\usepackage{amsmath, amssymb, amsfonts, amsthm}
\usepackage{overcite}
\usepackage{epsfig}
\usepackage{stmaryrd}
\usepackage{multicol}
\usepackage{pdfsync}
\usepackage{dsfont}
\usepackage{comment}
\usepackage{algorithmic}
\usepackage{algorithm}
\usepackage[font=footnotesize,format=plain,labelfont=bf]{caption}
\usepackage{psfrag}
\usepackage{subfig}
\usepackage[dvipsnames]{xcolor}
\usepackage{graphics}
\usepackage{enumitem}
\usepackage{mathtools}
\usepackage{fullpage}
\usepackage{booktabs}
\usepackage{multirow}
\usepackage{siunitx}

\graphicspath{{./Figures/}}

\newcommand{\DNSsub}{{\text{\textsc{dns}}}}
\newcommand{\LESsub}{{\text{\textsc{les}}}}
\newcommand{\SGSsub}{{\text{\textsc{sgs}}}}

\newcommand{\norm}[1]{\left\lVert #1 \right\rVert}

\newcommand{\fDelta}{\ol{\Delta}}

\newcommand{\vtheta}{\vec{\theta}\,}
\newcommand{\vphi}{\vec{\phi}\,}
\newcommand{\vtau}{\vec{\tau}\,}
\newcommand{\vr}{\vec{r}}

\newcommand{\hu}{\hat{u}}
\newcommand{\hw}{\hat{w}}
\newcommand{\hhu}{\hat{\hu}}
\newcommand{\hhw}{\hat{\hw}}
\newcommand{\ba}{\mathbf{a}}
\newcommand{\bb}{\mathbf{b}}
\newcommand{\bg}{\mathbf{g}}

\newcommand{\bh}{\mathbf{h}}

\newcommand{\bu}{\mathbf{u}}

\newcommand{\bx}{\mathbf{x}}
\newcommand{\by}{\mathbf{y}}

\newcommand{\eps}{\varepsilon}
\usepackage[colorinlistoftodos, color=blue!20!white, bordercolor=gray,
  textsize=tiny,textwidth=0.8in]{todonotes}

\newcommand{\Dpartial}[2]{\frac{\partial #1}{\partial #2}}
\newcommand{\Dparttwo}[2]{\frac{\partial^2 #1}{\partial #2^2}}

\theoremstyle{definition}

\theoremstyle{remark}

\numberwithin{equation}{section}

\newcommand{\pp}[2]{\frac{\partial #1}{\partial #2}}

\newcommand{\ol}[1]{\overline{#1}}

\newcommand{\avg}[1]{\langle {#1} \rangle}
\newcommand{\olu}{\overline{u}}
\newcommand{\olbu}{\overline{\mathbf{u}}}
\newcommand{\olxi}{\overline{\xi}}

\definecolor{lc1}{HTML}{CE282A}
\definecolor{lc2}{HTML}{255BCB}
\definecolor{lc3}{HTML}{2DCE65}
\definecolor{lc4}{HTML}{A21D8E}
\definecolor{lc5}{HTML}{F1AB65}
\definecolor{lc6}{HTML}{89B6D4}
\definecolor{lc7}{HTML}{A0A0A0}

\title{\Large Embedded training of neural-network sub-grid-scale turbulence models}

\author{
  Jonathan F.~MacArt\footnote{Department of Aerospace and Mechanical Engineering,
    University of Notre Dame,  jmacart@nd.edu}, 
  Justin Sirignano\footnote{Mathematics,
    University of Oxford, justin.sirignano@maths.ox.ac.uk} \footnote{Department of Industrial \& Systems Engineering,
    University of Illinois at Urbana--Champaign}, and
  Jonathan B.~Freund\footnote{Department of Aerospace Engineering,
    University of Illinois at Urbana--Champaign, jbfreund@illinois.edu}
}

\date{\today}

\begin{document}

\maketitle

\begin{abstract}

  The weights of a deep neural network model are optimized in
  conjunction with the governing flow equations to provide a model for
  sub-grid-scale stresses in a temporally developing plane turbulent
  jet at Reynolds number $Re_0=6\,000$. The objective function for
  training is first based on the instantaneous filtered velocity
  fields from a corresponding direct numerical simulation, and the
  training is by a stochastic gradient descent method, which uses the
  adjoint Navier--Stokes equations to provide the end-to-end
  sensitivities of the model weights to the velocity fields. In-sample
  and out-of-sample testing on multiple dual-jet configurations show
  that its required mesh density in each coordinate direction for
  prediction of mean flow, Reynolds stresses, and spectra is half that
  needed by the dynamic Smagorinsky model for comparable accuracy. The
  same neural-network model trained directly to match filtered
  sub-grid-scale stresses---without the constraint of being embedded
  within the flow equations during the training---fails to provide a
  qualitatively correct prediction. The coupled formulation is generalized to
  train based only on mean-flow and Reynolds stresses, which are more
  readily available in experiments. The mean-flow training provides a robust model,
  which is important, though a somewhat less accurate prediction for
  the same coarse meshes, as might be anticipated due to the reduced
  information available for training in this case. The anticipated
  advantage of the formulation is that the inclusion of resolved
  physics in the training increases its capacity to extrapolate. This
  is assessed for the case of passive scalar transport, for
  which it outperforms established models due to improved mixing
  predictions.
 
\end{abstract}

\section{Introduction} \label{Intro}

The attractiveness of sub-grid-scale models that reduce the resolution
needs of large-eddy simulation (LES) methods is well
understood.\cite{Meneveau2000} However, the required resolution to
achieve sufficiently universal sub-grid-scale behavior, which
simplifies modeling, remains higher than might be
desired,\cite{Ghosal1999} and, even for finely resolved LES, the
modeling challenge is not necessarily simple.\cite{Langford1999}
Flexibility to produce diverse sub-grid-scale behaviors is, in a
sense, at odds with cleanly expressed models, which are necessarily
simpler than the diverse range of phenomenology that they seek to
represent. Models with little empiricism, perhaps most notably dynamic
models,\cite{Germano1991,Lilly1992} are obviously attractive for their
anticipated convergence and concomitant portability, and are expected
to see continued use indefinitely, though models with greater
parametric flexibility might afford opportunities for further relaxing
mesh resolution needs. We examine such an approach here and introduce
sub-grid-scale models with many parameters. The hope is that the
greater empiricism will reduce the resolution requirements yet still
provide useful extrapolation to new configurations, beyond the
training data. Maybe most importantly for motivation, we also seek
procedures that are sufficiently flexible to include additional
sub-grid-scale physical effects, beyond that expressed in the flow
equations and possibly beyond that which has been well-described
to-date by governing equations. The attractiveness of the dynamic
procedure depends upon the turbulence physics that underlie its
spectral extrapolation procedure. Cases with coupled physics, such as
turbulent combustion, do not lend themselves directly to decoupled
sub-grid-scale predictions. We therefore also seek to use the
additional empiricism to increase flexibility in the proposed models
to facilitate incorporation of observations that lack as complete of a
mathematical description as turbulence.

The approach we take uses the formulations, optimization procedures,
and efficient implementations available for deep neural networks. As
for any large-eddy simulation, the $\,\bar \cdot\,$ filtered equations
will be solved numerically, 
\begin{align}
  \Dpartial{\olu_i}{t} + \olu_j\Dpartial{\olu_i}{x_j} &=
  -\frac{1}{\rho}\Dpartial{\ol{p}}{x_i} + \nu \Dparttwo{\olu_i}{x_j} +
  \Dpartial{h_{ij}(\olbu;\vtheta)}{x_j}
    \label{e:LESmomentumNNintro} \\
  \Dpartial{\olu_j}{x_j} &= 0,
  \label{e:LESmassNNintro}
\end{align}
with the closure $h_{ij} = h_{ij}(\olbu;\vtheta)$, with
model parameters $\vtheta$, dependent on the velocity and its
derivatives. The dimension of $\vtheta$ is small for established
sub-grid-scale models and nominally zero for models based on a
dynamic procedure. For large-eddy simulation, the simplest approach,
here referred to as the direct approach, is to minimize
\begin{equation}
L(\vtheta) = D\left(\tau_{ij}^\SGSsub(\olbu_e,\ol{\bu_e^\top\bu_e}), h_{ij}(\olbu_e;\vtheta)\right),
\label{e:Ldirect}
\end{equation} 
where $D$ is an appropriate distance
\begin{equation}
D(\ba,\bb) = \int_{0}^T \int_{\Omega} \norm{\ba-\bb}\; d\bx dt
\end{equation}
and $\tau_{ij}^\SGSsub$ and $\bu_e$ are \textit{a priori} trusted data, likely from a
direct numerical simulation (DNS) or possibly from advanced
experimental methods.
Accelerating optimization of $\vtheta$ by using the gradient
\begin{equation}
\Dpartial{L}{\vtheta} = \Dpartial{L}{\bh}\cdot\Dpartial{\bh}{\vtheta}
\label{e:dLdirect}
\end{equation}
is straightforward, with the $\cdot$ dot operator here representing an
appropriate inner product as needed. The first factor in
(\ref{e:dLdirect}) follows directly from the selected measure of
mismatch (\ref{e:Ldirect}), and the second is the usual sensitivity
gradient available from a backpropagation training algorithm, as is
available in any neural network software.\cite{PyTorchNIPS} One
drawback is the availability of $\tau_{ij}^\SGSsub$, which is a
complex quantity in a turbulent flow. Even when $\tau_{ij}^\SGSsub$
and $\bu_e$ are available from advanced diagnostics or direct
numerical simulation, it is alarming that the optimization problem
(\ref{e:Ldirect}) ignores the governing equations
(\ref{e:LESmomentumNNintro}) and (\ref{e:LESmassNNintro}), so the
trained model is at most indirectly constrained by the known and
resolved physics expressed in the unclosed governing equations, which
risks diminishing its capacity for extrapolative prediction.

Such a direct approach, using a variant of (\ref{e:Ldirect}), has been
employed with some success in Reynolds-averaged Navier--Stokes (RANS)
modeling to close the mismatch with trusted data.\cite{Ling2} Of
course, in Reynolds-averaged descriptions, the mismatch is relatively
simple to infer from mean flow fields.  In contrast, for large-eddy
simulation, it is time-dependent, three-dimensional, and
smaller-scale, so it is difficult to design a training regimen that
would with confidence close the mismatch\cite{Gamahara2017a,Beck2019}.
Still more challengingly, it is unclear how to anticipate the limits
of model validity: How far from the training data will the model
successfully extrapolate?  For this question, how even to define
distance is not obvious.  Training on the mismatch (the direct
approach) has the additional challenge that it requires data
specifically aligned with that mismatch.  Such correspondingly rich
datasets likely do not exist in many machine learning applications to
fluid mechanics.\cite{FreundMLPersp:2019}

The approach we pursue is designed to both reduce the need for
training data and remove the requirement of obtaining the challenging
data for $\tau_{ij}^\SGSsub$. To constrain the models by the physics
in (\ref{e:LESmomentumNNintro}) and (\ref{e:LESmassNNintro}), we
minimize
\begin{equation}
L(\vtheta) = D\left(F(\olbu_e),F(\olbu)\right),
\label{e:Lindirect}
\end{equation}
where $F$ is a function of the flow field, $\olbu_e$ indicates that
$F(\olbu_e)$ is derived from a trusted flow field, even if $\olbu_e$
is not necessarily known in detail, and $\olbu$ is the \textit{a posteriori} solution of
(\ref{e:LESmomentumNNintro}) and (\ref{e:LESmassNNintro}). Thus,
minimization of (\ref{e:Lindirect}) is a PDE-constrained optimization
problem, unlike (\ref{e:Ldirect}). If $\olbu_e$ is available, at
least in some $\Omega$ part of the domain, then $F(\olbu) = \olbu$ can
be simply used, though any convenient and effective quantity of
interest can, in principle, be selected.

Unfortunately, this indirect loss function (\ref{e:Lindirect}) greatly increases the training challenge because any new set of parameters $\vtheta$ requires
a new flow solution to determine $\olbu$ and thus $F(\olbu)$.  
Gradient-based acceleration of the optimization, in particular, is more
challenging to implement due to the dimension of $\vtheta$. The
gradient of \eqref{e:Lindirect} for the discretized form of (\ref{e:LESmomentumNNintro})
satisfies
\begin{equation}
\Dpartial{L}{\vtheta} = \Dpartial{L}{\olbu}\cdot\Dpartial{\olbu}{\bh}\cdot\Dpartial{\bh}{\vtheta},
\label{e:dLindirect}
\end{equation}
the middle factor of which describes the change in the flow solution
due to changes in the closure term $\bh$. The $\cdot$ is a
dot-product, and the dimension of $\Dpartial{L}{\olbu}$ matches the
number of space and time degrees of freedom in the discretization. For
the continuous case of (\ref{e:LESmomentumNNintro}),
(\ref{e:dLindirect}) would be replaced by functional derivatives that
satisfy an adjoint PDE. Adjoint-based methods are well-understood to
efficiently provide such high-dimensional gradient information, and we
employ them here. Most of the details of the numerical implementation
are available elsewhere,\cite{DPM-JCP} where they were demonstrated
for the limited case of $F(\olbu) = \olbu$ for isotropic turbulence. A
similar approach has also been demonstrated in two-dimensional
aerodynamics applications,\cite{Duraisamy} based on a similarly
motivated framework previously demonstrated on illustrative model
applications.\cite{Parish2016} The same challenges also motivate
recent efforts to achieve more-robust learning via more sophisticated
network models, rather than building the governing equations directly
into the learning procedure.\cite{Novati2020} Another approach uses
reverse-mode differentiable programming to implement adjoint
PDEs\cite{Rackauckas2020}, similar to algorithmic
differentiation\cite{Baydin2018}, and could be useful in
solving \eqref{e:dLindirect} for complex systems.

Building on the success for isotropic turbulence\cite{DPM-JCP},
several temporally developing planar turbulent shear flows with
Reynolds number $6\,000$ are used to assess the method for the more
complex case of free shear flow, including its scalar mixing, and
multiple training regimens. The numerical simulation methods are
summarized in section~\ref{s:sims}. The turbulent 
flows considered are introduced in
section~\ref{s:flows}. The neural-network model is standard; for
completeness, its particulars are summarized in section~\ref{s:nn}. We
also consider non-trivial $F$ in (\ref{e:Lindirect}) and portability
of trained $\bh$ to other flows (section~\ref{s:variants}). The form
of the trained $\bh$ is analyzed in section~\ref{s:h}, and how it
better represents turbulence on relatively coarse meshes is analyzed
in section~\ref{ss:heffect}. The extrapolative capacity of the method
is demonstrated in section~\ref{s:scalar}, where a model trained on
sub-grid-scale stresses is shown to be more effective than established
models when applied directly---without additional training---to scalar
transport. Finally, in section~\ref{s:num} we consider the
implications of the mesh resolution. Conclusions are discussed in the
context of
future directions in section~\ref{s:summary}.

\section{Flow Simulations}
\label{s:sims}

\subsection{Governing equations}
\label{s:numerics}

To provide the direct numerical simulation training and testing data, the incompressible-fluid flow equations, including transport of a passive scalar $\xi$,
\begin{align}
  \Dpartial{u_i}{t} + u_j\Dpartial{u_i}{x_j} &=
  -\frac{1}{\rho}\Dpartial{p}{x_i} + \nu \Dparttwo{u_i}{x_j}
    \label{e:DNSmomentum} \\
    \Dpartial{u_j}{x_j} &= 0 \label{e:DNSmass} \\
    \Dpartial{\xi}{t} + u_j\Dpartial{\xi}{x_j} &=
    D\Dparttwo{\xi}{x_j}, \label{e:DNSscalar}
\end{align}
are discretized with sufficient resolution such that all turbulence scales are sufficiently resolved.  

\begin{figure}
  \centering
  \includegraphics[width=0.48\textwidth]{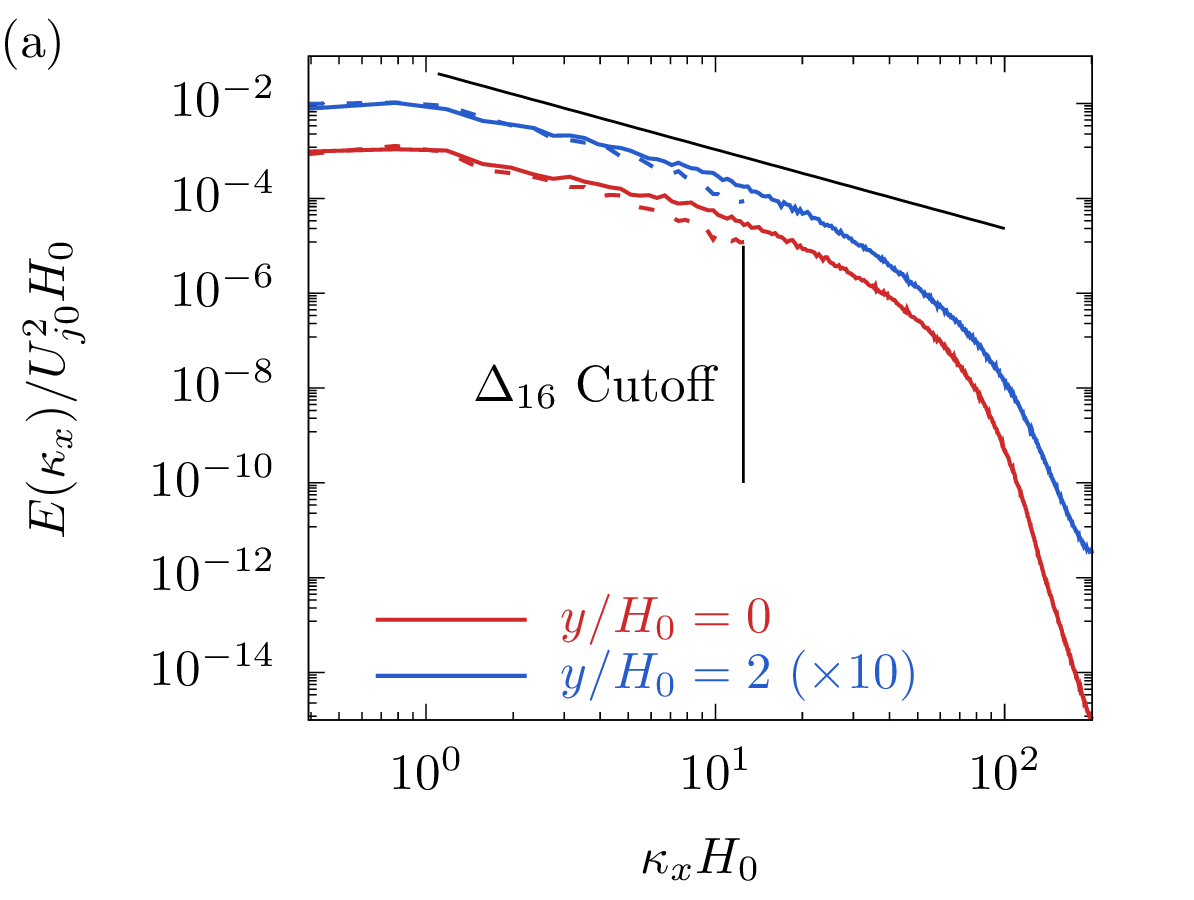}
  \includegraphics[width=0.48\textwidth]{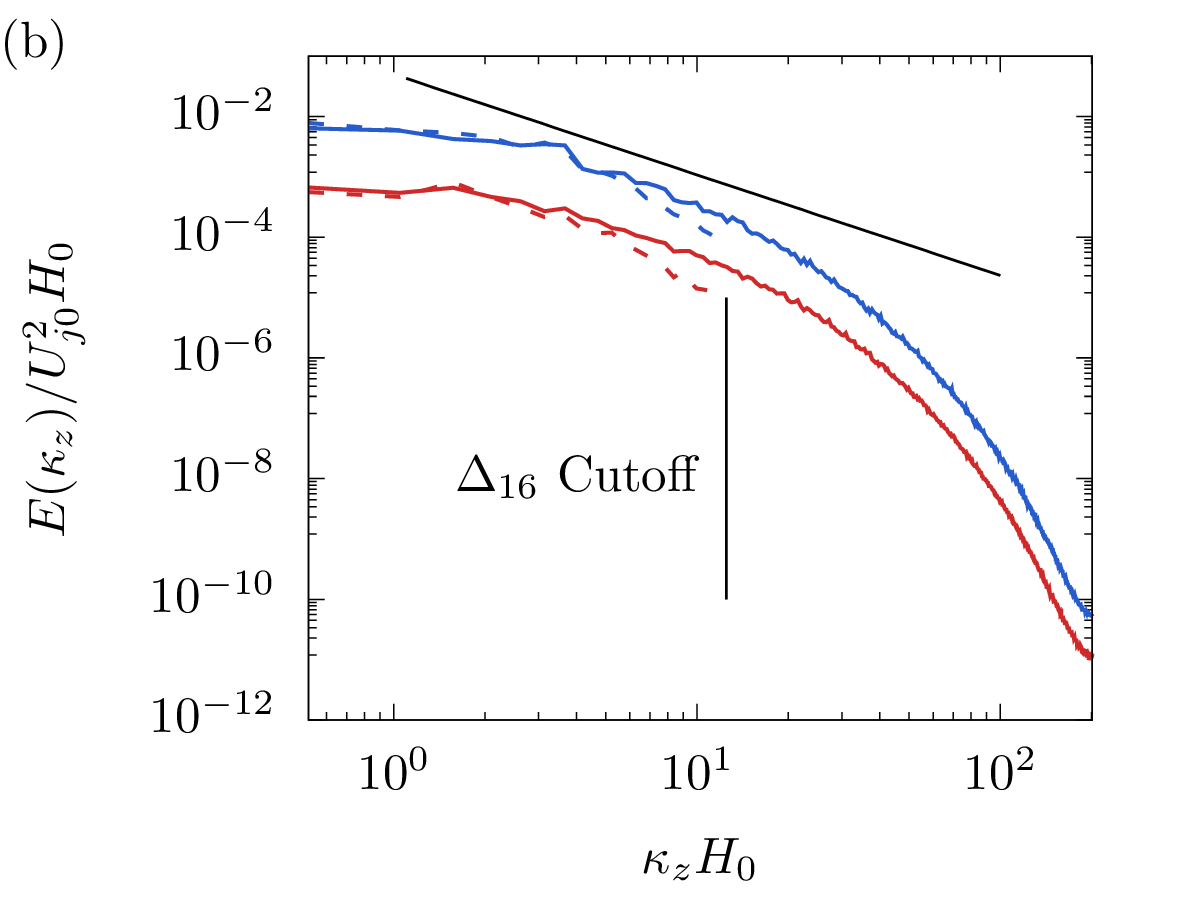}
  \caption{One-dimensional energy spectra from direct numerical
    simulation of a temporally evolving jet (case~A; see
    table~\ref{tab:jet_IC}) at the centerline ($y/H_0=0$) and in the
    shear layer ($y/H_0=2$, shown offset by a factor of 10), at time
    $t=62.5t_{j0}$: (a) streamwise and (b) spanwise spectra. Dashed
    lines are the filtered direct numerical simulation data, with
    vertical lines indicating the $\Delta_{16}$ filter-cutoff wavenumber.  Reference $\sim\kappa^{-5/3}$ lines are shown.}
  \label{fig:DNS_jet_spect}
\end{figure}

For the large-eddy simulations, (\ref{e:DNSmomentum}) is replaced with either
\begin{equation}
  \Dpartial{\olu_i}{t} + \olu_j\Dpartial{\olu_i}{x_j} =
  -\frac{1}{\rho}\Dpartial{\ol{p}}{x_i} + \nu \Dparttwo{\olu_i}{x_j} - \Dpartial{\tau^r_{ij}}{x_j}
  \label{e:LESmomentumTaur},
\end{equation}
for comparison with established $\tau_{ij}^r$ closure approaches, or
(\ref{e:LESmomentumNNintro}), with the fitted neural network closure
$h_{ij}$.  For the anisotropic residual stress
$\tau_{ij}^r=\tau_{ij}^\SGSsub - \tau_{kk}^\SGSsub\delta_{ij}/3$, we
consider the Smagorinsky model\cite{Smagorinsky1963,Rogallo1984} with
constant coefficient $C_s = 0.18$, the dynamic Smagorinsky
model\cite{Germano1991,Lilly1992}, and the model of Clark \textit{et
  al.}\cite{Clark1979}. Differently from a previous
demonstration\cite{DPM-JCP}, where $\bh$ modeled $\nabla\cdot\vtau^r$,
here $h_{ij}$ models $\tau^r_{ij}$ directly for smoother optimization
and greater modeling flexibility
(\S\ref{ss:symmetric}). Moreover, it enables us to compare the
functional form of $h_{ij}$ to traditional models (section~\ref{s:h}).

The filter underlying the large-eddy simulation formulation is 
\begin{equation}
  \ol{\phi}(\bx,t) = \int_\Omega G(\by,\bx)\phi(\bx-\by,t)\,d\by,
  \label{e:filter}
\end{equation}
though this is only explicitly used for comparison with direct
numerical simulations and for generating large-eddy simulation initial
conditions. For this, $G(\by,\bx)$ has unit support in a
${\fDelta}$-cube centered on $\mathbf{x}$~\cite{Clark1979}.

\subsection{Numerical discretization}
\label{ss:numerics}

Second-order centered differences are used on a staggered
mesh~\cite{Harlow1965}, except for convective terms in the scalar
equation \eqref{e:DNSscalar}, which were discretized using a
third-order weighted essentially non-oscillatory (WENO)
scheme~\cite{Jiang1996}.  Time integration is by a fractional-step
method~\cite{Kim1985} and a linearized trapezoid method in an
alternating-direction implicit (ADI) framework for the diffusive
terms, as implemented by Desjardins \textit{et
  al.}\cite{Desjardins2008} Mesh densities for the direct numerical
simulation cases listed in table~\ref{tab:jet_IC} were selected so
that the Kolmogorov scale $\eta=(\nu^3/\eps)^{1/4}$, based on the
calculated viscous dissipation rate $\eps$, is resolved with uniform
mesh spacing $\Delta x_\DNSsub$ such that $\Delta x_\DNSsub/\eta <
1.8$ in the turbulence regions, which matches successful resolutions
of turbulent jets at similar Reynolds
number\cite{Freund2000d,Pantano2002,Knaus2009,Hawkes2012,MacArt2018}.
Spectra in figure~\ref{fig:DNS_jet_spect} confirm that several decades
of turbulence energy decay are resolved.  The time step $\Delta
t_\DNSsub$ was fixed, with initial CFL number 
about 0.37 that decreased to about 0.18 by
the end of the simulation. 
\textit{A posteriori} large-eddy simulations used the same numerical
methods together with sub-grid-scale models or the deep learning model
discussed subsequently in section~\ref{s:nn}. Analogous filtered
spectra in figure~\ref{fig:DNS_jet_spect} with
$\Delta_{16}$ show a relatively high energy at the cutoff wavenumber. This substantially increases the modeling challenge, as is shown in section~\ref{ss:baseline_sgs}.

The adjoint-based neural network model training, which is described in
section~\ref{TrainingAlgorithm}, uses the same numerical methods for
the forward solution with the exception of time integration, which is
by the fourth-order Runge--Kutta method for simplicity; thus the
temporal adjoint solution is approximate. Since the overall
order-of-accuracy is limited by the pressure-projection step, it is
comparable to that of the forward solution. We further note that the
adjoint solution is computed over only short time horizons
(\S\ref{TrainingAlgorithm}), for which the slightly different
energy-conservation properties of the forward and adjoint time
integrators do not become appreciable. An exact (in space only)
discrete adjoint was solved.

\section{Configurations}
\label{s:flows}

Free-shear-flow turbulence from multiple 
temporally developing planar jets, some with multiple streams, will be used for
training and testing.  These are introduced in the following
subsections.

\subsection{Turbulent plane jets}
\label{s:jets}

The baseline turbulent plane jet has a rounded top-hat initial streamwise mean velocity, 
\begin{equation}
  \hu(y) = \frac{U_{j0}}{2}\left[ \tanh\frac{y/H_0 + 1/2}{\delta_L} - \tanh\frac{y/H_0 - 1/2}{\delta_L} \right],
  \label{e:jetic}
\end{equation}
of width $\approx H_0$ bounded by shear layers of thickness $\approx \delta_L=0.1 H_0$.
The Reynolds number is $Re_0 = U_{j0} H_0/\nu = 6\,000$.

To seed transition to turbulence, spatially periodic fluctuations in the $x$--$z$ plane are added to the streamwise and cross-stream velocities:
\begin{align}
  \hhu(y,z) &= a\, \hu(y)\cos\!\left[\frac{16\pi z}{L}\right] + \hat u(y)\\
  \hhw(x,y) &= a\, \hu(y)\cos\!\left[\frac{16\pi x}{W}\right],
\end{align}
where $a=0.1$, and $L$ and $W$ are the domain sizes in the periodic
$x$- and $z$-directions, respectively (see
table~\ref{tab:jet_IC}). Three-dimensional random perturbations are
superimposed on top of these at each mesh point:
\begin{align}
  u(x,y,z) &=  \left[1 + a(r_x(x,y,z)-0.5)\right] \hhu(y,z) \\
  w(x,y,z) &=  \left[1 + a(r_z(x,y,z)-0.5)\right] \hhw(x,y),
\end{align}
where $r_x$ and $r_z \in[0,1]$ are uniformly distributed. The initial cross-stream velocity is $v(x,y,z) = 0$.  This single-jet baseline case~is case~A in table~\ref{tab:jet_IC}.

Additional dual-jet cases, which are anticipated to involve entrainment and merging processes not seen in the single jet,\cite{Miller1960} are used for out-of-sample tests.  These additional processes occur over significantly longer duration than the development of approximate self-similarity in the single jet.  Three specific dual-jet cases are considered: symmetric parallel jets (case~B), asymmetric parallel jets (case~C), and asymmetric anti-parallel jets (case~D).  These have initial streamwise mean velocity profile
\begin{equation}
\begin{split}
  \hat u(y) & = \frac{U_{j0}}{2}\left[ \tanh\frac{(y + H_{s,0})/H_0 + 1/2}{\delta_L} - \tanh\frac{(y + H_{s,0})/H_0 - 1/2}{\delta_L} \right]
 \\
  &+ \frac{U_{j1}}{2}\left[ \tanh\frac{(y - H_{s,1})/H_1 + 1/2}{\delta_{L,1}} - \tanh\frac{(y - H_{s,1})/H_1 - 1/2}{\delta_{L,1}} \right],
  \label{e:dualjetic}
\end{split}
\end{equation}
where $H_s=\min(H_0,H_1)$, $H_{s,0}=(H_s+H_0)/2$,
$H_{s,1}=(H_s+H_1)/2$, and $\delta_{L,1}=(H_1/H_0)\delta_L$.  The
primary and secondary jets are set to have the same Reynolds number ($Re_0=Re_1$) based on their speed and width.  Perturbations are superimposed following the same long-wavelength and random perturbations as for the single jet. Specific $H_1/H_0$ and $U_{j1}/U_{j0}$ ratios are listed in table~\ref{tab:jet_IC}.

\begin{table}
  \centering
  \begin{tabular}{c c c c c c c}
    \toprule
    Case & Configuration & $Re_0 = U_{j0} H_0/\nu$ & $H_1/H_0$ & $U_{j1}/U_{j0}$ &
                                                                  $(L,H,W)/H_0$
    & Uniform mesh \\
    \midrule
    A & Single jet    & \multirow{4}{*}{6\,000} & --   & --       & \multirow{4}{*}{$16,20,12$} & \multirow{4}{*}{$1024\times1280\times768$} \\
    B & Dual jets     & & 1.0  & 1.0    & & \\
    C & Dual asym.    & & 2.0  & $+0.5$    & & \\
    D & Dual anti-para.   & & 2.0  & $-0.5$ & & \\
    \bottomrule
  \end{tabular}
  \caption{Conditions for the initial mean velocity profiles
    (\ref{e:jetic}) for case~A and (\ref{e:dualjetic}) for cases B--D.}
  \label{tab:jet_IC}
\end{table}

\subsection{Jet evolution}
\label{ss:jet_evo}

Reynolds averages are over the periodic streamwise ($x$) and transverse ($z$) directions and are functions of the cross-stream ($y$) coordinate and time ($t$):
\begin{equation}
  \avg{\phi}(y,t) = \frac{1}{L_xL_z}\int_0^{L_x}\int_{-L_z/2}^{L_z/2} \phi(x,y,z,t)\; dz dx,
\end{equation}
with discrete analog
\begin{equation}
  \avg{\phi}(y_j,t^n) = \frac{1}{N_xN_z}\sum_{i=1}^{N_x}\sum_{k=1}^{N_z} \phi(x_i,y_j,z_k,t^n),
  \label{e:discrete_avg}
\end{equation}
where $N_x$ and $N_z$ are the number of mesh points in the $x$- and $z$-directions. Subsequently presented large-eddy simulation results were also ensemble averaged in time using at least two realizations initialized with different random perturbations sampled from the same distributions.

\begin{figure}
  \centering
  \includegraphics[width=0.48\textwidth]{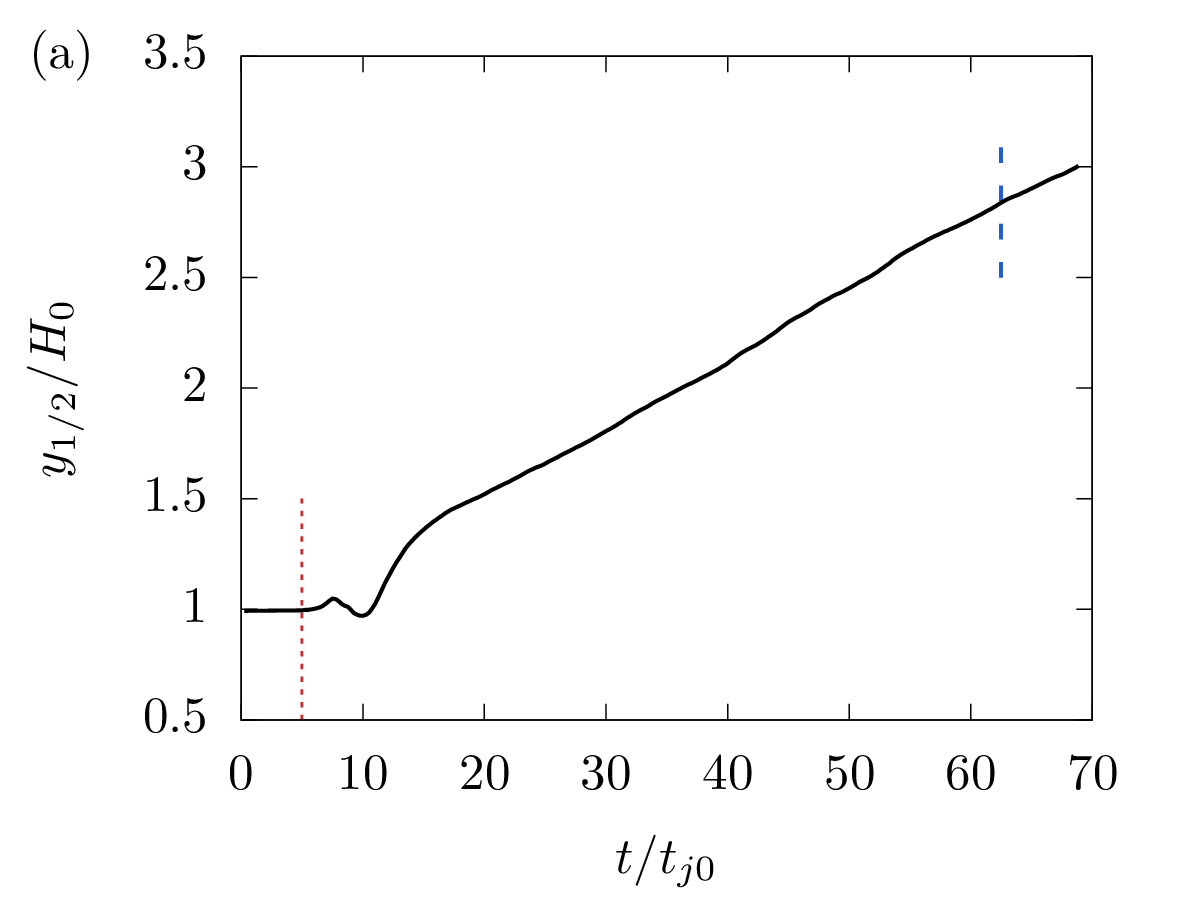}
  \includegraphics[width=0.48\textwidth]{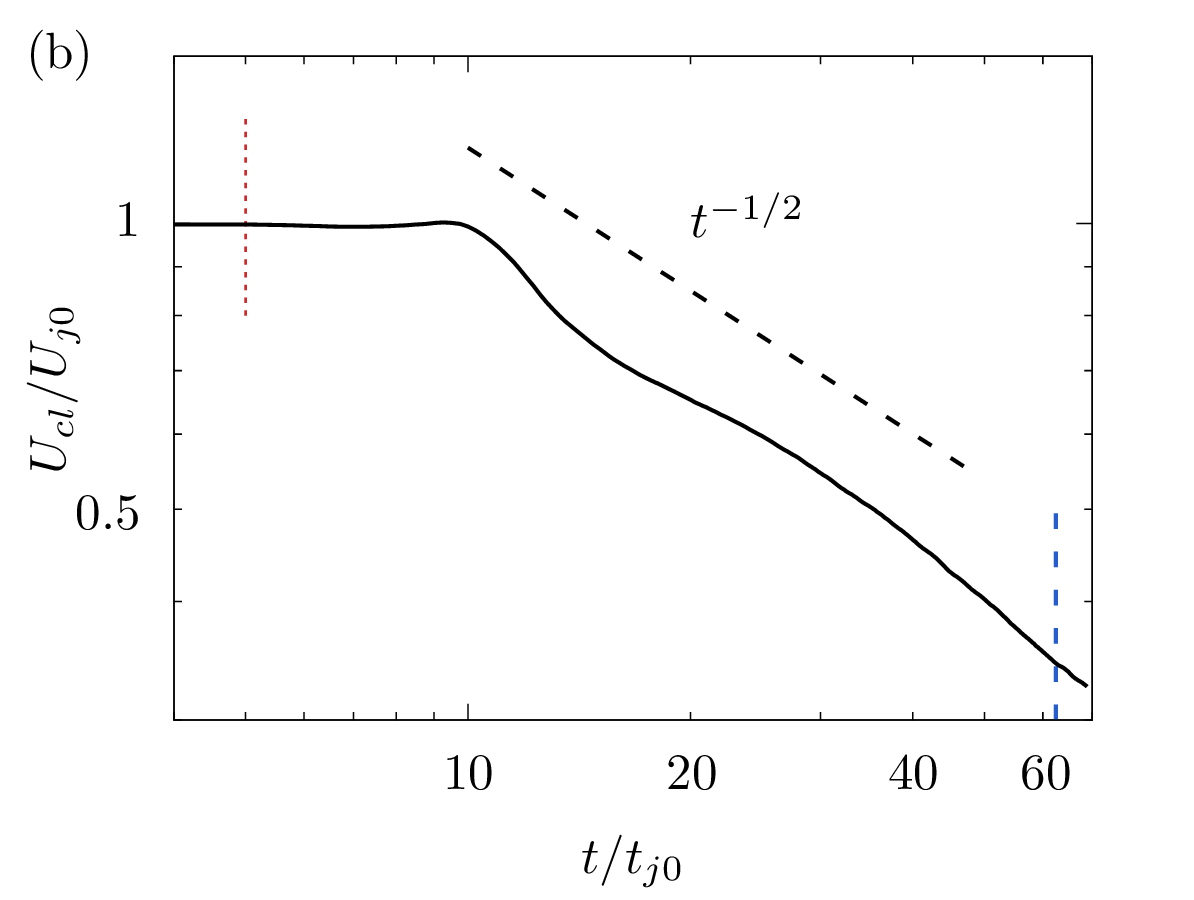}
  \caption{(a) Half-width and (b) centerline velocity of the case~A
    jet direct numerical simulation.  Subsequently discussed
    large-eddy simulations are initialized at $t =5 t_{j0}$, indicated
    by dotted lines.  Comparisons for this case are carried out at $t
    =62.5 t_{j0}$, indicated by dashed lines.}
  \label{fig:jet_evo}
\end{figure}

\begin{figure}
  \centering
  \includegraphics[width=1.0\textwidth]{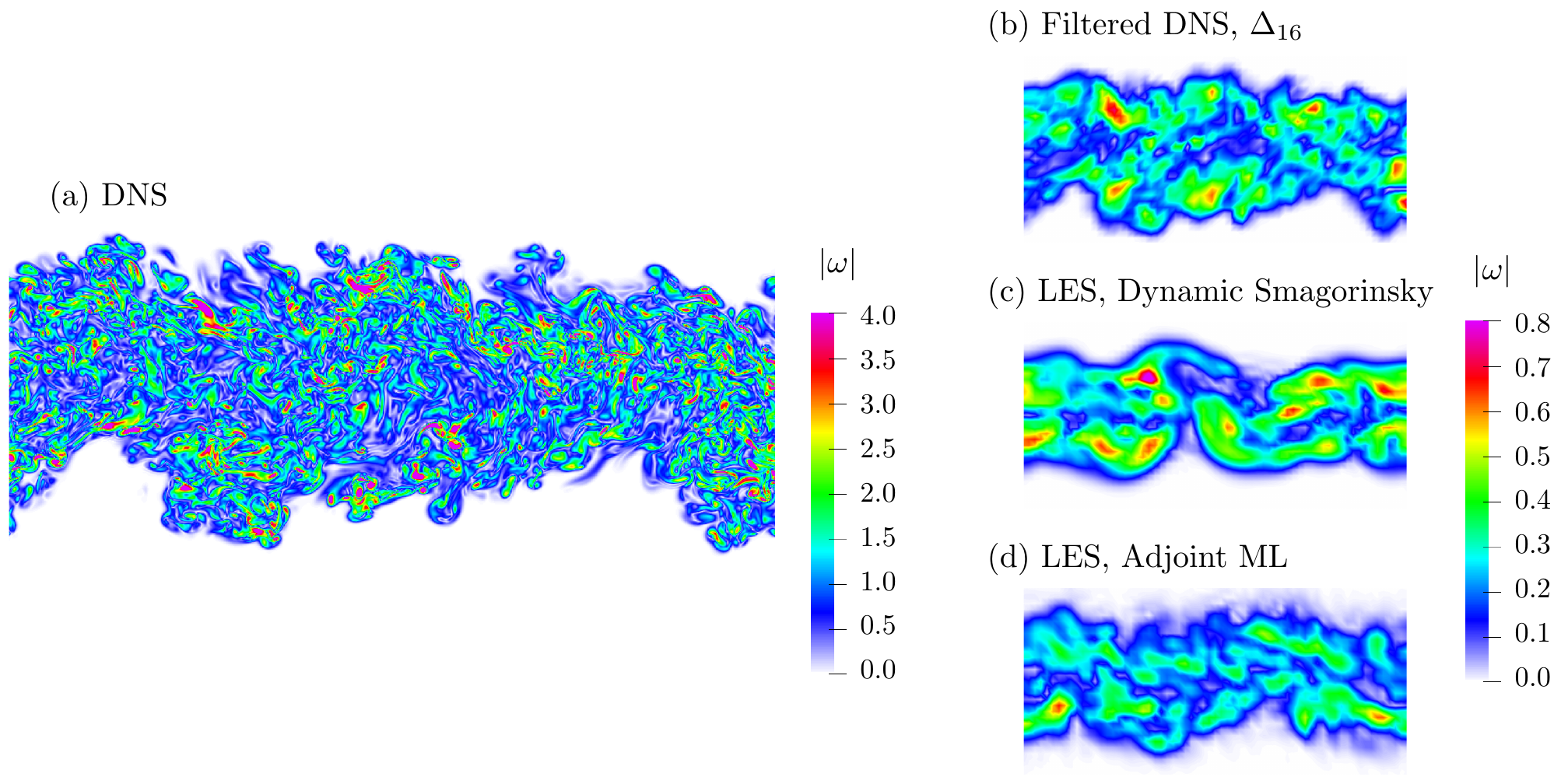}
  \caption{Vorticity magnitude at $z=0$ for $t=62.5 t_{j0}$: (a)
    direct numerical simulation case~A ($1024\times1280\times768$ mesh
    points); (b) the same field as (a) filtered using
    $\fDelta/\Delta_\DNSsub=16$ for an effective mesh of
    $64\times80\times48$ points; (c) large-eddy simulation using the
    dynamic Smagorinsky sub-grid-scale model, initialized at $t=5
    t_{j0}$ (see section~\ref{ss:jet_evo}), on $64\times80\times48$
    mesh points; and (d) the analogous $64\times80\times48$ simulation
    on using the adjoint-trained deep learning model
    (section~\ref{s:nn}).  }
  \label{fig:jet_vorticity}
\end{figure}

The evolving centerline velocity $U_{cl}(t)\equiv\avg{u}(0,t)$ and
half-width $y_{1/2}(t)$, defined by $\avg{u}(y_{1/2},t) =
U_{cl}(t)/2$, are shown in figure~\ref{fig:jet_evo} for the single jet (case~A).  There is an adjustment from the initial non-self-similar plug-like flow to an approximately linear spreading with $\sim t^{-1/2}$ centerline velocity decay for $t \gtrsim 15 t_{j0}$, $t_{j0}=H_0/U_{j0}$, consistent with an onset of similarity.  Spectra at $t=62.5t_{j0}$, such as are used subsequently for model evaluation, are shown in figure~\ref{fig:DNS_jet_spect}. The vorticity magnitude at $t=62.5 t_{j0}$, as visualized in figure~\ref{fig:jet_vorticity}~(a), shows a range of large- and small-scale turbulence.

The large-eddy simulations are initialized by filtering the direct
numerical simulation fields at $t=5t_{j0}$. As seen in
figure~\ref{fig:jet_evo}, this is prior to the period of approximate
self-similarity, which presumably increases the difficulty of matching
statistics at later times. This initialization is also more
representative of realistic LES, for which fully developed
filtered-DNS initial conditions are typically unavailable. Primary comparisons for the single-jet case are made at $t =62.5 t_{j0}$, which is approximately four times the time for first achieving approximate self-similarity based on the mean velocity.  At this time, the jet has grown to approximately 2.75 times its original width. As an example, the vorticity magnitude of filtered DNS data is compared to large-eddy simulations using the dynamic Smagorinsky and adjoint-trained machine learning model in figures~\ref{fig:jet_vorticity}~(b--d). On this coarse grid, the learned model, described in section~\ref{s:nn}, produces turbulence that qualitatively better matches the filtered direct numerical simulation.  Quantitative comparisons of these and other model predictions are presented in section~\ref{s:variants}.

\begin{figure}
  \centering
  \includegraphics[width=0.48\textwidth]{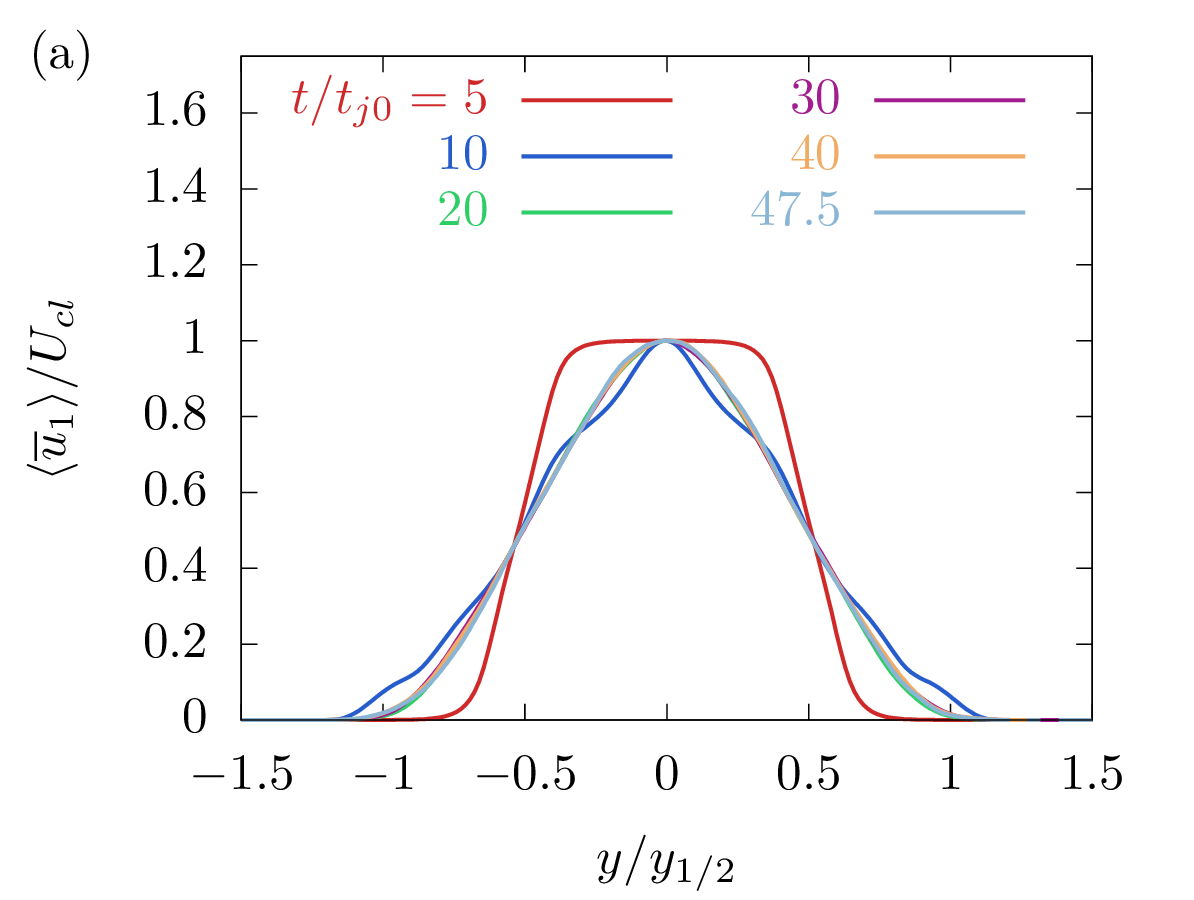}
  \includegraphics[width=0.48\textwidth]{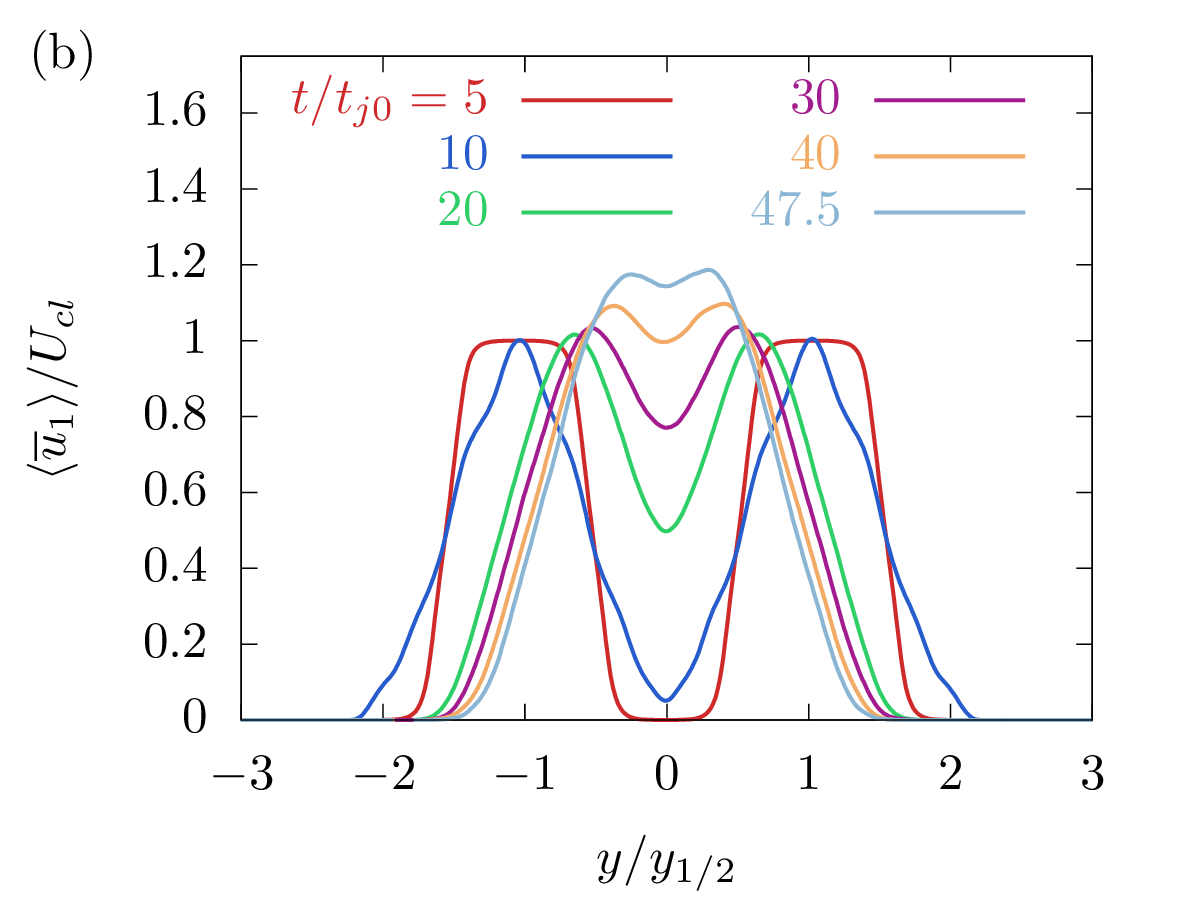}
  \includegraphics[width=0.48\textwidth]{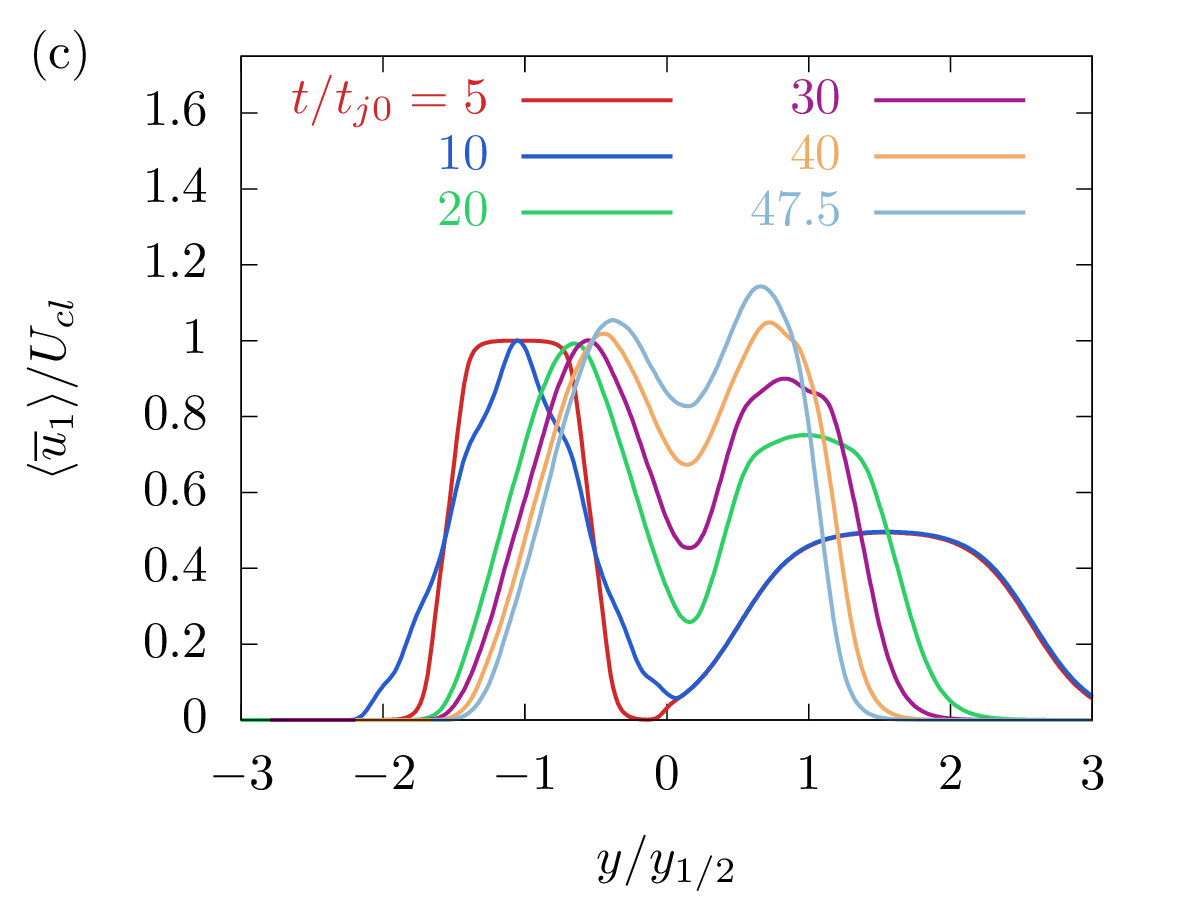}
  \includegraphics[width=0.48\textwidth]{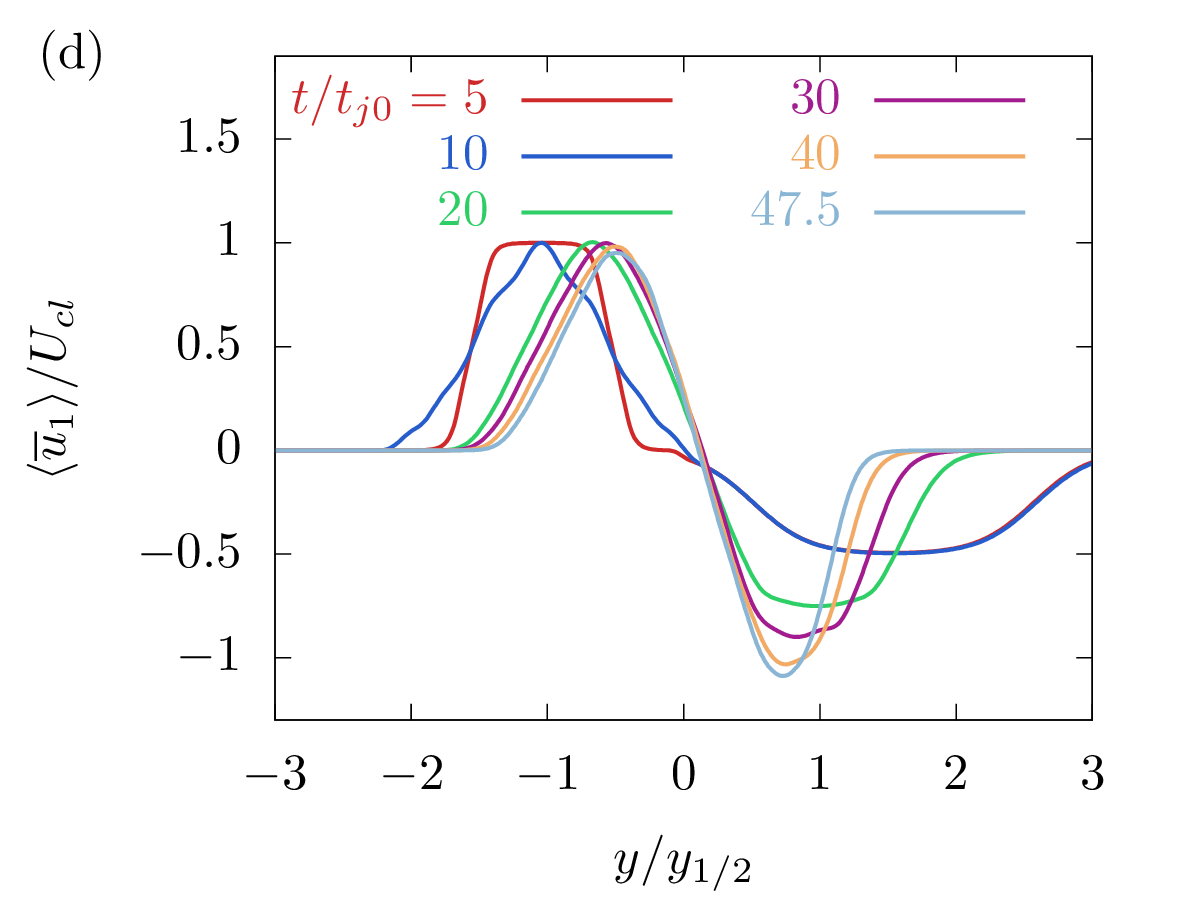}
  \caption{Mean streamwise velocity for (a) single jet (case~A), (b)
    symmetric parallel jets (case~B), (c) asymmetric parallel jets
    (case~C), and (d) asymmetric anti-parallel jets (case~D), all normalized by case~A similarity variables. Profiles are shown starting at $t=5 t_{j0}$. }
  \label{fig:jet_collapse}
\end{figure}

Mean velocities for all cases are shown in figure~\ref{fig:jet_collapse}. As expected, the single-jet profiles exhibit approximately self-similar evolution for $t \gtrsim 15 t_{j0}$. However, the dual-jet cases do not, so a successful model will need to reproduce a presumably more challenging non-self-similar evolution. Subsequent quantitative comparisons for the dual-jet cases are at $t=50t_{j0}$ (case~B) and $t=42.5t_{j0}$ (cases~C and D).

\subsection{Baseline sub-grid-scale models}
\label{ss:baseline_sgs}

The deep learning sub-grid-scale model is analyzed and compared primarily against a standard dynamic Smagorinsky model~\cite{Germano1991,Lilly1992}. Baseline performance of the dynamic Smagorinsky model is established by comparing with filtered direct numerical simulations. Since there is not explicit filtering during the large-eddy simulation, the nominal filter width and mesh resolution are equivalent: $\fDelta\equiv\Delta_N = N\Delta_\DNSsub$~\cite{Rogallo1984}. 

\begin{figure}
  \centering
  \includegraphics[width=0.5\textwidth]{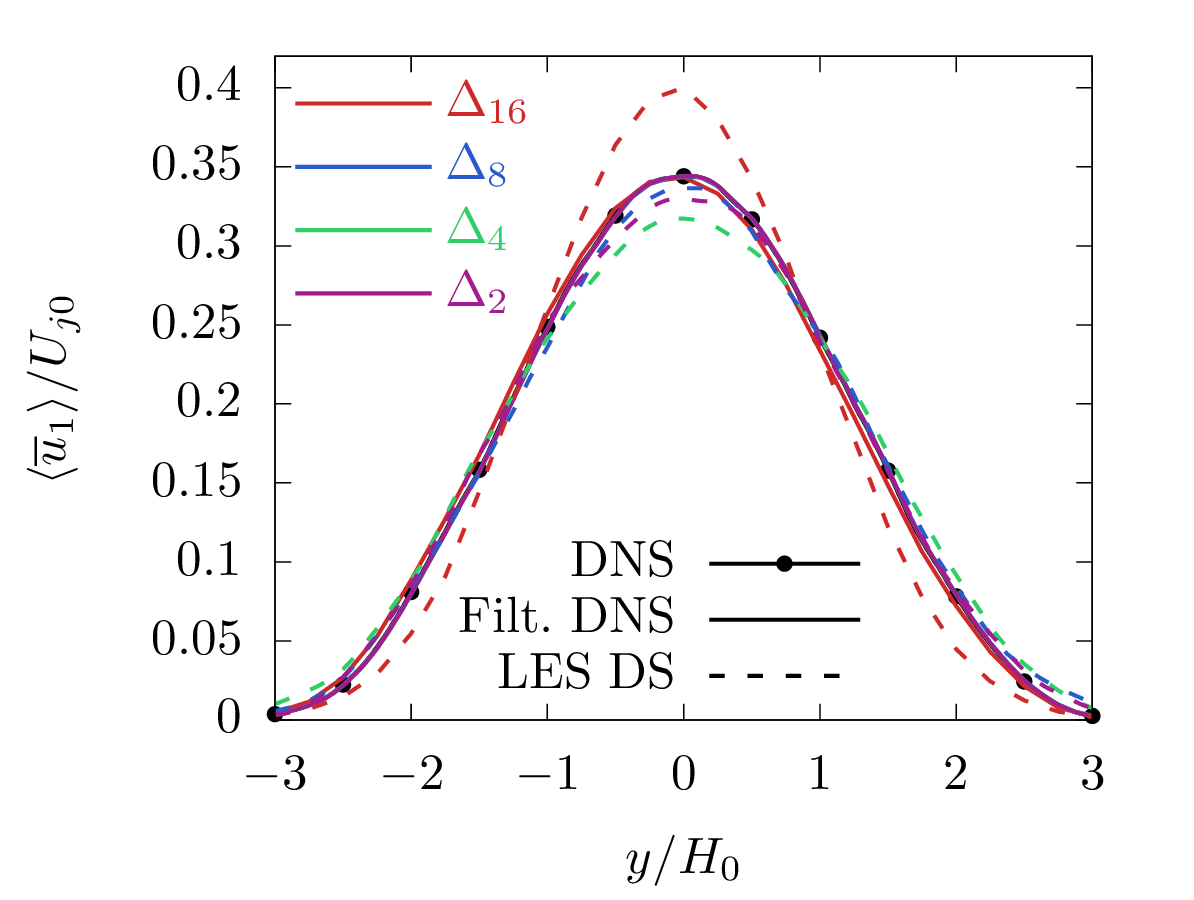}
  \caption{Direct numerical simulation, filtered DNS, and dynamic
    Smagorinsky (DS)\cite{Germano1991} large-eddy simulation mean
    streamwise velocity $\ol{u}$ for the single jet (case~A) at
    $t=62.5t_{j0}$ for filter sizes indicated: $\Delta_{16,8,4,2}$. }
  \label{fig:jet_vel_DS_only}
\end{figure}

Figure~\ref{fig:jet_vel_DS_only} shows mean streamwise velocity
predictions of the single-jet case~(A) for $\Delta_{N=16,8,4,2}$. The
$\Delta_{16}$ case overpredicts the centerline velocity and
underpredicts the jet spreading, though, as expected, agreement
improves with finer meshes $\Delta_{8,4,2}$. This increases the
separation between the dynamic-model test-filter scale and the large
turbulence scales, thereby improving the realism of the isotropy assumption. The DNS mean filtered velocity is also shown in figure~\ref{fig:jet_vel_DS_only}; this deviates only slightly from the unfiltered mean velocity.

Budgets of turbulence kinetic energy provide a measure of the role of the models' resolved-to-subgrid scale energy transfer versus viscous dissipation. Upon filtering, energy may be decomposed into resolved $E_f$ and residual (sub-grid-scale) $k_r$ components as
\begin{equation}
  \ol{E} \equiv \frac{1}{2}\ol{u_iu_i} = \underbrace{\frac{1}{2}\ol{u}_i\ol{u}_i }_{E_f} + \underbrace{\frac{1}{2}\left(\ol{u_iu_i} - \ol{u}_i\ol{u}_i\right)}_{k_r}.
\end{equation}
A gage of model importance can be inferred from the relative amounts viscous dissipation of $E_f$ by the resolved (filtered) velocity field 
\begin{equation}
\eps_f\equiv 2\nu\ol{S}_{ij}\ol{S}_{ij}\label{e:epsf}
\end{equation}
and the production rate of residual kinetic energy 
\begin{equation}
P_r\equiv -\tau_{ij}^r\ol{S}_{ij}.
\label{e:Pr}
\end{equation}
For both of these, $\ol{S}_{ij}$ is the filtered strain-rate. Both $\eps_f$ and $P_r$ represent sink terms when positive.

\begin{figure}
  \centering 
  \includegraphics[width=0.48\textwidth]{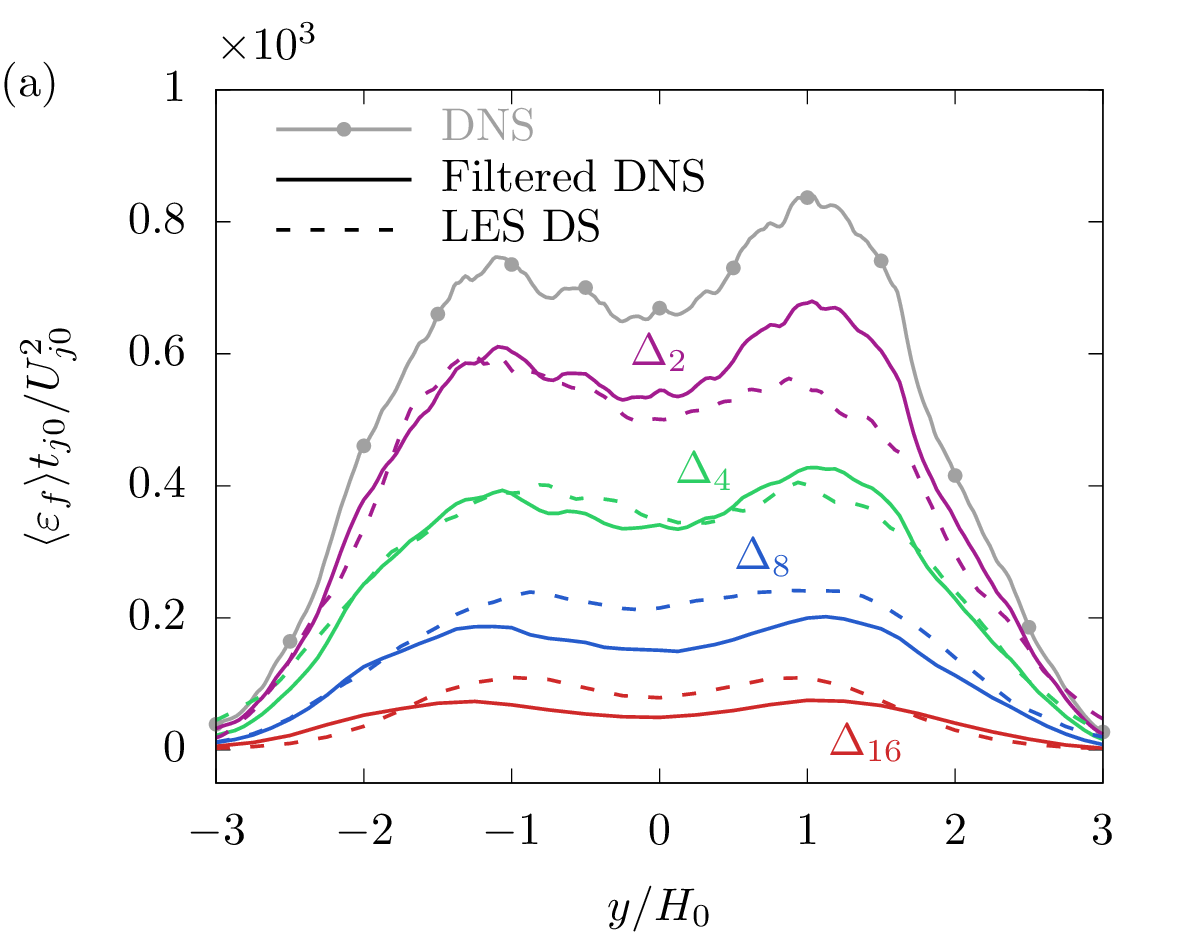}
  \includegraphics[width=0.48\textwidth]{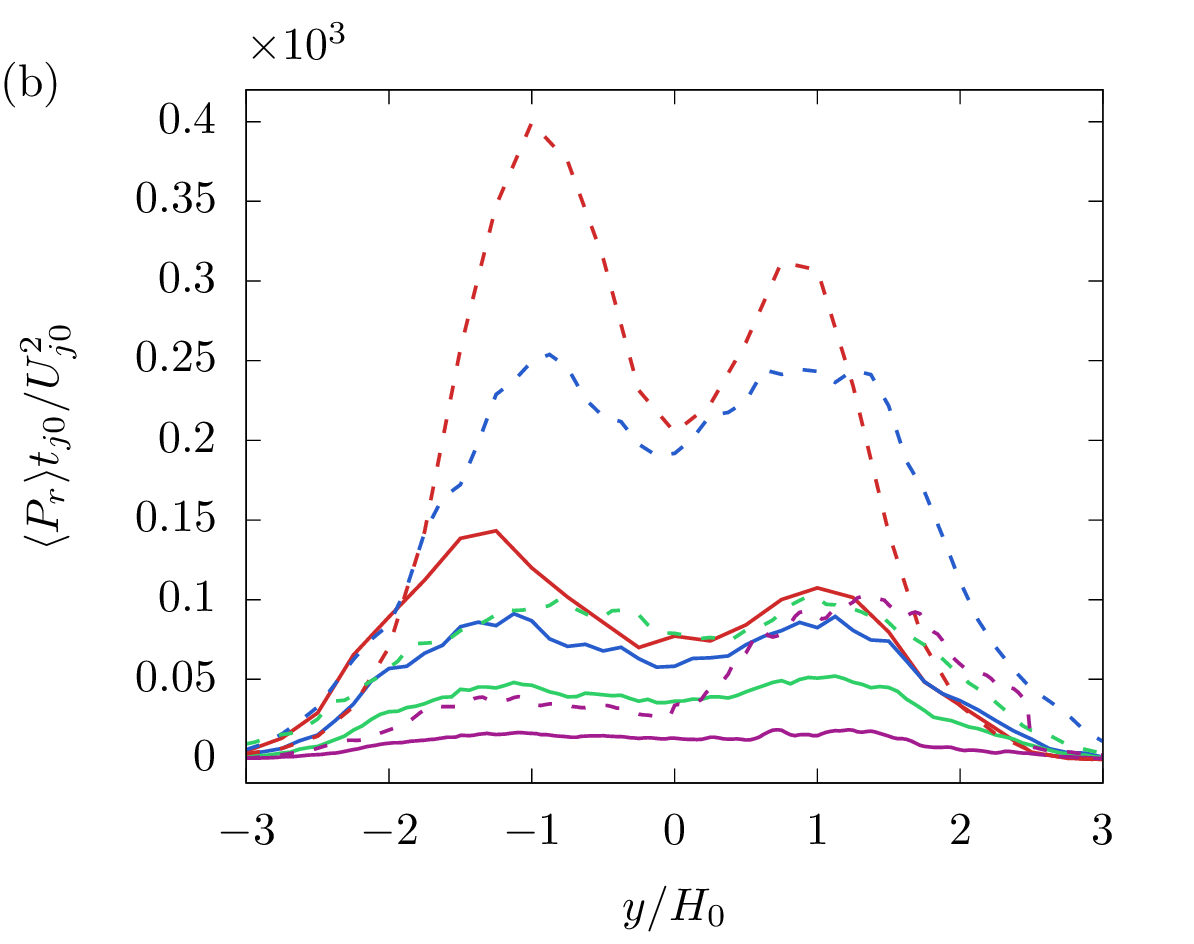}
  \caption{Evaluation of the dynamic Smagorinsky (DS) model for the
    baseline single jet (case~A):  (a) resolved viscous dissipation $\eps_f$ (\ref{e:epsf}) and (b) residual kinetic energy production rate $P_r$ (\ref{e:Pr}) at $t=62.5t_{j0}$ for filter sizes indicated.}
  \label{fig:jet_KE_DS_only}
\end{figure}

Figure~\ref{fig:jet_KE_DS_only} compares the large-eddy simulation and
filtered direct numerical simulation $\eps_f$ and $P_r$ for the
baseline jet. Ideally, $\eps_f$ and $P_r$ should match the filtered
direct numerical simulation values for the same filter size. As filter
size increases, $\eps_f$ is expected to decrease and $P_r$ to
increase, and both do exhibit these trends. However, the dynamic
Smagorinsky model increasingly overpredicts both $\eps_f$ and $P_r$
with increasing filter size. This means that there is excessive
removal of kinetic energy from the resolved scales.  For $\Delta_4$,
$P_r$ is a small fraction of $\eps_f$, so the model for $\tau_{ij}^r$
plays only a minor role. The $\Delta_4$ filtered $\eps_f^\DNSsub$ is
about half the unfiltered dissipation $\eps^\DNSsub$
(figure~\ref{fig:jet_KE_DS_only}~a), so it dominates the evolution of
the resolved kinetic energy.  However, the balance is different for $\Delta_{16}$.  Here, the sub-grid-scale production dominates the resolved dissipation, making the model essential. Such a challenge affords opportunities for improvement, so we focus on $\Delta_{16}$ for subsequent deep-learning modeling.

\section{Neural-network Model and Training} 
\label{s:nn}

The deep learning model architecture, hyperparameters, and training algorithm are described in this section. Section~\ref{ss:model_arch} presents the neural network model. Section~\ref{TrainingAlgorithm} discusses the training algorithm.

\subsection{Model architecture}
\label{ss:model_arch}

We use a deep neural network $\mathbf{G}(\vphi(\bx,t); \vtheta)$, where $\vphi$ represents space- and time-dependent input data from the running simulation, with the following architecture:
\begin{equation}
\begin{split}
  H^1 &= \sigma(W^1 \vphi + b^1)   \\
  H^2 &= \sigma(W^2 H^1 + b^2)    \\
  H^3 &=  G^1 \odot H^2 \qquad\qquad \text{with} \quad  G^1 = \sigma(W^5 z + b^5) \\
  H^4 &= \sigma(  W^3 H^3 + b^3)  \\
  H^5 &=   G^2 \odot H^4 \qquad\qquad \text{with} \quad G^2 = \sigma(W^6 z + b^6) \\
  \mathbf{G}(\vphi; \vtheta) &= W^4 H^5  + b^4 
\end{split}\label{e:NN}
\end{equation}
where $\sigma$ is a $\tanh()$ element-wise nonlinearity, $\odot$ denotes element-wise multiplication, and the parameters are $\theta = \{ W^1, W^2, W^3, W^4, W^5, W^6, b^1, b^2, b^3, b^4, b^5, b^6 \}$.  The gated units $G^1$ and $G^2$ are chosen for their general capability to represent stiff, nonlinear behavior. Each layer includes $N_\textsc{H} = 100$ hidden units, all initialized using a standard Xavier initialization~\cite{Xavier}. Inputs to the neural network are normalized by the same set of constants for all cases.  

In (\ref{e:LESmomentumNNintro}), $\bh(\bx,t; \vtheta) = \Psi(y) \mathbf{G}(\vphi(\bx,t)); \vtheta)$, 
where in the implementation $\Psi$ simply sets the neural network to zero at the $y = \pm H/2$ boundary mesh points. 

We note that the discrete stress tensor $\bh^{i,j,k}$ is not
constrained to be exactly symmetric on the staggered mesh at finite
resolution.  Similarly, the symmetric sub-grid-scale stress components
(\textit{e.g.}, $\tau^r_{12}$ versus $\tau^r_{21}$) appear in
different equations ($x$- versus $y$-momentum), which are not exactly
constrained by the same symmetries as the (unfiltered) Navier--Stokes
equations.\cite{Ghosal1999} Thus, further model flexibility in the
finite-resolution limit may be achieved by relaxing the symmetry
requirement. The same argument applies for corresponding tensor
invariants; they will not hold for finite resolution, so are not
imposed on the network. To be clear, as for any well-posed
sub-grid-scale model, symmetry and invariance properties will be
achieved with increasing resolution. Symmetry is revisited in
section~\ref{ss:symmetric} and invariants in section~\ref{ss:heffect}.

The output $h^{i,j,k}_{ll}$ is at the cell center $\bx = (i \Delta + \frac{\Delta}{2}, j \Delta + \frac{\Delta}{2}, k \Delta + \frac{\Delta}{2} )$. For $l \neq m$, the output $h^{i,j,k}_{lm}$ is at the cell corners
\begin{equation}
\bx =  
\begin{cases}
      (i \Delta , j \Delta, k \Delta + \frac{\Delta}{2} ), & \textrm{if} \phantom{...} (l,m) = (1,2)    \\
      (i \Delta , j \Delta + \frac{\Delta}{2}, k \Delta), & \textrm{if} \phantom{...} (l,m) = (1,3)    \\
      (i \Delta + \frac{\Delta}{2} , j \Delta, k \Delta  ), & \textrm{if} \phantom{...} (l,m) = (2,3) .   
\end{cases} 
\end{equation}
The derivatives $\frac{\partial h_{ij} }{\partial x_j}$ are then calculated at the cell faces using central differences with mesh-size $\Delta$, for example
\begin{align}
\frac{\partial h_{11} }{\partial x}(i \Delta , j \Delta + \tfrac{\Delta}{2}, k \Delta + \tfrac{\Delta}{2} ) &= \frac{h_{11}^{i,j,k} - h_{11}^{i-1,j,k}}{\Delta}, \notag \\
\frac{\partial h_{12} }{\partial y}(i \Delta , j \Delta + \tfrac{\Delta}{2}, k \Delta + \tfrac{\Delta}{2} ) &= \frac{h_{12}^{i,j+1,k} - h_{12}^{i,j,k}}{\Delta}, \notag \\
\frac{\partial h_{13} }{\partial y}(i \Delta , j \Delta + \tfrac{\Delta}{2}, k \Delta + \tfrac{\Delta}{2} ) &= \frac{h_{13}^{i,j,k+1} - h_{13}^{i,j,k}}{\Delta}. 
\end{align}

Models were implemented and tested using two different sets of input
variables. We first tested using $\vphi = (\ol{\bu},
\partial_{x_l}\ol{\bu}, \partial_{x_lx_l}\ol{\bu})$ as an input,
although this closure model is not Galilean invariant.  For the
simpler case of isotropic turbulence, for which this approach was
first tested,\cite{DPM-JCP} the model appeared to learn Galilean
invariance.  However, for the jets here, not surprisingly, it was
found to be unstable in \emph{a posteriori} simulations, at least for
the amount of training evaluated. Therefore, a Galilean invariant form
with $\vphi = ( \partial_{x_l}\ol{\bu}, \partial_{x_lx_l}\ol{\bu})$
was used here.

The input $\vphi$ to \eqref{e:NN} for a large-eddy simulation grid
cell $(i,j,k)$ includes the variables $\vphi^{i,j,k}$ for $(i,j,k)$ as
well as the 6 adjacent cells. Velocity-gradient inputs are calculated using the same stencils as the convective and viscous operators. Thus, on-diagonal derivatives $\olu_{l,l}$ are evaluated at cell centers, and off-diagonal derivatives $\olu_{l,m},\ m\neq l$ are evaluated at cell faces. As an example,
\begin{align}
\ol{u}_{1,x}(i \Delta + \tfrac{\Delta}{2} , j \Delta + \tfrac{\Delta}{2} , k \Delta + \tfrac{\Delta}{2} ) &= \frac{ \ol{u}_1(i \Delta + \Delta , j \Delta + \frac{\Delta}{2}, k \Delta + \frac{\Delta}{2} ) -  \ol{u}_1(i \Delta , j \Delta + \frac{\Delta}{2}, k \Delta + \frac{\Delta}{2} )  }{\Delta}, \notag \\
\ol{u}_{1,y}(i \Delta , j \Delta , k \Delta + \tfrac{\Delta}{2} ) &= \frac{ \ol{u}_1(i \Delta , j \Delta + \frac{\Delta}{2}, k \Delta + \frac{\Delta}{2} ) -  \ol{u}_1(i \Delta , j \Delta - \frac{\Delta}{2}, k \Delta + \frac{\Delta}{2} )  }{\Delta}, \notag \\
\ol{u}_{1,z}(i \Delta , j \Delta  + \tfrac{\Delta}{2}, k \Delta ) &= \frac{ \ol{u}_1(i \Delta , j \Delta + \tfrac{\Delta}{2}, k \Delta + \frac{\Delta}{2} ) -  \ol{u}_1(i \Delta , j \Delta + \tfrac{\Delta}{2}, k \Delta - \tfrac{\Delta}{2} )  }{\Delta}. 
\end{align}

\subsection{Training algorithm} \label{TrainingAlgorithm}

A stochastic gradient descent-type scheme is employed, in which the adjoint partial differential equation is solved on random time sub-intervals $m = 1,\ldots, M$ during each optimization iteration, each of which can be evaluated in parallel. The training algorithm is summarized below. We first consider the case in which we seek to minimize \eqref{e:Lindirect} as discussed in section~\ref{Intro} with $F(\olbu) = \olbu(\bx,t)$. The extension to training on mean statistics is described in section~\ref{ss:meanflow}.

\begin{itemize}
\item Initialize the parameters $\vtheta^0$ with samples from the Xavier distribution \cite{Xavier}.
\item For training iteration $k = 1, 2, \ldots, K$:
\begin{itemize}[label={$\circ$}]

\item Each sub-iteration $m = 1, \ldots, M$ selects a random time $t^m$; the corresponding filtered direct numerical simulation data for $[t^m, t^m + T_s]$ is taken as the training target: $F(\olbu_e)=\olbu^\DNSsub$.

\item Equations (\ref{e:LESmomentumNNintro}) and (\ref{e:LESmassNNintro}) are solved over $[t^m, t^m + T_s]$ with the filtered direct numerical simulation data at $t^m$  projected onto a divergence-free manifold as the initial condition.\cite{DPM-JCP}

\item For each $m$, the adjoint is solved on $[t^m + T_s,t^m]$ with objective function, rewritten from \eqref{e:Lindirect},
\begin{equation}
  L^m(\theta) = \int_{\Omega} \norm{\ol u^\DNSsub(t^m + T_s,x)- \ol{u}(t^m + T_s,x)}\, d\bx.
  \label{e:instObjFn}
\end{equation}
\item The adjoint solution provides the gradient $\nabla_{\vtheta} L^m(\theta)$ for $m = 1, \ldots, M$. 
\item These $M$ gradients are averaged as $G = \displaystyle \frac{1}{M} \sum_{m=1}^M \nabla_{\theta} L^L(\theta)$. 
\item The parameter $\vtheta$ for the deep learning large-eddy simulation model is updated using the RMSprop algorithm\cite{Goodfellow} with the gradient $G$:
\begin{align}
\vr^{k+1} &= \rho \vr^{k} + (1 - \rho) G \odot G, \notag \\
\vtheta^{k+1} &= \vtheta^{k} - \frac{ \alpha^{k} }{ \sqrt{\vr^{k+1}} + \epsilon} \odot G,
\end{align}
where $\alpha^{k}$ is the learning rate, and we take $\rho = 0.99$ and $\epsilon = 10^{-8}$.  The $\sqrt{\vr^{\ell+1}}$ operation is element-wise. The learning rate $\alpha^{k}$ is decreased by a factor of $\frac{1}{2}$ every $K$ iterations.
\end{itemize}

\end{itemize}
Taking $K = 250$ and $M = 12$ was found to be effective, though of
course the optimal choice of such hyperparameters will generally
depend upon the application and the model architecture.  The time
segments used had $T_s = 5\Delta t_\LESsub$, with the large-eddy
simulation time step $\Delta t_\LESsub=10\Delta t_\DNSsub$.

\subsection{Training on the Reynolds-averaged statistics}
\label{ss:meanflow}

As discussed in section~\ref{Intro}, direct numerical simulation data,
or correspondingly resolved experimental measurements, are not
available in all cases.  Therefore, as a relevant example of limited
availability of data, we train a model only on the mean velocity and
resolved Reynolds stresses: $F(\ol{\bu}) =
(\avg{\olbu};\avg{{\olbu'\olbu'}})(y,t)$. We obtain target data
$F(\olbu_e) = (\avg{\olbu^\DNSsub};\avg{(\olbu'\olbu')^\DNSsub})$ from
the filtered case~A, though in general measured data may be
substituted without modification. In this case, the objective function
becomes
\begin{equation}
L(\theta) =  \frac{1}{L_y} \frac{1}{(T_2 - T_1)} \int_{T_1}^{T_2}  \int_0^{L_y} \sum_{i=1}^3  \frac{c_1}{U_{j0}^2}  \bigg( \avg{ \olu_i^{\DNSsub} - \olu_i }^2 + 
\frac{c_1}{U_{j0}^2} \sum_{j=i}^3\avg{\olu_{i}^{\DNSsub} \olu_{j}^{\DNSsub}  - \olu_i   \olu_j }^2\bigg)\; dy dt,
\label{AverageObjFunction}
\end{equation}
with $c_1 = U_{j0}^2 = 467.4$ found to be effective for the current
application.  Training this objective function is
computationally more challenging than with \eqref{e:instObjFn}, since
\eqref{AverageObjFunction} requires solving the flow and its adjoint
on the full $t\in [T_1=5,T_2=67.5]t_{j0}$ case~A time horizon for each
optimization iteration. Despite solving over a relatively long time
length, the adjoint equation remains well-behaved.

%
\section{Model Predictions}
\label{s:variants}

The objective functions, training data, and testing data for all cases
are listed in table~\ref{tab:traintest}. We start in
section~\ref{s:apriori} with an \textit{a priori} $\tau_{ij}^\SGSsub$
model, trained directly based  on the sub-grid-scale stress mismatch \eqref{e:Ldirect}.  This simpler approach is demonstrated to be poor compared with the adjoint-based training in the following sections: in-sample results for the baseline case~A jet (section~\ref{s:geo_jet}) and out-of-sample analysis of cases~B, C, and D (section~\ref{s:geo_doublejet}). The performance for the averaged objective function (\ref{AverageObjFunction}) is assessed in section~\ref{ss:mean_flow}.

\begin{table}
  \begin{center}
    \begin{tabular}{cccc}\toprule
      \textsc{Objective Function} ($F$) & \textsc{Training} & \textsc{Testing} & \textsc{Section} \\\midrule
      
      $\tau_{ij}^\SGSsub(\olbu)$
      &\multirow{2}{*}{Jet (\ref{e:jetic})} & \multirow{2}{*}{Jet (\ref{e:jetic})}  &
      \multirow{2}{*}{\S\ref{s:apriori}} \\
      (\footnotesize\textit{a priori} ML) \\
      \midrule
      
      \multirow{3}{*}{$F(\ol{\bu}) = \ol{\bu}(\bx,t)$}
      &\multirow{4}{*}{Jet (\ref{e:jetic})} & Jet (\ref{e:jetic})  & \S\ref{s:geo_jet} \\
      \multirow{3}{*}{\footnotesize (filtered velocity)}
      &  & Dual jets (\ref{e:dualjetic}) & \S\ref{s:geo_doublejet} \\
      &  & Dual asymmetric (\ref{e:dualjetic})&  \S\ref{s:geo_doublejet}\\
      &  & Dual anti-parallel (\ref{e:dualjetic}) & \S\ref{s:geo_doublejet} \\ \midrule

      \multirow{3}{*}{$F(\ol{\bu}) = (\avg{\olbu};\avg{{\olbu'\olbu'}})_{(y,t)}$}
      & \multirow{4}{*}{Jet (\ref{e:jetic})}  & Jet (\ref{e:jetic})  & \multirow{4}{*}{\S\ref{ss:mean_flow}} \\
      \multirow{3}{*}{\footnotesize ($x$ and $z$ averaged)}
      &  & Dual jets (\ref{e:dualjetic}) &\\
      & & Dual asymmetric (\ref{e:dualjetic})&  \\
      &  & Dual anti-parallel (\ref{e:dualjetic}) & 
      \\
\midrule
     \multirow{3}{*}{$F(\ol{\bu}) = \ol{\bu}(\bx,t)$}
      & \multirow{4}{*}{Jet (\ref{e:jetic})}  & Jet (\ref{e:jetic})  & \multirow{4}{*}{\S\ref{ss:symmetric}} \\
      \multirow{3}{*}{\footnotesize (filtered velocity; $h_{ij}=h_{ji}$)}
      &  & Dual jets (\ref{e:dualjetic}) & \\
      & & Dual asymmetric (\ref{e:dualjetic})&  \\
      &  & Dual anti-parallel (\ref{e:dualjetic}) & 

      \\\bottomrule
    \end{tabular}
    \caption{Objective functions, training data, and testing data for
      primary cases studied.}\label{tab:traintest}
  \end{center}
\end{table}

\subsection{\textit{A priori} training}
\label{s:apriori}
\begin{figure}
  \centering
  \includegraphics[width=0.48\textwidth]{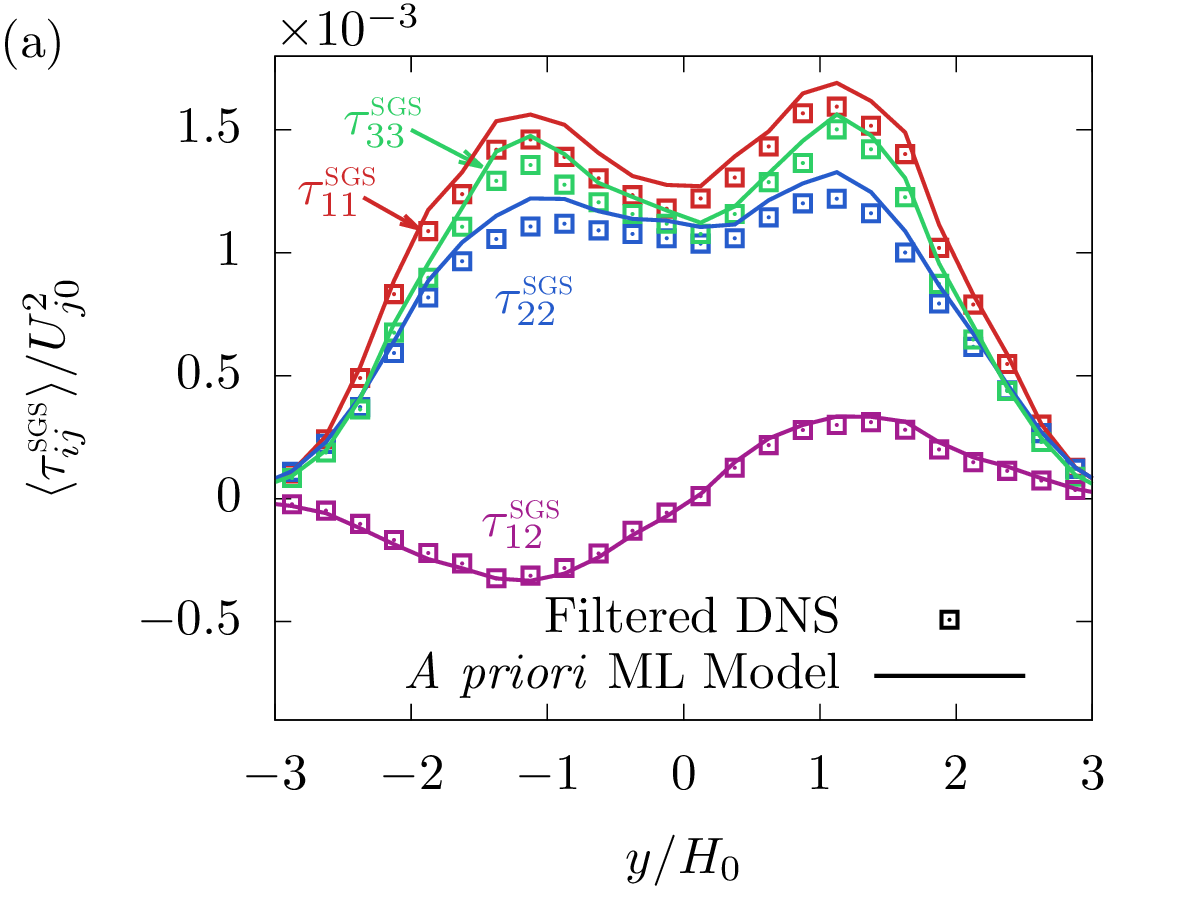}
  \includegraphics[width=0.48\textwidth]{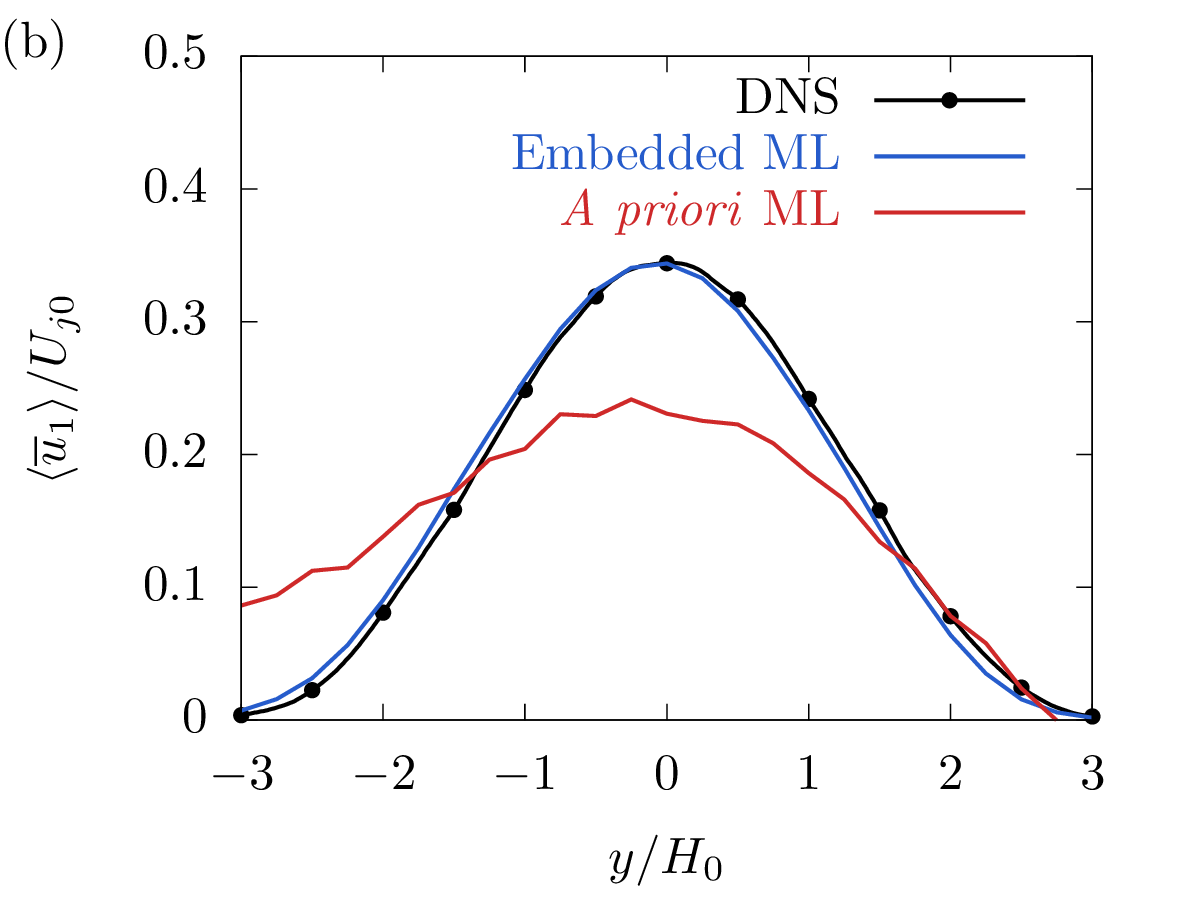}
  \caption{The \textit{a priori}-trained deep learning model for case~A for $\Delta_{16}$ at $t=62.5 t_{j0}$: (a) residual stress agreement from filtered direct numerical simulations and the \textit{a priori}-trained model, and (b) predicted mean streamwise velocity.}
  \label{fig:apriori}
\end{figure}

The model of section~\ref{ss:model_arch} is trained to minimize the
\textit{a priori} mismatch between the modeled and filtered direct
numerical simulation $\tau_{ij}^\SGSsub$. This is the simple, direct
approach:  training requires only 5.5~hours on 24~Nvidia K20X
GPU nodes. The \textit{a priori} $\tau_{ij}^\SGSsub$ agreement in
figure~\ref{fig:apriori}~(a) is good, as expected since deep neural
networks are well-understood to fit data well. However, this fit is
unconstrained by the how the model interacts with the governing
equations, so it is unsurprising that the \textit{a posteriori}
performance in a large-eddy simulation is poor
(figure~\ref{fig:apriori} b), even for the mean velocity. Simply
minimizing the mismatch is insufficient, even for this relatively
simple free-shear flow.  In figure~\ref{fig:apriori}~(b), the
adjoint-trained neural network with $F(\olbu)=\olbu$, which is
discussed subsequently, performs far better for the same network
model.

\subsection{Single-jet case (in-sample)}
\label{s:geo_jet}

\begin{figure}
  \centering
  \includegraphics[width=0.48\textwidth]{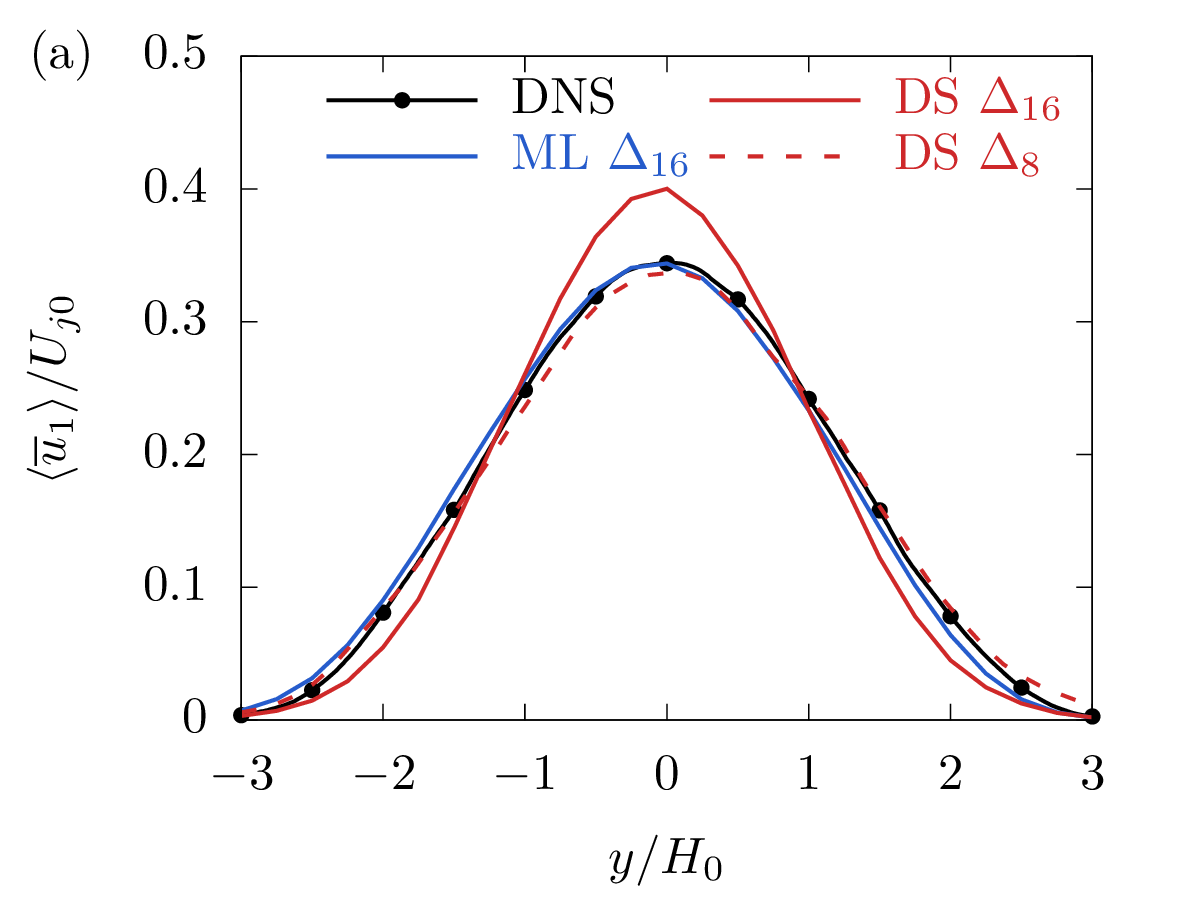}
  \includegraphics[width=0.48\textwidth]{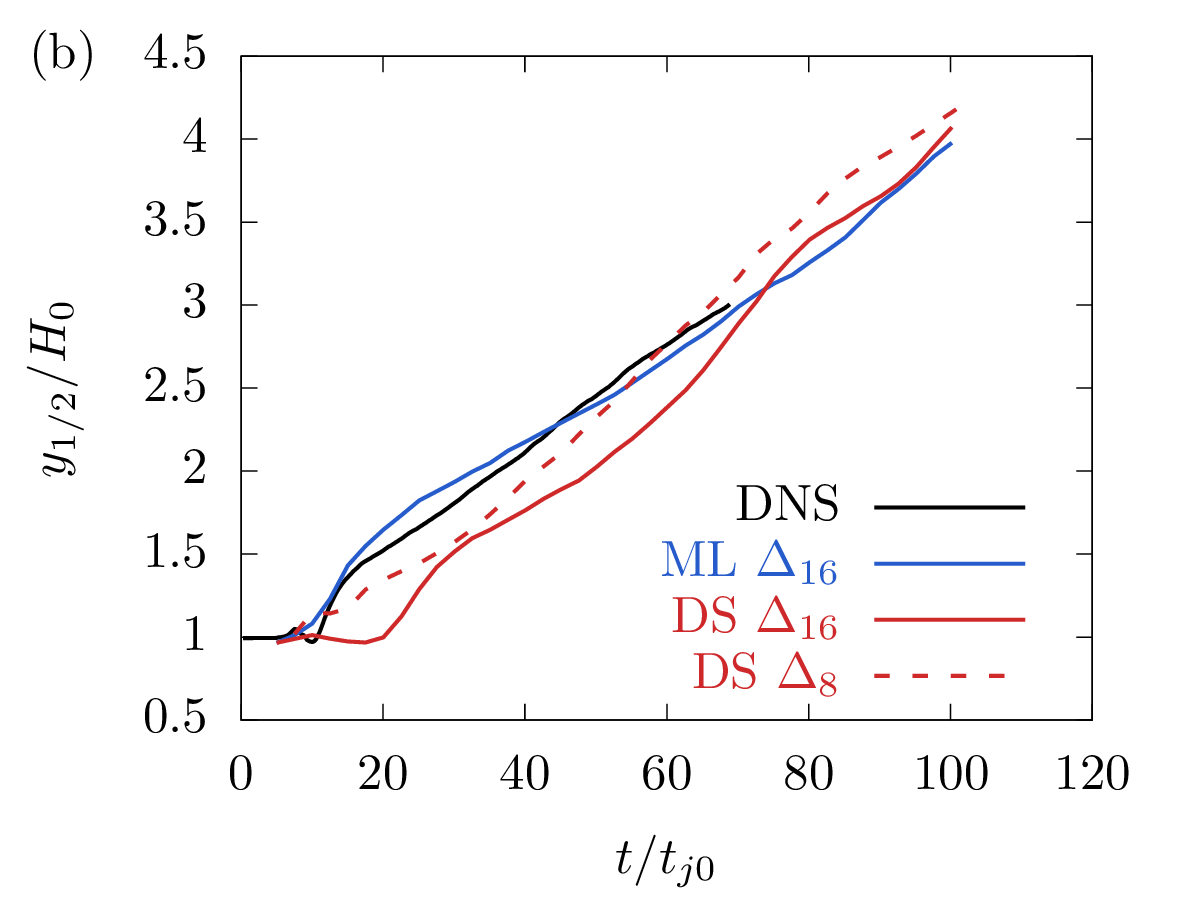}
  \caption{Single jet (case~A) in-sample comparison for learning (ML)
    and dynamics Smagorinsky (DS) models:  (a) mean
    streamwise velocity $\ol{u}_1$ at $t=62.5t_{j0}$ and (b) half-width $y_{1/2}$ evolution for the indicated filter sizes.  The direct numerical simulation data are included for comparison.}
  \label{fig:jet_vel}
\end{figure}

In this case, the model is trained on filtered instantaneous velocity fields ($F(\olbu)=\olbu(\bx,t)$) from case~A and tested on the same configuration. In figure~\ref{fig:jet_vel}~(a), we see that for $\Delta_{16}$ the deep learning model nearly perfectly reproduces the mean velocity, doing better than the dynamic Smagorinsky (DS) model for the coarse $\Delta_{16}$ case and comparably to it for $\Delta_8$.  In figure~\ref{fig:jet_vel}~(b), it also better tracks the evolution of $y_{1/2}$:  for both $\Delta_8$ and $\Delta_{16}$, the dynamic Smagorinsky model underpredicts jet spreading at early times. 

\begin{figure}
  \centering
  \includegraphics[width=0.48\textwidth]{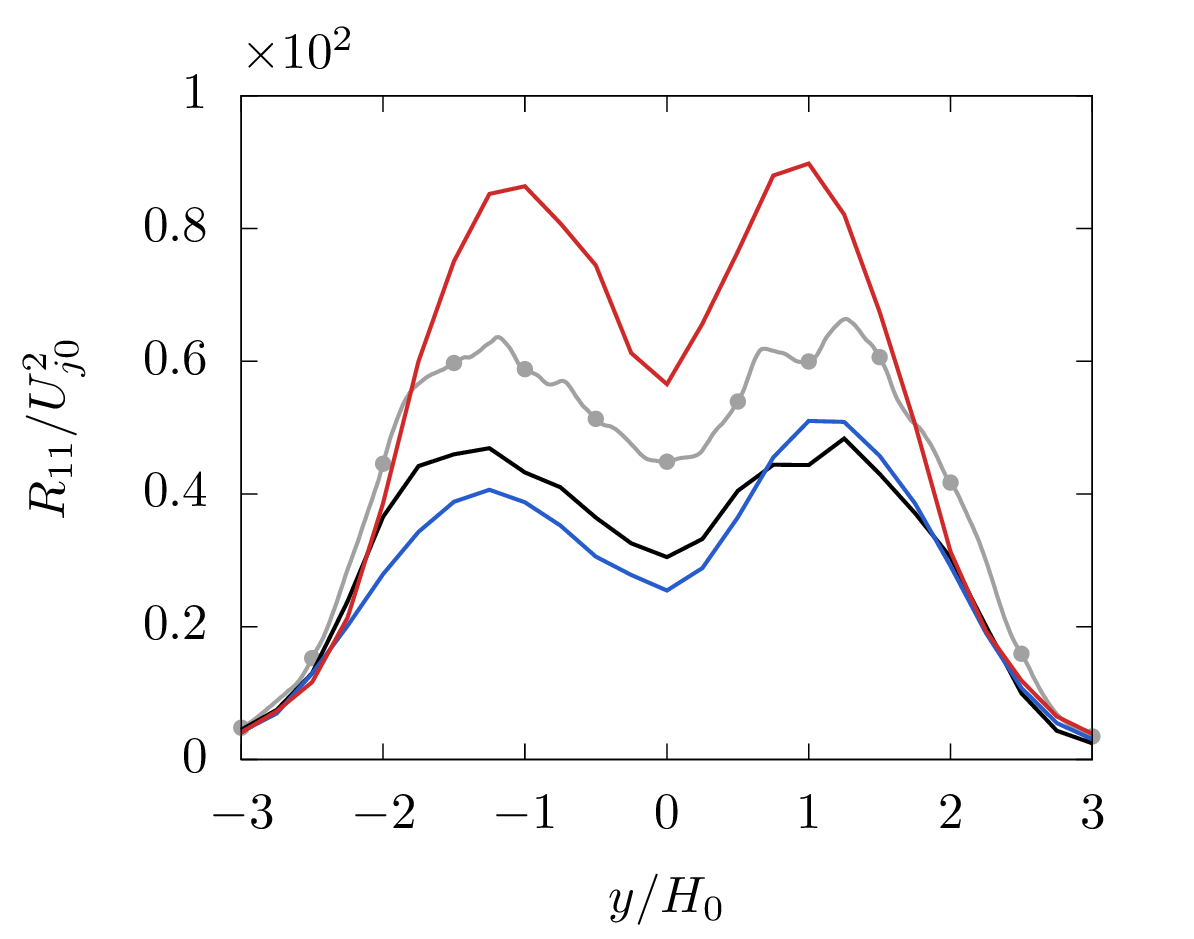}
  \includegraphics[width=0.48\textwidth]{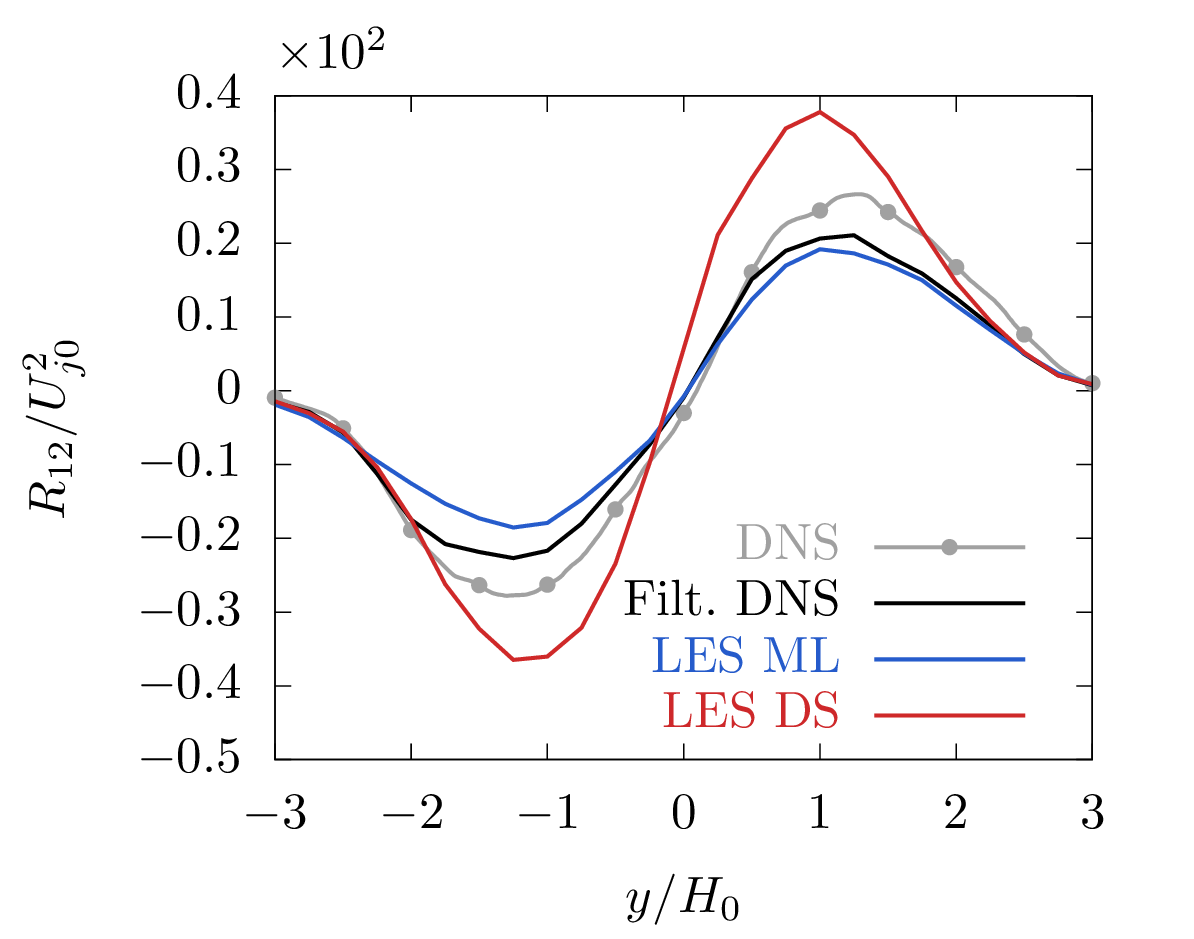}
  \includegraphics[width=0.48\textwidth]{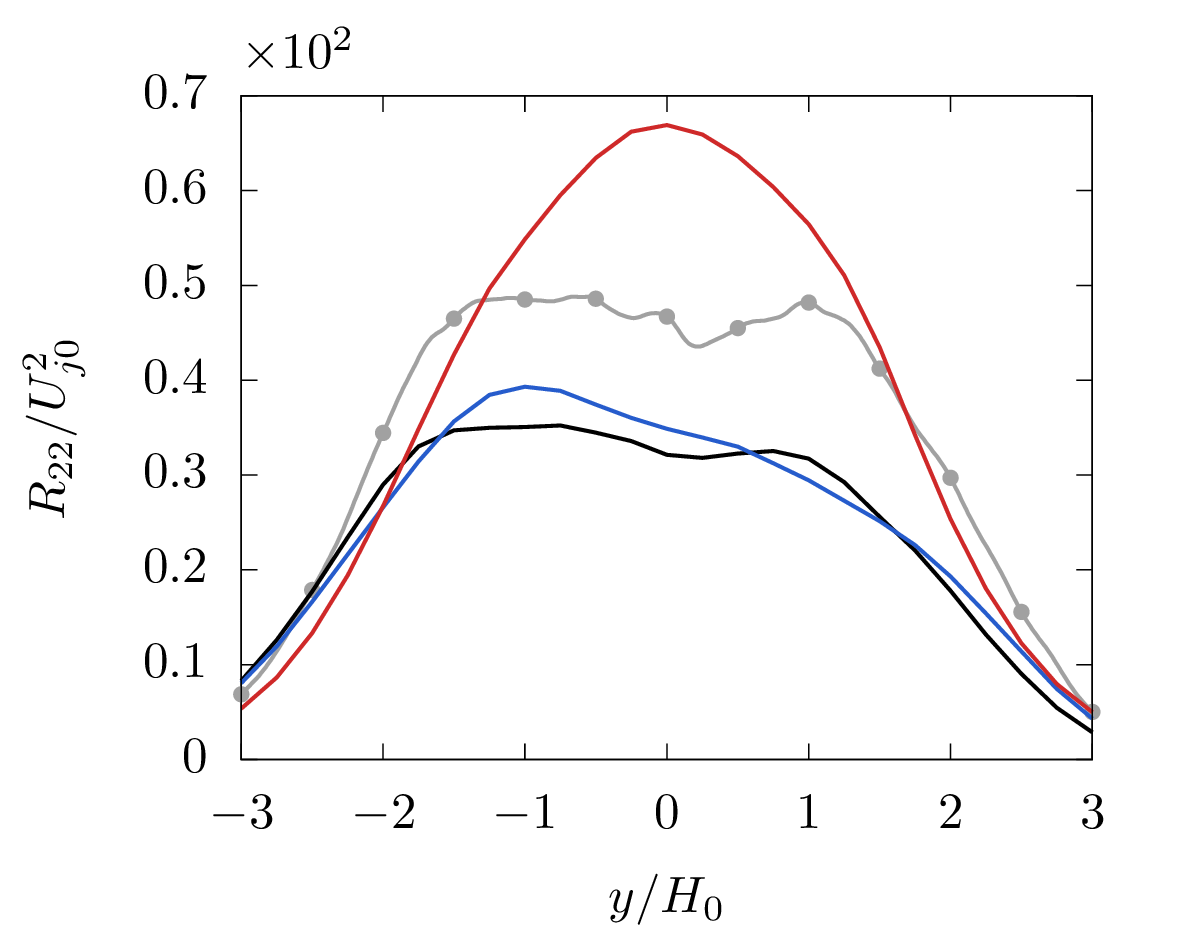}
  \includegraphics[width=0.48\textwidth]{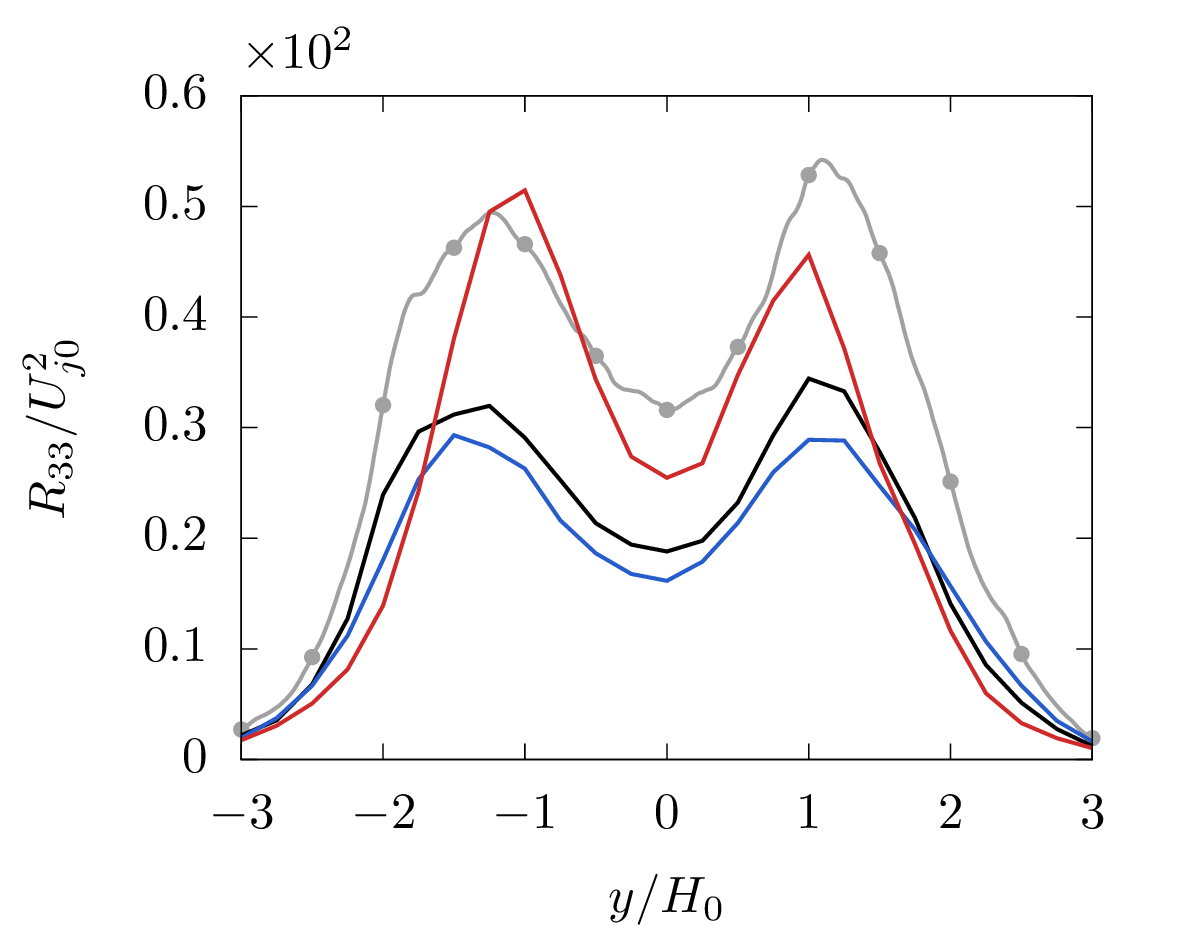}
  \caption{Resolved Reynolds stress components as labeled in the single-jet case~A at $t=62.5t_{j0}$ for $\Delta_{16}$:  the new learning model (ML), dynamic Smagorinsky (DS), and the filtered direct numerical simulation (DNS).   The unfiltered Reynolds stress components are also shown for reference.}
  \label{fig:jet_R}
\end{figure}

\begin{figure}
  \centering
  \includegraphics[width=0.48\textwidth]{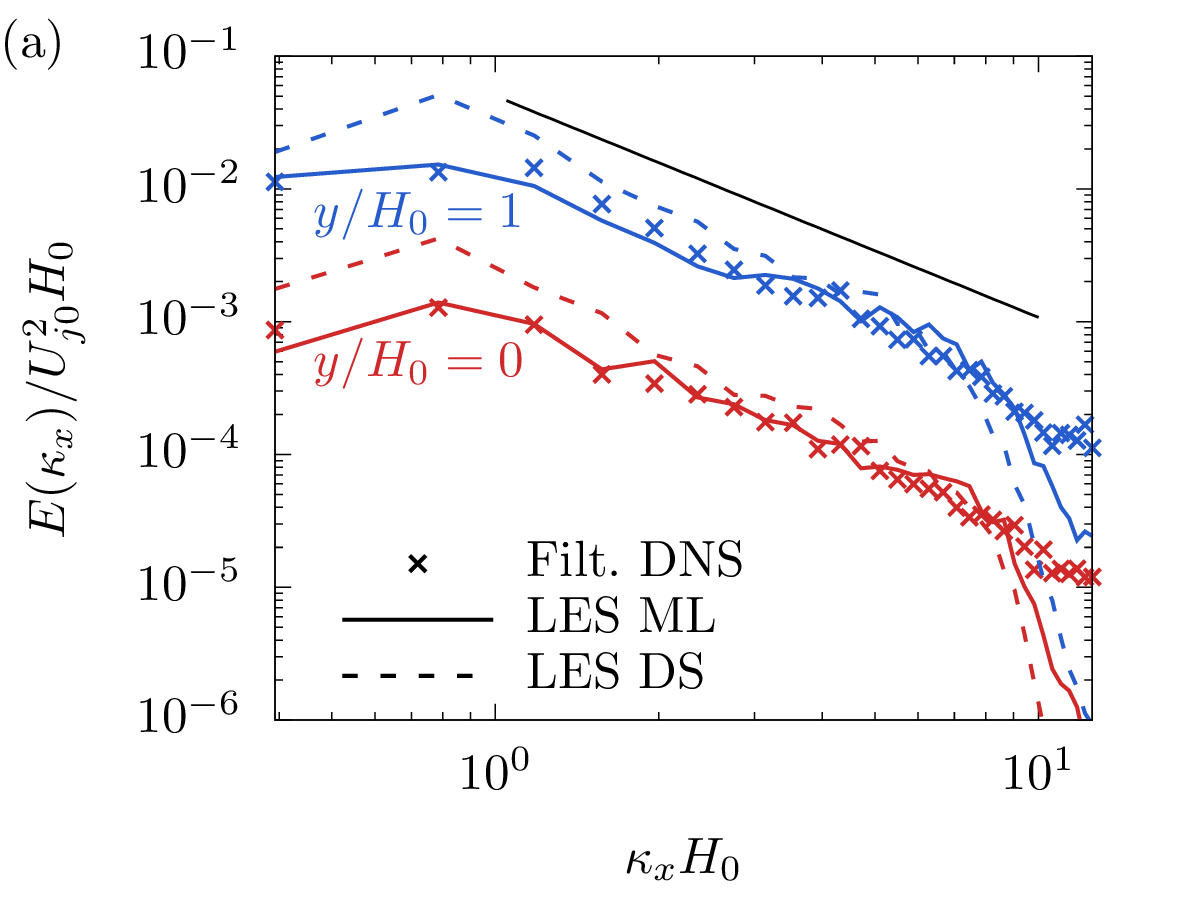}
  \includegraphics[width=0.48\textwidth]{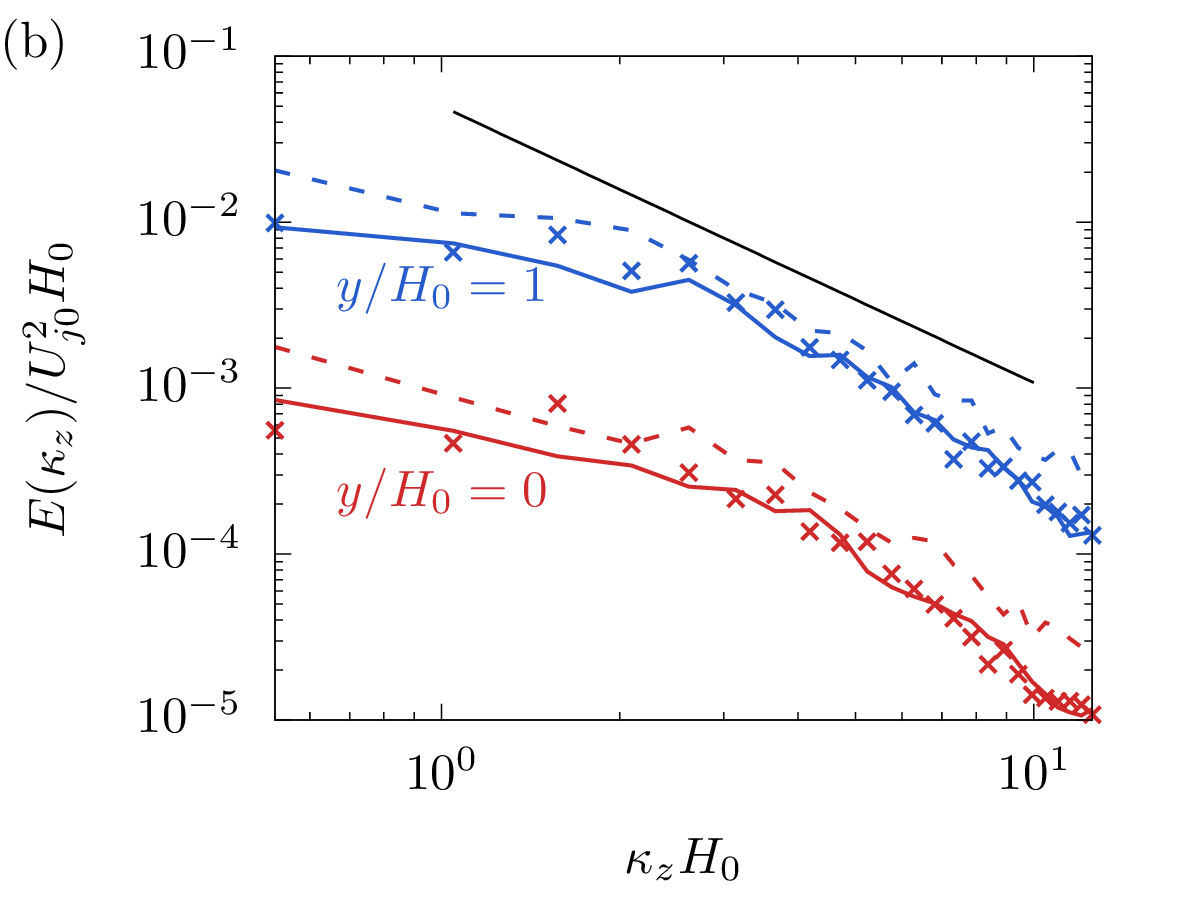}
  \caption{(a) Streamwise and (b) spanwise energy spectra for the single-jet case~A with $\Delta_{16}$ at $t=62.5t_{j0}$: filtered (DNS), machine learning model (ML), and dynamic Smagorinsky (DS). The straight lines have $\kappa^{-5/3}$ slope. Spectra are shown at the centerline ($y/H_0=0$) and at $y/H_0=1$, which corresponds to the cross-stream location of maximum resolved kinetic energy is shown offset by a factor of 10.
}
  \label{fig:jet_spect}
\end{figure}

Resolved Reynolds stress components $R_{ij}\equiv \avg{\left(\ol{u}_i-\avg{\ol{u}_i}\right)\left(\ol{u}_j-\avg{\ol{u}_j}\right)}$ are compared in figure~\ref{fig:jet_R}.  As expected, based on the mean-flow evolution, the trained model significantly outperforms dynamic Smagorinsky at this coarse resolution.  The complete Reynolds stress from the direct numerical simulation, prior to filtering to match the large-eddy simulation model, is also shown to quantify the sub-grid-scale contribution to the turbulence stresses. One-dimensional energy spectra in figure~\ref{fig:jet_spect} confirm that the deep learning model more accurately reproduces the spectral energy distribution than the dynamic Smagorinsky model, and this is particularly evident at high wavenumbers.

\subsection{Dual-jet cases (out-of-sample)}
\label{s:geo_doublejet}

\begin{figure}
  \centering
  \includegraphics[width=0.45\textwidth]{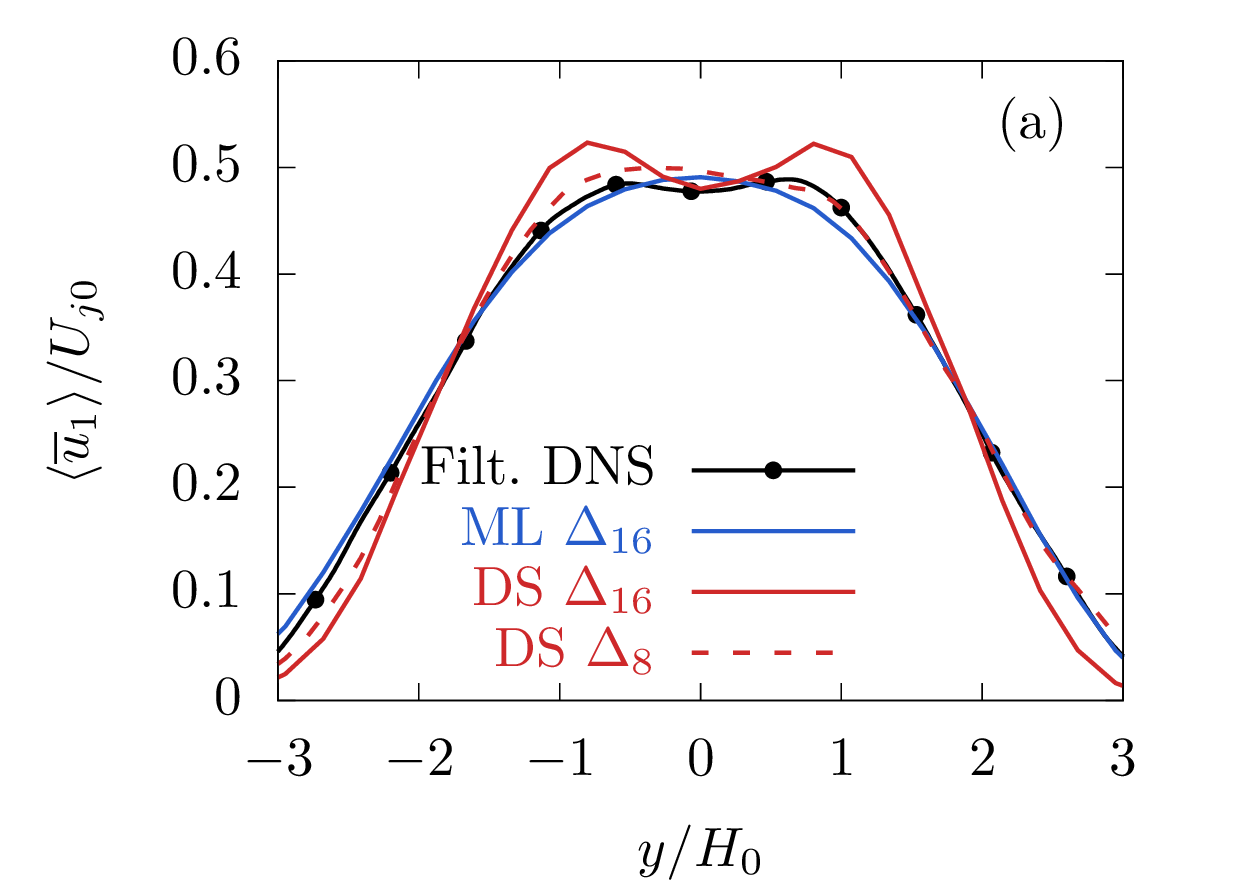}
  \includegraphics[width=0.45\textwidth]{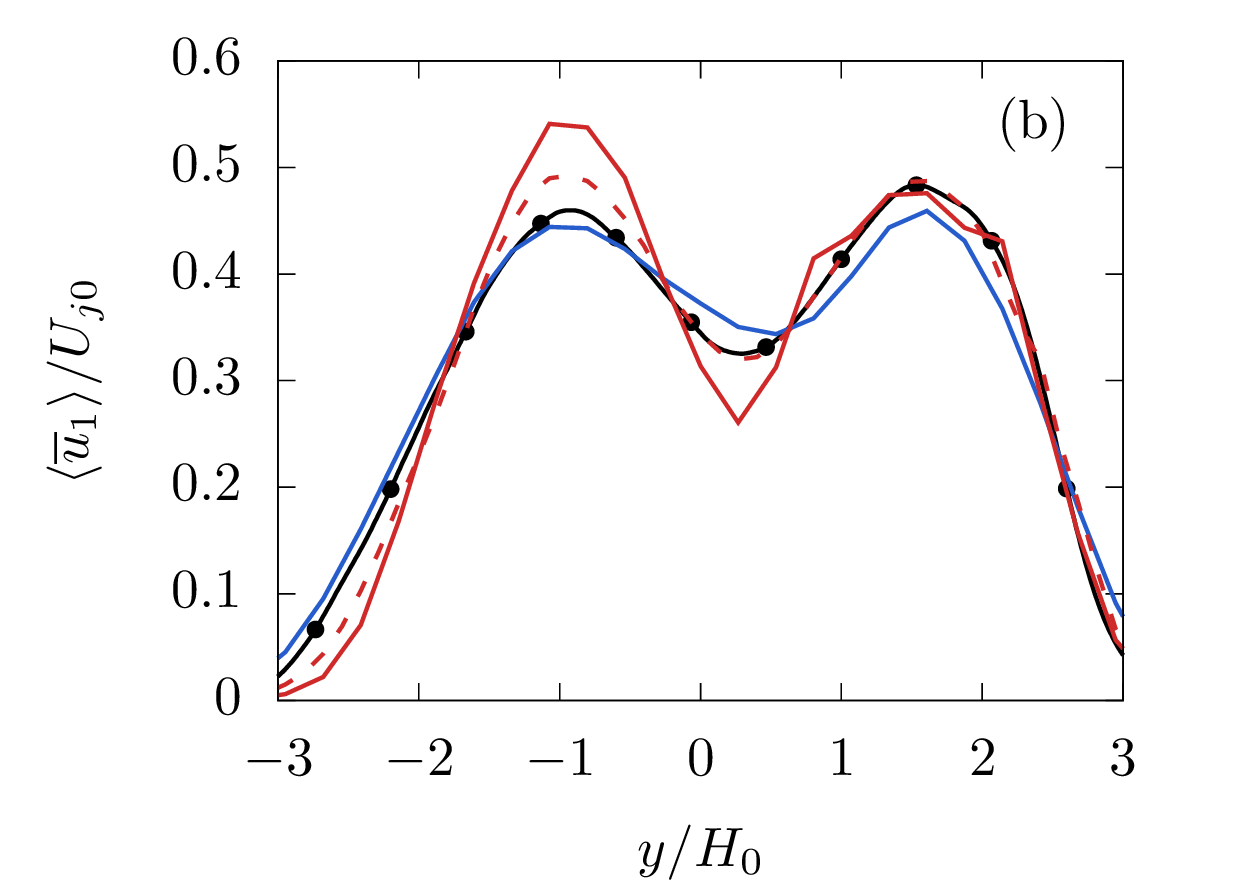} \\
  \includegraphics[width=0.45\textwidth]{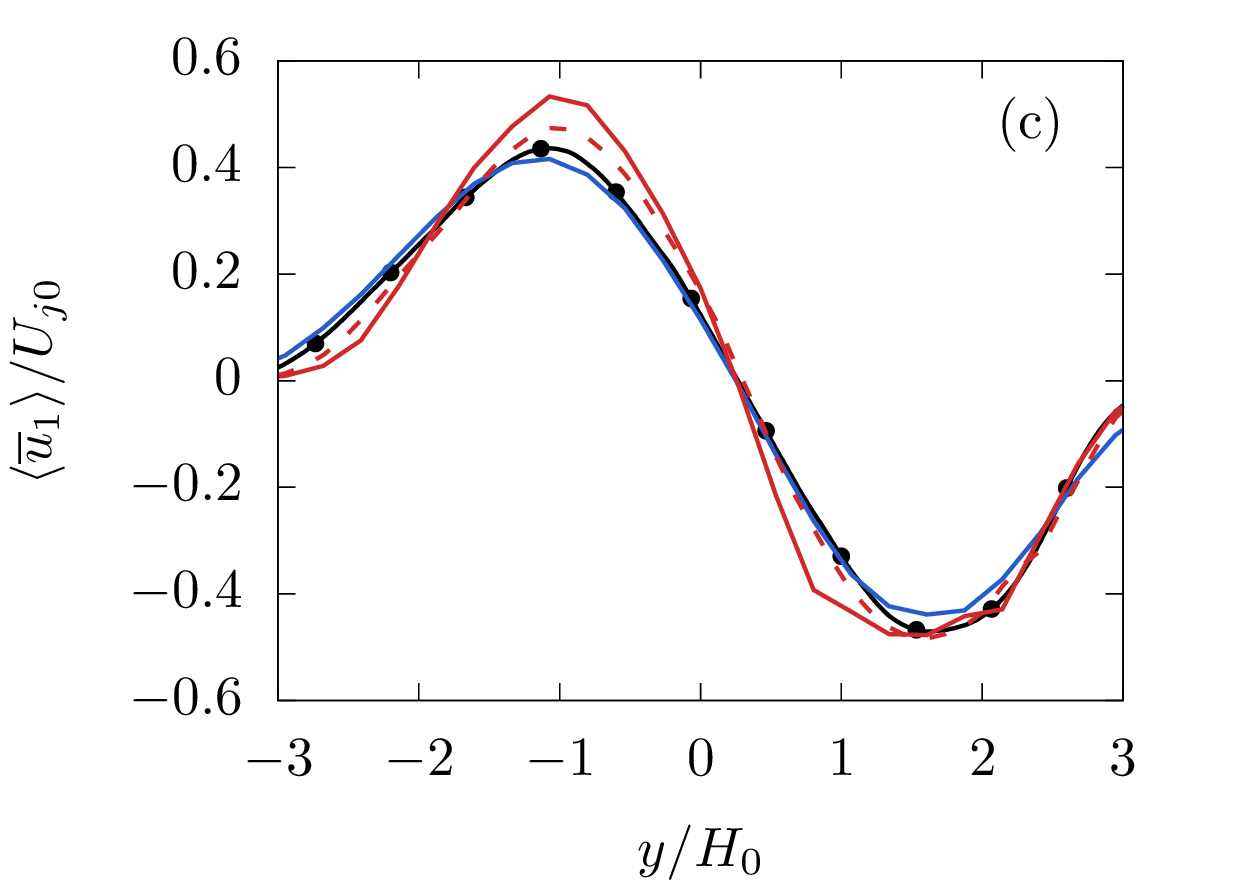}
  \hspace{-0.7cm}
  \caption{Out-of-sample mean streamwise velocity $\ol{u}_1$ for
    $\Delta_{16}$ and the deep learning (ML) and dynamic Smagorinsky
    (DS) models:  (a) case~B at $t=50t_{j0}$, (b) case~C at
    $t=42.5t_{j0}$, and (c) case~D at $t=42.5t_{j0}$. The $\Delta_8$
    case is shown for reference.}
  \label{fig:doublejet_vel}
\end{figure}

\begin{figure}
  \centering
  \includegraphics[width=0.45\textwidth]{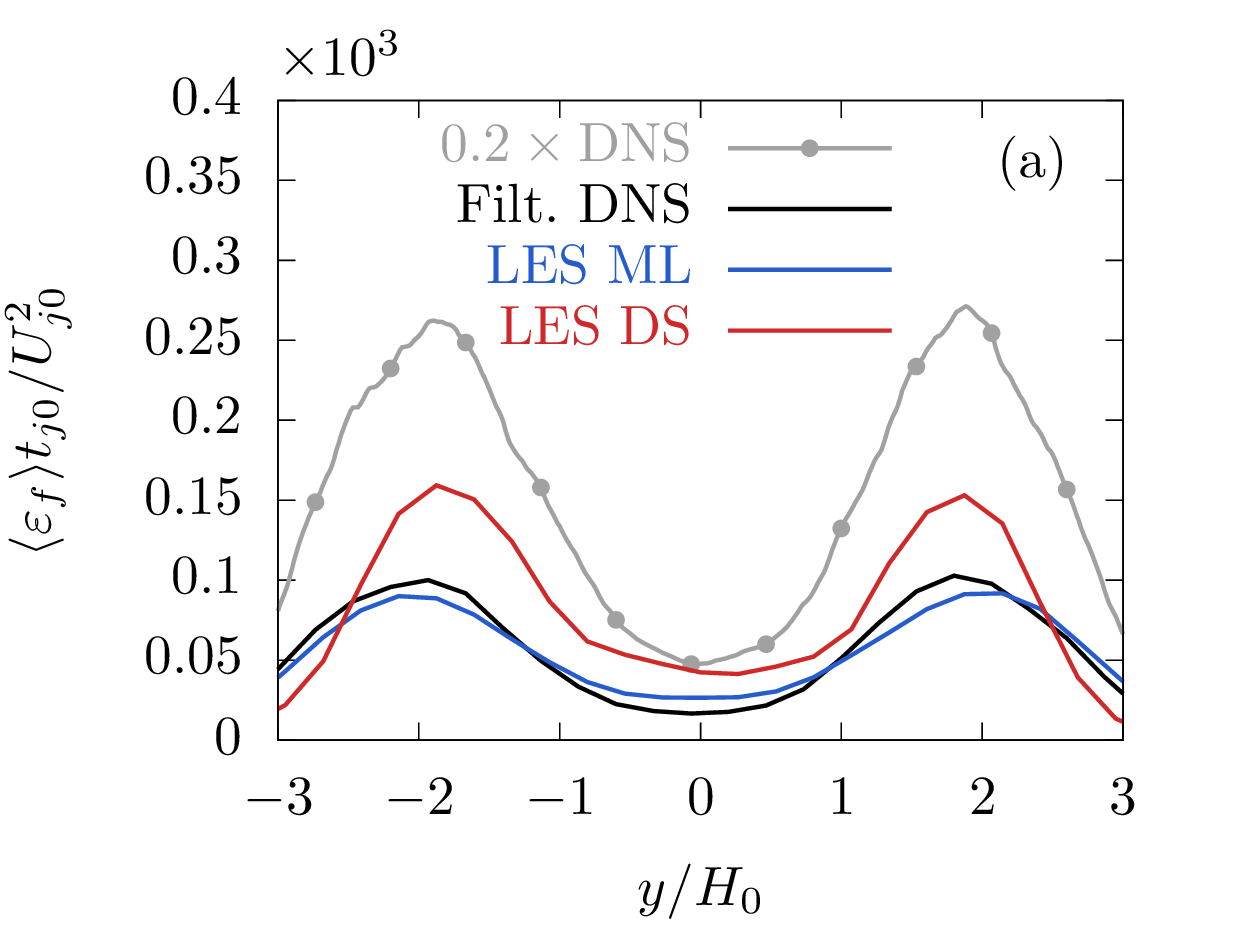}
  \includegraphics[width=0.45\textwidth]{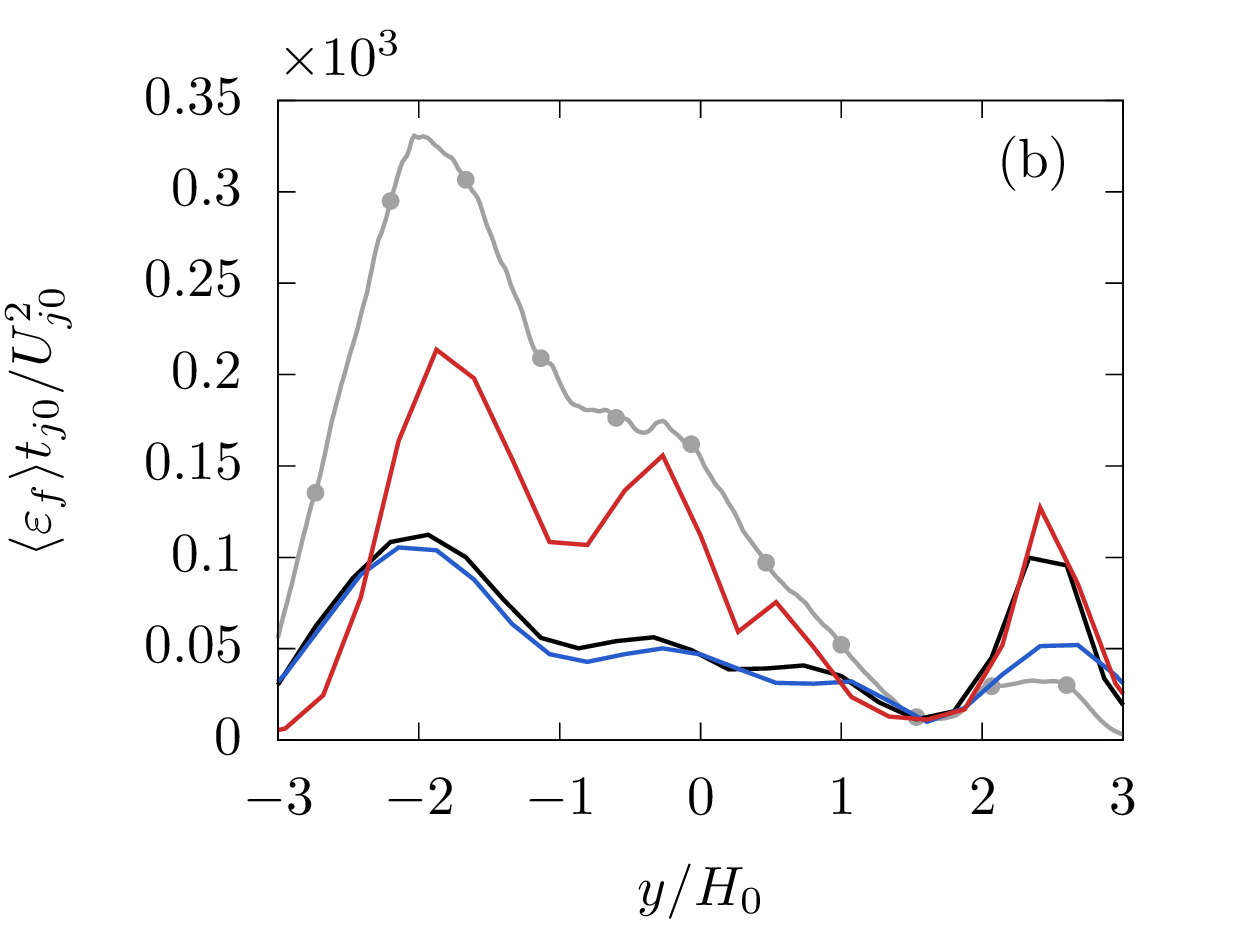}
  \includegraphics[width=0.45\textwidth]{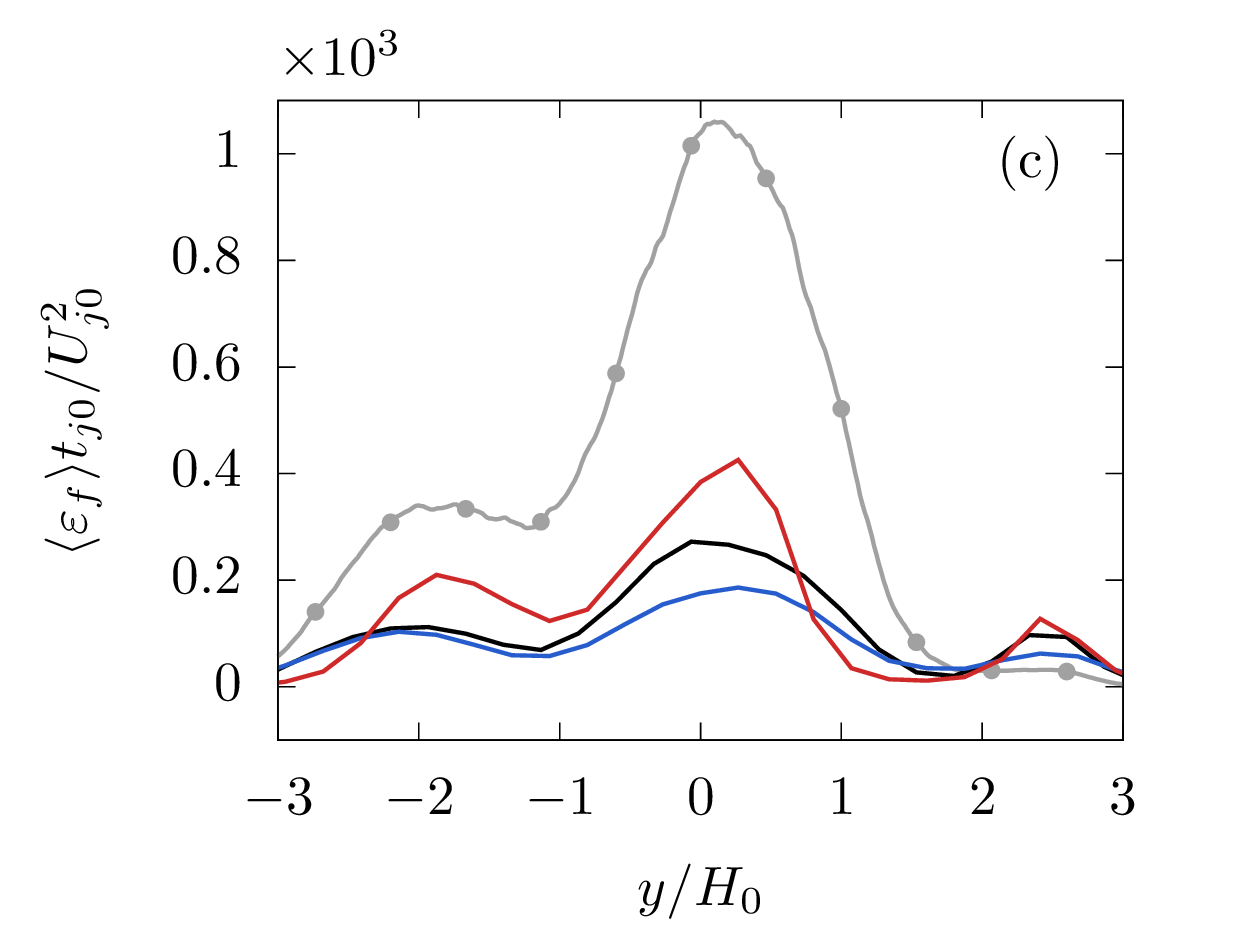}
  \caption{Out-of-sample resolved viscous dissipation $\eps_f$ for the
    deep learning (ML) and dynamic Smagorinsky (DS) models for
    $\Delta_{16}$: (a) case~B at $t=50t_{j0}$, (b) case~C at
    $t=42.5t_{j0}$, and (c) case~D at $t=42.5t_{j0}$.  The unfiltered
    direct numerical simulation viscous dissipation rate $\times 0.2$ is shown for comparison.
}
  \label{fig:doublejet_KE}
\end{figure}

Out-of-sample mean-flow results are shown in figure~\ref{fig:doublejet_vel} for all three dual-jet cases of table~\ref{tab:jet_IC}. The corresponding resolved viscous dissipation rates are shown in figure~\ref{fig:doublejet_KE}. 

For the symmetric parallel jets case~B, the deep learning model on $\Delta_{16}$ has accuracy comparable to the dynamic Smagorinsky model on $\Delta_8$ and predicts the resolved viscous dissipation rate.  It is thus predicts the merging process for which it was not trained.  Conversely, for this resolution, the dynamic Smagorinsky model on the coarse $\Delta_{16}$ mesh substantially underpredicts the jet merging, showing two distinct peaks of mean velocity. Similarly to case~A, this corresponds to over-prediction of $\eps_f$ near the high-shear regions.

The overall performance for the asymmetric parallel jets (case~C) is similarly good: the coarse-grid deep learning model more accurately reproduces the jet spreading than the dynamic Smagorinsky model for both $\Delta_{16}$ and $\Delta_8$ (figure~\ref{fig:doublejet_vel} b). Of the three models, the deep learning model also most accurately reproduces the peak velocity of the lower jet ($y\approx-H_0$), though at the same time it less accurate for the peak velocity of the slower-speed upper jet ($y \approx 2H_0$).  This might be because the lower jet more closely matches the training case~A, whereas the upper jet is slower ($U_1=U_0/2$) and wider ($H_1=2H_0$), as seen in table~\ref{tab:jet_IC} and figure~\ref{fig:jet_collapse}. These trends are repeated in figure~\ref{fig:doublejet_KE}~(b) for the resolved dissipation rate.

The deep learning model similarly outperforms the dynamic Smagorinsky model in the asymmetric, anti-parallel jets (case~D) (figures~\ref{fig:doublejet_vel} c and~\ref{fig:doublejet_KE} c), for both $\Delta_{16}$ and $\Delta_{8}$.  Unlike previous cases, this configuration has much larger $\eps_f$ in the high-shear, inter-jet region ($y \approx 0$) than in the jets themselves. The deep learning model performs especially well in this region.

\begin{figure}
  \centering
  \includegraphics[width=0.55\textwidth]{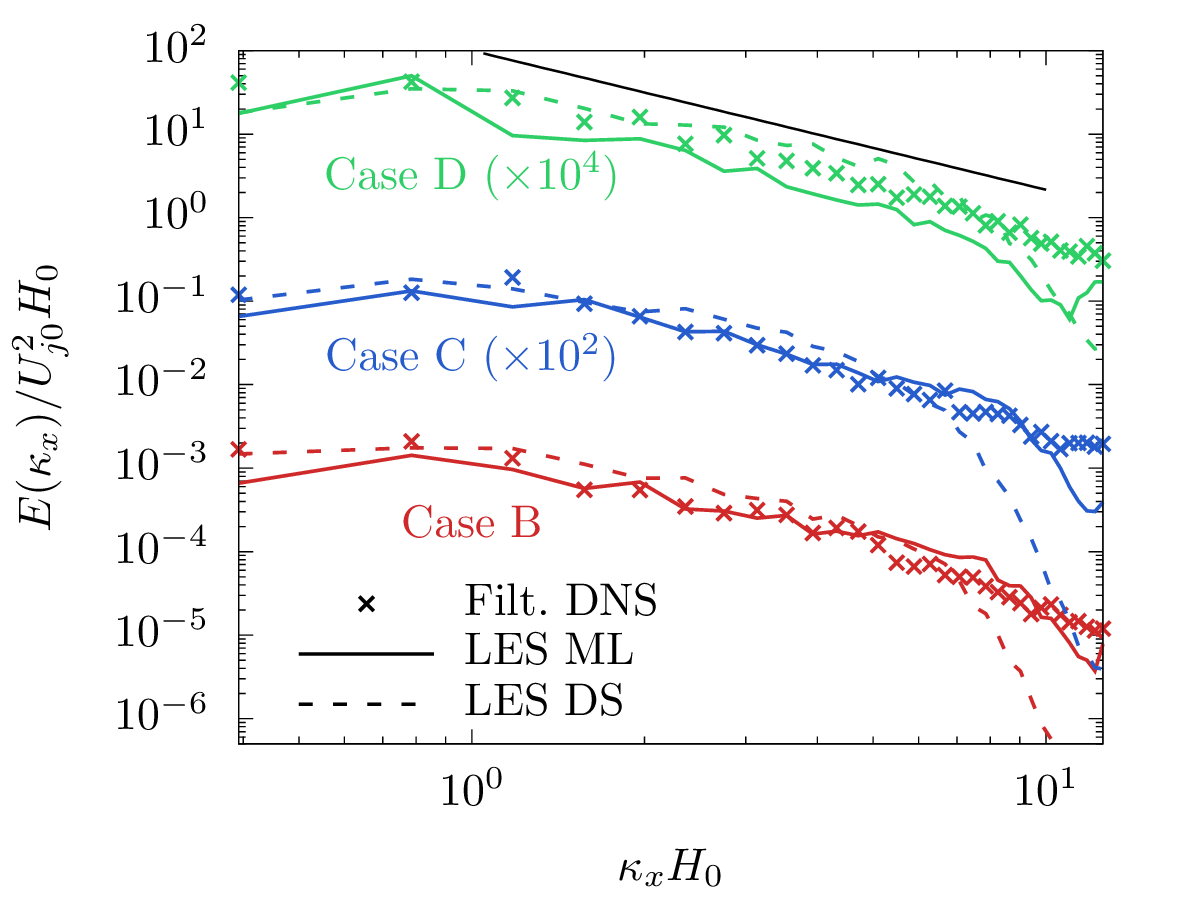}
  \caption{Out-of-sample one-dimensional streamwise energy spectra at
    the point of peak resolved turbulence kinetic energy for
    $\Delta_{16}$ at
    $t=50t_{j0}$ for case~B, at $t=42.5t_{j0}$ for case~C, and at
    $t=42.5t_{j0}$ for case~D. The straight line has $\kappa^{-5/3}$ slope. 
}
  \label{fig:doublejet_spect}
\end{figure}

One-dimensional energy spectra in figure~\ref{fig:doublejet_spect} show that the deep learning model performs particularly well at high wavenumbers. This might follow from the ability of the deep learning model to compensate for discretization errors, as considered for isotropic turbulence\cite{DPM-JCP}. As can be seen in figure~\ref{fig:doublejet_spect}, the deep learning model nearly eliminates unphysical near-cutoff-wavenumber dissipation.

\subsection{Training for mean statistics }
\label{ss:mean_flow}

It is, of course, expected that the reduced information content in this case will make training sub-grid-scale turbulence models more challenging.  We take  
$F(\olbu)=(\avg{\olbu};\avg{{\olbu'\olbu'}})_{(y,t)}$, leading to the loss function (\ref{AverageObjFunction}) developed in section~\ref{ss:meanflow} trained on the single-jet case~A.  This corresponds to the third set of tests in table~\ref{tab:traintest}.  The same neural network architecture (section~\ref{ss:model_arch}) and training algorithm (section~\ref{TrainingAlgorithm}) are used as for the previous demonstrations.

\begin{figure}
  \centering
  \includegraphics[width=0.48\textwidth]{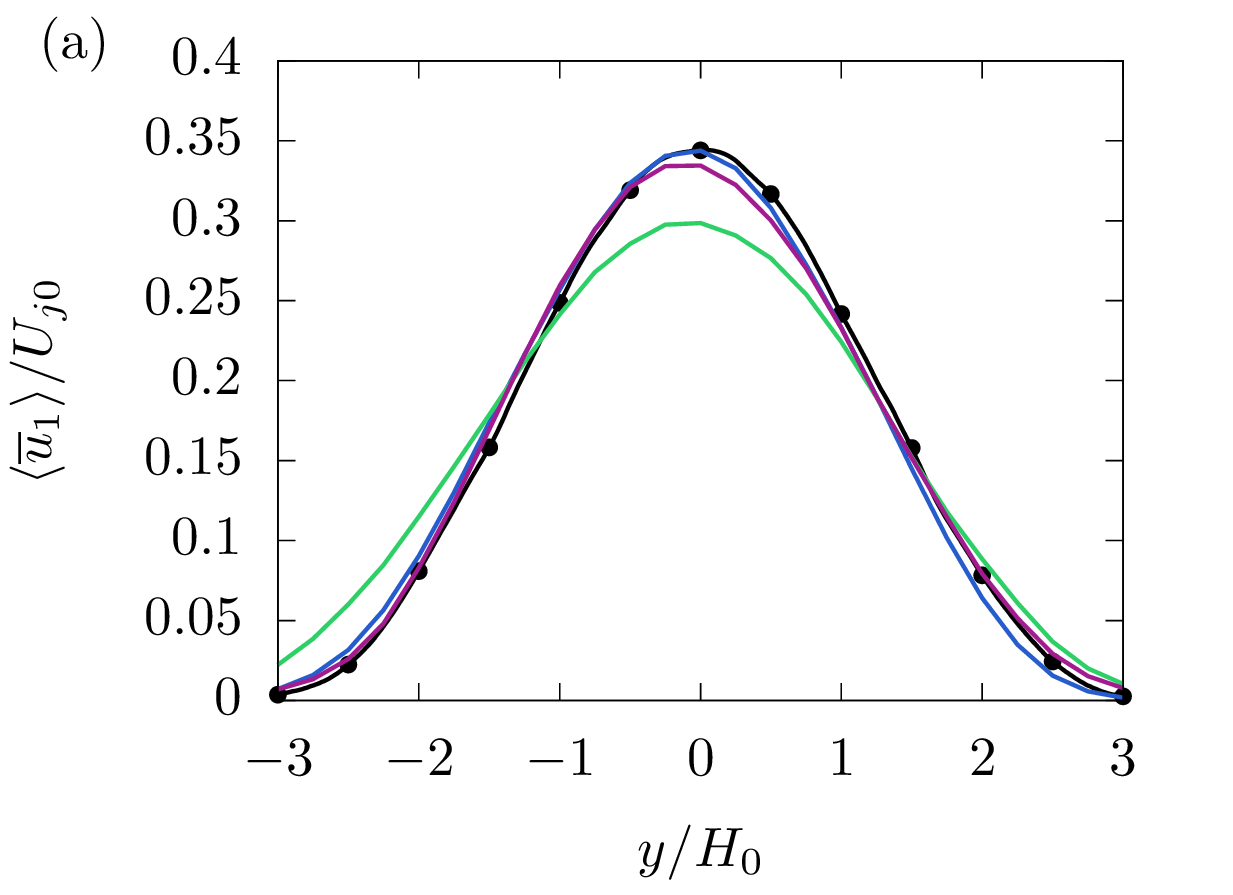}
  \includegraphics[width=0.48\textwidth]{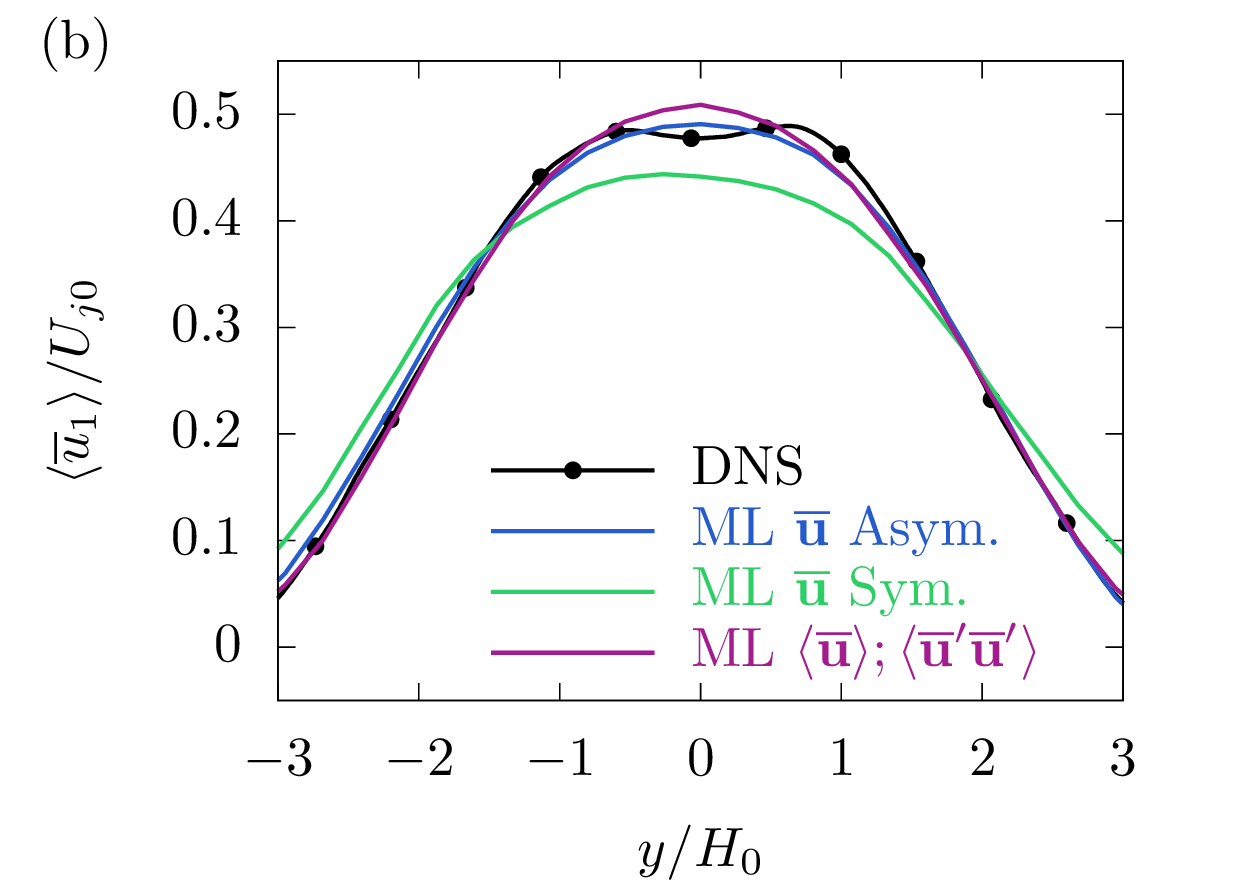}
  \includegraphics[width=0.48\textwidth]{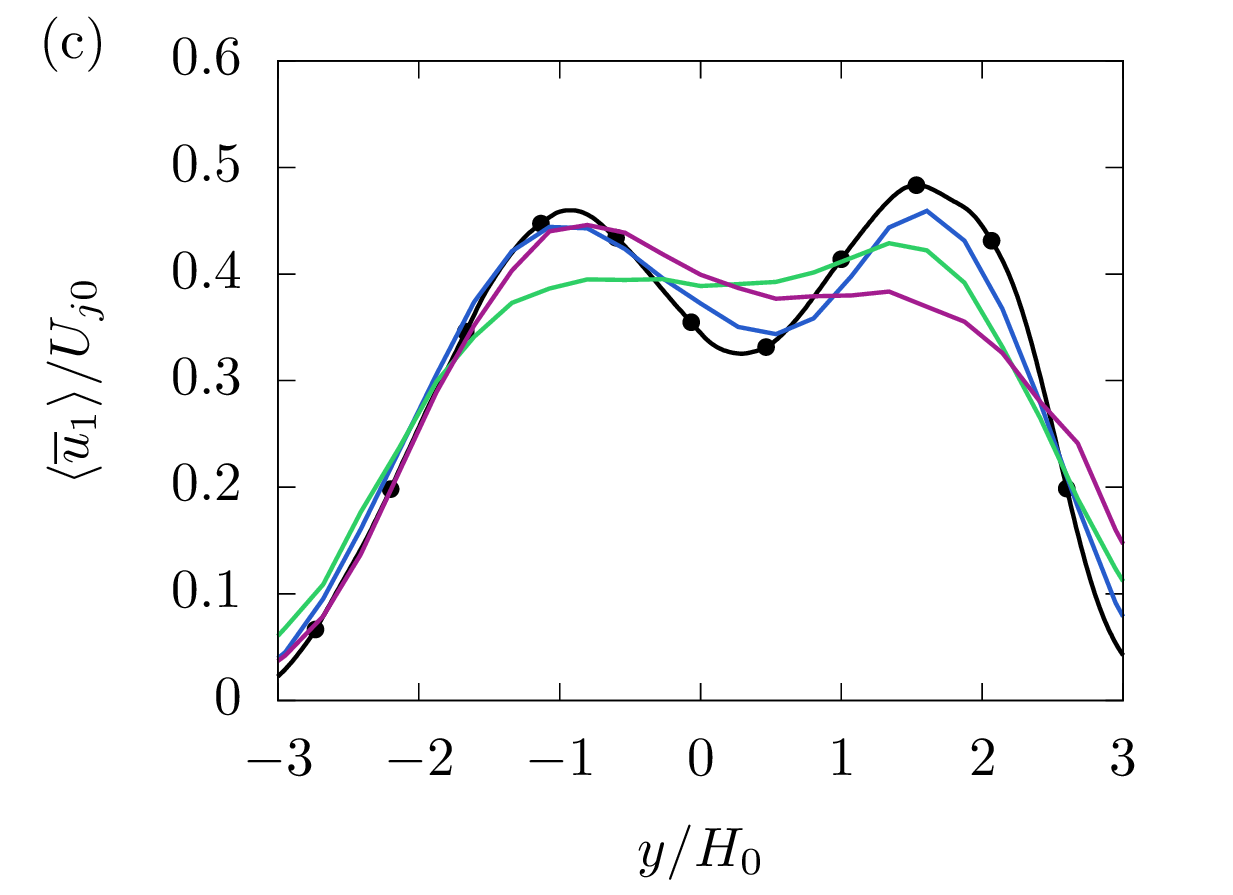}
  \includegraphics[width=0.48\textwidth]{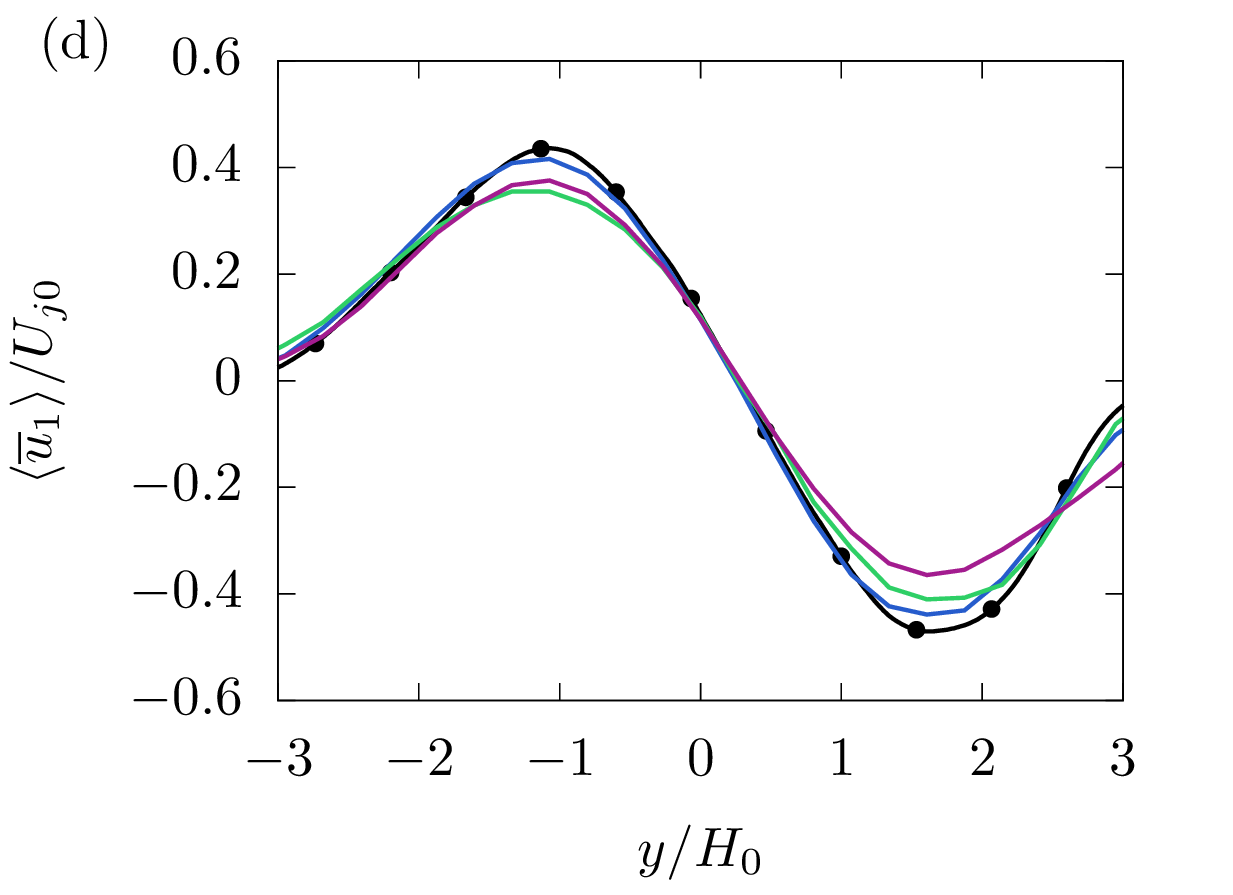}
  \caption{Mean-flow development for training with
    (\ref{AverageObjFunction}) mean statistics (see
    table~\ref{tab:traintest}) for $\Delta_{16}$:  in-sample (a) case~A at
    $t=62.5t_{j0}$ and out of sample (b) case~B at
    $t=50t_{j0}$, (c) case~C at
    $t=42.5t_{j0}$, and (d) case~D at $t=42.5t_{j0}$. Results for the
    baseline instantaneous field training (ML $\ol{\bu}$ Asym.) and its symmetrically
    constrained variant of section~\ref{ss:symmetric} (ML $\ol{\bu}$ Sym.) are also
    shown. }
  \label{fig:jet_vel_Rstress}
\end{figure}

The results for the mean flow are shown in figure~\ref{fig:jet_vel_Rstress}.  In all cases, the model produces a reasonable prediction, unlike the \textit{a priori} training on the mismatch of section~\ref{s:apriori}.  And, as expected, it performs well in-sample, which is expected since it was trained specifically to match its Reynolds stress data.  For the out-of-sample symmetric dual-jet (case~B), is also performs comparably to the model trained on the filtered data.  However, with its less rich training set, it is not expected to extrapolate as well as the model trained with the richer three-dimensional filtered data.
For the asymmetric out-of-sample jets (cases~C and D), it only closely matches the data on the sides of the dual-jets that are similar to the case~A training data ($y<0$; see table~\ref{tab:jet_IC} and figure~\ref{fig:jet_collapse}). On the other sides, the model overpredicts spreading, though not catastrophically, despite the huge reduction of training data size: the mean flow statistics have $N_xN_z$ times fewer degrees of freedom per \eqref{e:discrete_avg} than the instantaneous velocity.  It might be anticipated that expanded training data sets are likely necessary for this approach to be more accurate.  Additional implications of the mean-statistics training are discussed in section~\ref{ss:heffect}.

\subsection{A symmetric constraint}
\label{ss:symmetric}

As introduced in section~\ref{ss:model_arch}, the models discussed up
to this point were not constrained to produce symmetric stress
outputs, which is not a rigid constraint for typical discretizations
of the governing equations. Corresponding results for a
model that is trained to match the filtered velocity from case~A
($F(\olbu)=\olbu$), but now exactly constrained to produce symmetric
stresses ($h_{ij} = h_{ji}$), are shown figure~\ref{fig:jet_vel_Rstress} for both
in-sample and out-of-sample cases. Its performance is markedly worse
than either previous model in most circumstances.  This highlights
performance gained by relaxing the symmetry constraint, which is not
consistent with the discretized form of the governing equations.

\section{The $\bh$  Closure }
\label{s:h}

With the trained model working effectively in testing cases, especially when the training is based on the filtered instantaneous data, it is natural to analyze its form.  As for any deep-learning model, it is difficult to infer the mode of its operation, which in turn risks hindrance of its physical interpretation for flows.\cite{FreundMLPersp:2019} However, we can make some assessment of it.  Here, we develop simplified functional forms based on $\bh$ from the trained neural network by calibrating candidate functional forms $\bg$ that minimize
\begin{equation}
  J(\vphi) = D\left(\bh(\olbu;\vtheta) , \bg(\olbu;\vphi)\right),
  \label{e:loss_hform}
\end{equation}
as determined by a linear least-squares fit.
Each candidate closure $\bg$ has a low-dimensional set of parameters $\vphi$ that are fitted to the $\bh$ trained for $F(\olbu)=\olbu$ based on evaluation at all mesh and time points in case~A.  Attempts to interpret the fitted models are then made, and several of these $\bh$-fitted models are implemented for \textit{a posteriori} tests.

\begin{figure}
  \centering
  \includegraphics[width=0.5\textwidth]{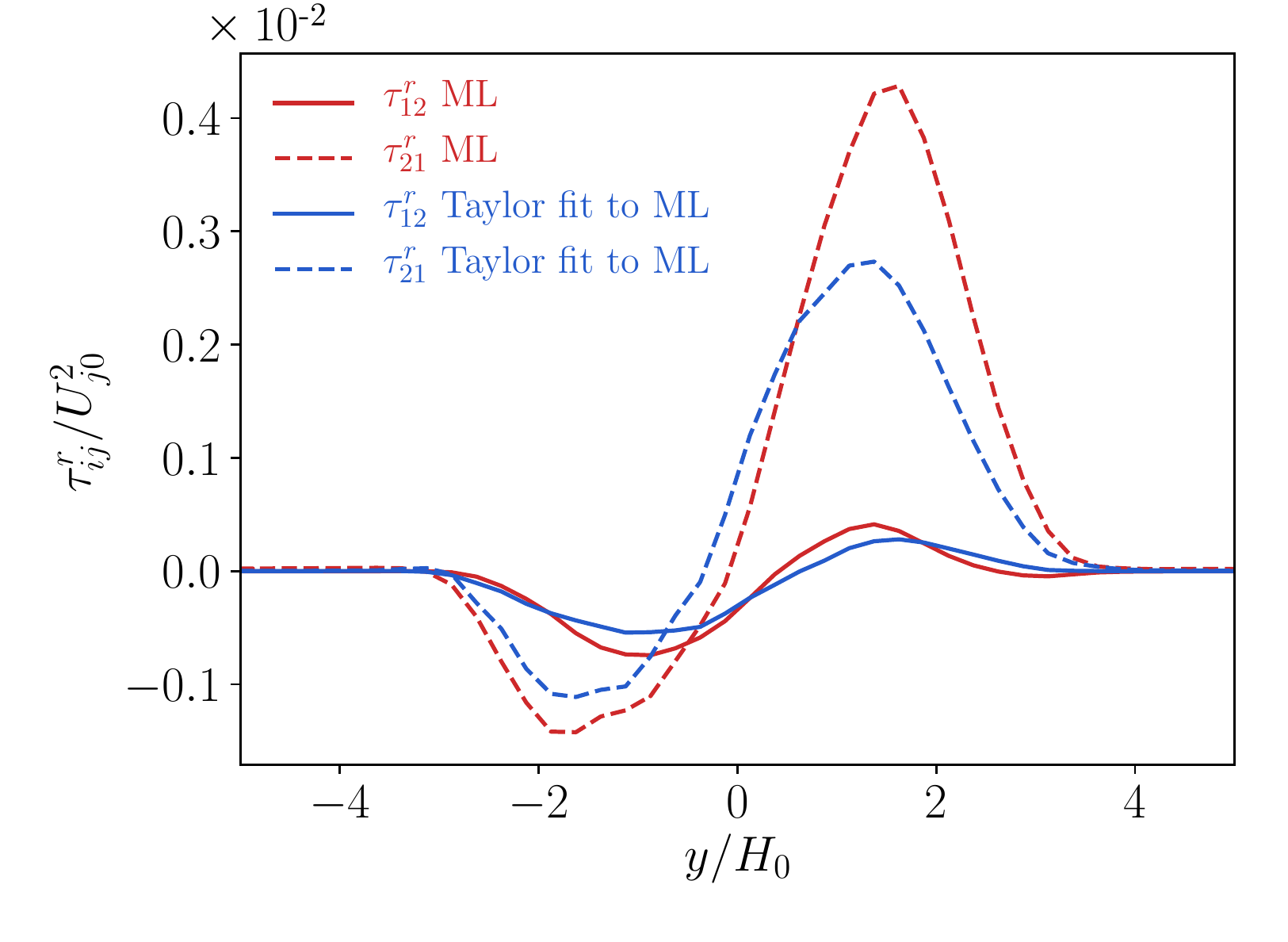}
  \caption{Taylor-series expansion $\bg$ from \eqref{e:tseries_hform} fitted to the deep learning (ML) model outputs $\bh$ for the $\tau_{12}$ and $\tau_{21}$ non-symmetric model components.}
  \label{fig:tseries_tau}
\end{figure}

We start with a Taylor series expansion in filter width including terms up to $O(\fDelta^2)$, which thus has 39 parameters per $\tau^r_{ij}$ component (351 total):
\begin{equation}
  g_{ij}(\ol{\mathbf{u}}) = a_{ijk}\ol{u}_k + \fDelta b_{ijkl}\pp{\ol{u}_k}{x_l} + \fDelta^2 c_{ijklm} \frac{\partial^2\ol{u}_k}{\partial x_l\partial x_m}.
  \label{e:tseries_hform}
\end{equation}
This expansion for two sub-grid-scale stress components is shown in
figure~\ref{fig:tseries_tau}. We focus on components for which the
deep learning model learns asymmetries (discussed in
section~\ref{s:nn}), with $\tau^r_{12}$ and $\tau^r_{21}$ shown in
figure~\ref{fig:tseries_tau} as an example, though we made analogous
observations for all other components. The Taylor-series model
\eqref{e:tseries_hform} is able to reproduce the asymmetries learned
by $\bh$ and, roughly, its variances. Other model forms, such as those
discussed subsequently based on the Smagorinsky and Clark models, are
unable to do so. There seem to be two aspects to the relative success
of \eqref{e:tseries_hform}: (1) assigning separate parameters to
asymmetric velocity gradient tensors (and not using the symmetric
strain-rate tensor $S_{ij}$) and (2) including all velocity
derivatives in $g_{ij}$, not just those with the same $i,j$ indices.
One would expect that asymmetries in decoupled closures (for example,
those based on the Boussinesq hypothesis) could be learned simply by
including the asymmetric velocity derivatives. However, the
improvements of including non-index-matching velocity derivatives
reinforces the idea that $\bh$ can learn inter-PDE couplings,
consistently with the overall discretization of the equations.

\begin{figure}
  \centering
  \includegraphics[width=0.48\textwidth]{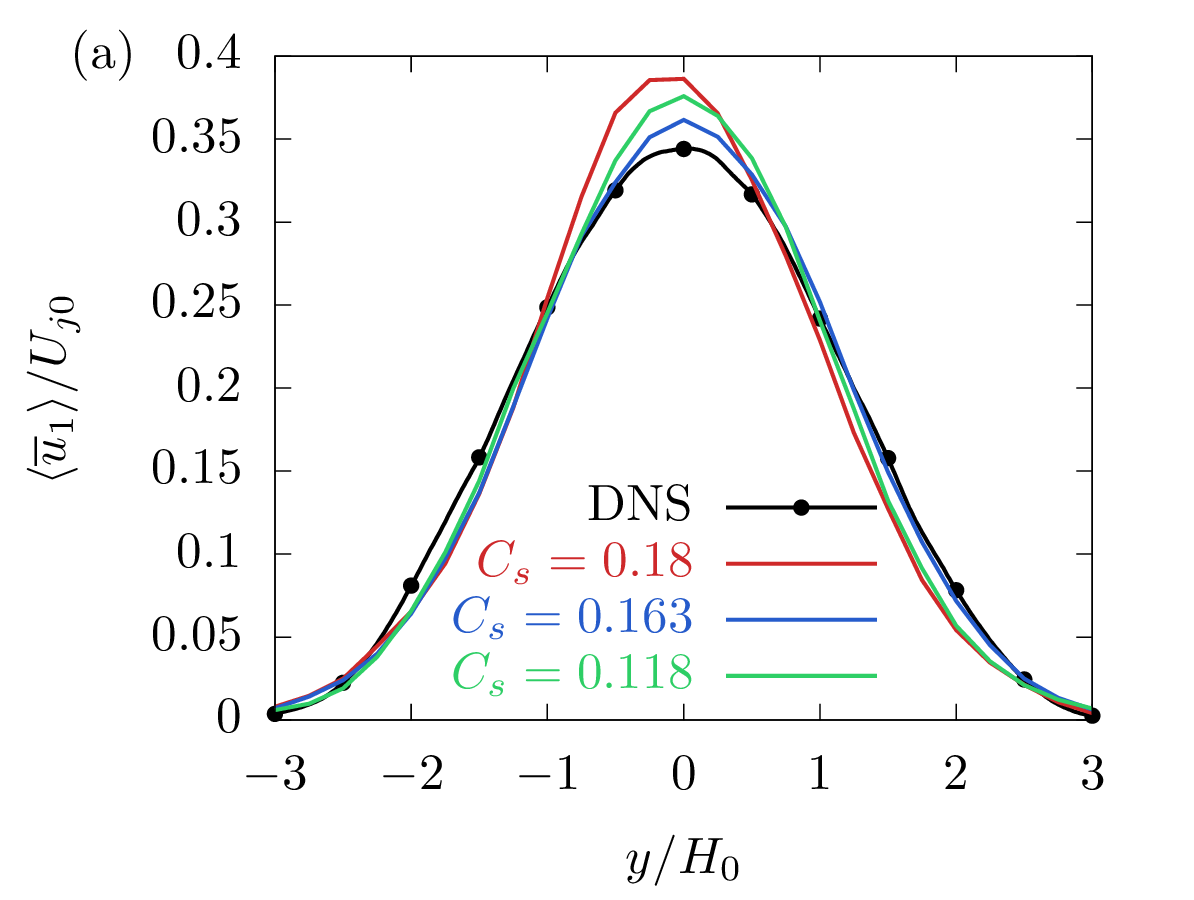}
  \includegraphics[width=0.48\textwidth]{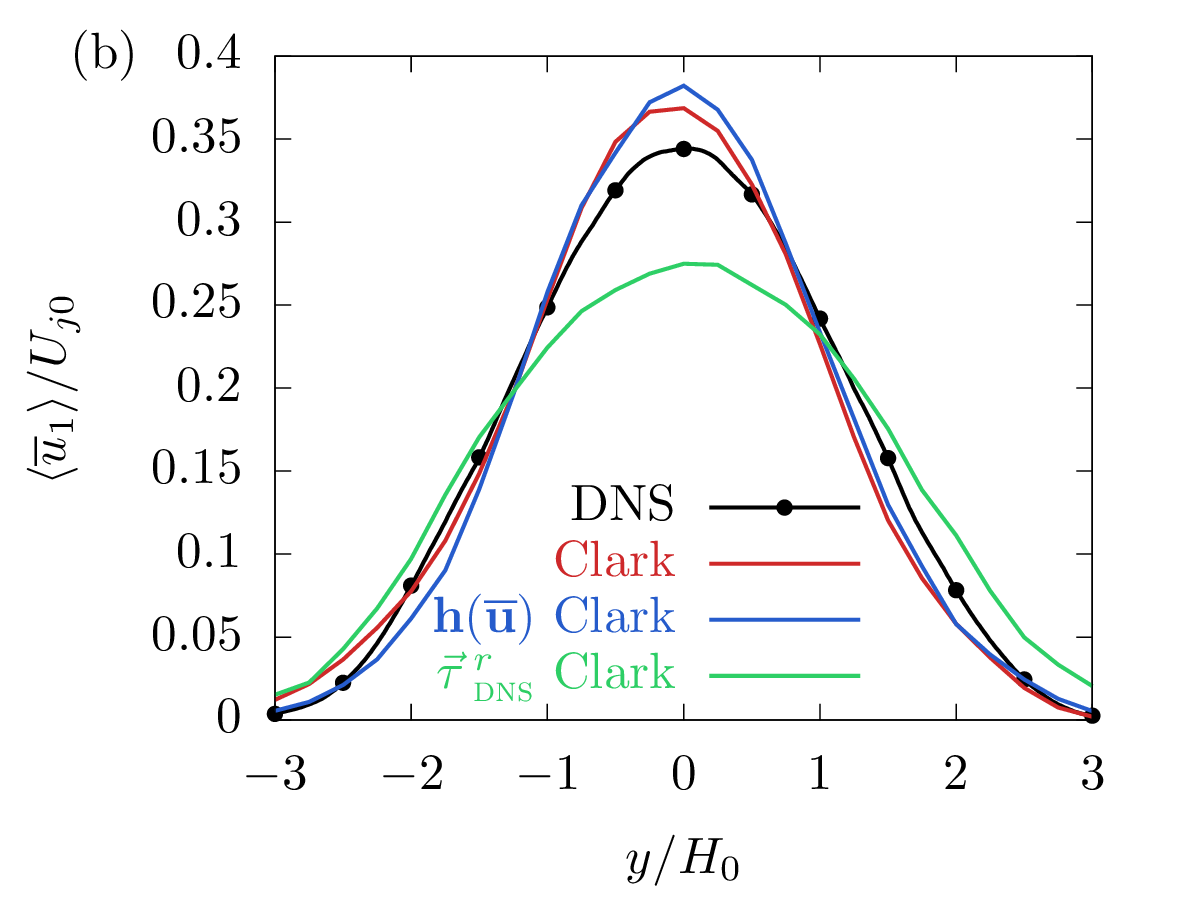}
  \caption{Performance on case~A of baseline, $\bh$-fitted, and $\ol{u}$-fitted models at $t=62.5t_{j0}$ and for $\Delta_{16}$: (a) the constant-coefficient Smagorinsky model ($C_s=0.163$: $\bh$-fit; $C_s=0.118$: $\vtau^r_\DNSsub$-fit), and (b) the Clark model (see text). }
  \label{fig:jet_vel_h_fit}
\end{figure}

We can also infer properties of $\bh$ by fitting it with the functional forms of standard low-degree-of-freedom turbulence models. We first consider the constant-coefficient Smagorinsky model,\cite{Smagorinsky1963,Rogallo1984}
\begin{equation}
  \tau^{r,\mathrm{Smag}}_{ij}=-2 C_s^2 \fDelta^2 |\ol{S}|\left(\ol{S}_{ij} - \frac{1}{3}\ol{S}_{kk}\delta_{ij}\right),
  \label{e:model_smag}
\end{equation}
where $|\ol{S}|=(2\ol{S}_{ij}\ol{S}_{ij})^{1/2}$ is the filtered strain-rate magnitude. The fitting parameter is $\vphi = [C_s^2]$. With all components of the model denoted $\vtau^{r,\mathrm{Smag}} = \ba(\olbu)\vphi$, we can obtain $C_s$ by minimizing \eqref{e:loss_hform}. For case~A, this yields $C_s=0.163$ for $\bh(\olbu;\vtheta)$ and $C_s=0.118$ for $\vtau^{r}_\DNSsub$, both of which are comparable to common values.  Simulation results for these candidate coefficients are shown in figure~\ref{fig:jet_vel_h_fit} (a). The coefficient learned from $\bh$, $C_s=0.163$, most accurately reproduces the mean flow of the three. The learned coefficients ($C_s=0.163$ and $C_s=0.118$) also more correctly reproduce jet spreading (not shown), but not as accurately as the full neural-network model.

We also consider the model of Clark \textit{et al.}\cite{Clark1979}, which is obtained via a Taylor-series expansions of $\vtau^r$ in filter size $\fDelta$:
\begin{align}
  \tau^{r,\mathrm{Clark}}_{ij} &= \tau^{r,\mathrm{Smag}}_{ij} + \tau^{r,\mathrm{grad}}_{ij} \notag \\
  &= \tau^{r,\mathrm{Smag}}_{ij} + \frac{1}{12}\left(\fDelta^2\pp{\olu_i}{x_k}\pp{\olu_j}{x_k} - \frac{\fDelta^2}{3}\pp{\olu_k}{x_k}\pp{\olu_k}{x_k}\delta_{ij}\right).
  \label{e:model_clark}
\end{align}
The second term on its own is known as the gradient model, which,
while accurate in \textit{a priori} evaluations because it
provides a good approximation for filter-scale perturbations in an
ideally filtered DNS field, can cause numerical instabilities in
\textit{a posteriori} evaluation.\cite{Vreman1996}

To fit coefficients with $\vphi=[C_s^2,C_g]^T$, we rewrite \eqref{e:model_clark} as
\begin{equation}
  \vtau^{r,\mathrm{Clark}} = C_s^2\,\ba_\mathrm{Smag}(\olbu) + C_g\,\ba_\mathrm{grad}(\olbu).
\end{equation}
With $\ba(\olbu) = [\ba_\mathrm{Smag},\ba_\mathrm{grad}]$, the model is $\vtau^{r,\mathrm{Clark}} = \ba(\olbu)\vphi$. Fitting to $\bh(\olbu)$ yields $ C_s = 0.151$ and $C_g = 0.422$, and fitting to $\vtau^{r}_\DNSsub$ yields $C_s = 0.041$ and $C_g = 1.288$.
Of these, only the fit to $\bh(\olbu)$ appears reasonable based on
typical values. This is a surprise, since one might expect that the
coefficients could be learned directly from the filtered data.
However, the same assumption is the reason \textit{a priori}
evaluation often incorrectly predicts a model's \textit{a posteriori}
performance.  \textit{A posteriori} results are shown in
figure~\ref{fig:jet_vel_h_fit} (b). With $\bh$-learned coefficients,
results are comparable to the unity-coefficient model, and, indeed,
the DNS-fitted coefficients are inaccurate.

These results suggest that high-degree-of-freedom models can be
interpreted in the context of pre-specified functional forms. In
particular, we observe that asymmetries in the modeled sub-grid-scale stress can
be codified using asymmetric velocity gradient tensors, and that $\bh$
might learn inter-PDE couplings between the different velocity
components. However, low-degree-of-freedom attempts to interpret
highly parameterized models will often be found wanting, and this can
be seen in attempts to fit standard turbulence-model parameters to the
neural network form. While these can produce better-performing fitted
(or at least not seriously performance degrading) parameters, they do
not match the \textit{a posteriori} accuracy of the full model.

\section{Predicted Turbulence Structure}
\label{ss:heffect}

The agreement of the deep learning model prediction with the usual
statistical measures (mean, turbulence stresses, energy spectrum)
suggests an improved representation of realistic turbulence at the
resolved scales.  Whether or not this agreement translates to an
improved representation of the turbulence structure, which would be
important for extrapolation beyond the training metrics, is quantified
with barycentric maps.  These track the eigenvalue invariants of the
normalized resolved Reynolds stress anisotropy $a_{ij} = R_{ij}/R_{kk}
- \delta_{ij}/3$,\cite{Banerjee2007} and as such quantify resolved
turbulence componentiality.  Though an obvious downside for neural
network models is the lack of proof about realizability and
boundedness, barycentric maps offer an empirical assessment of this:
the lines connecting the vertices of such maps are physical bounds,
and remaining within them is not guaranteed. One could envision that a
deep learning model might, despite a PDE constraint, sacrifice
physical realism for a good fit to data; this is tested in this
section. The degree to which a deep learning closure recovers the
ideal turbulence structure supports its extrapolative capacity and its
additional adherence to the physics beyond reproduction of the
training data.

\begin{figure}
  \centering
  \includegraphics[width=0.49\textwidth]{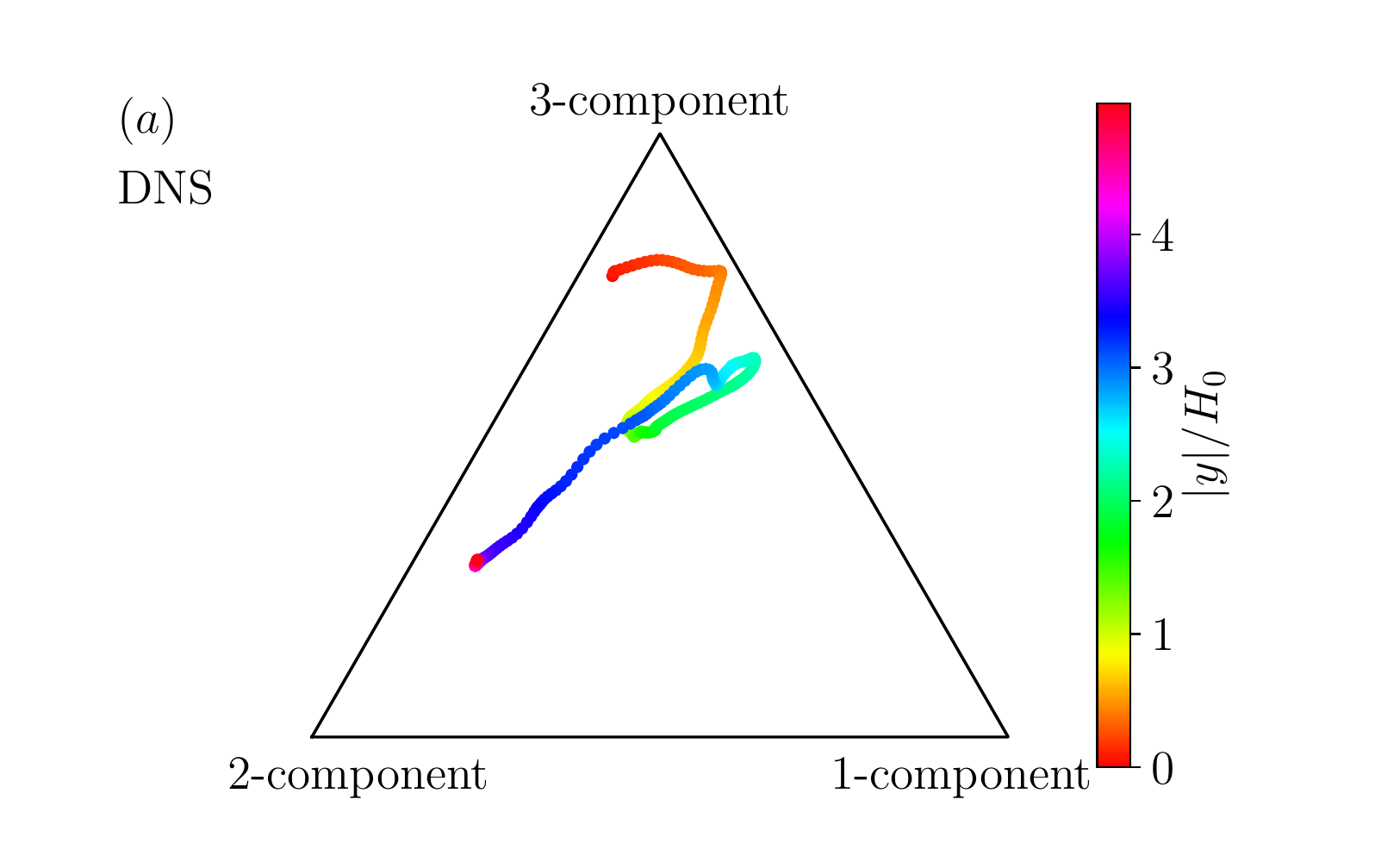}
  \includegraphics[width=0.49\textwidth]{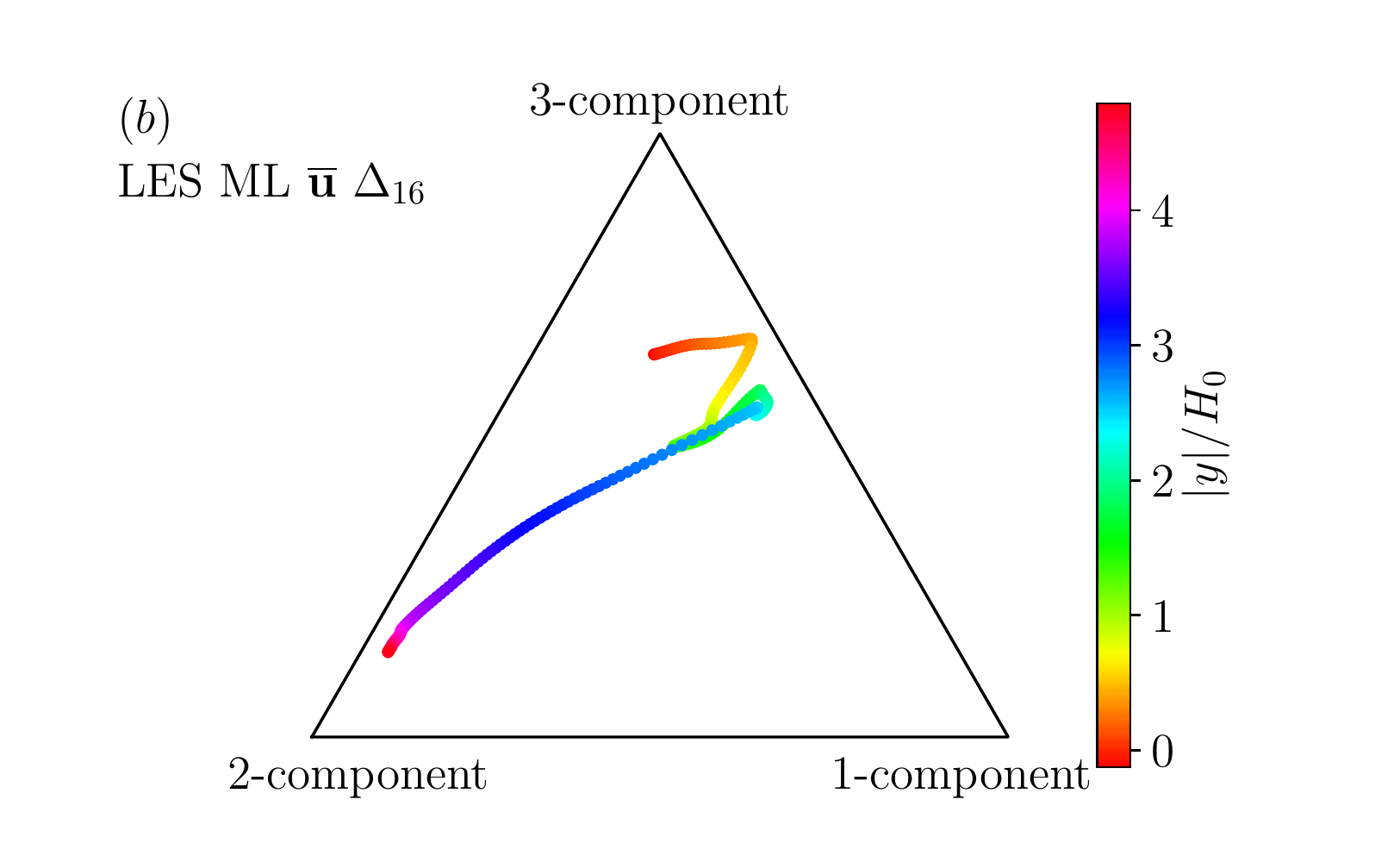} \\
  \includegraphics[width=0.49\textwidth]{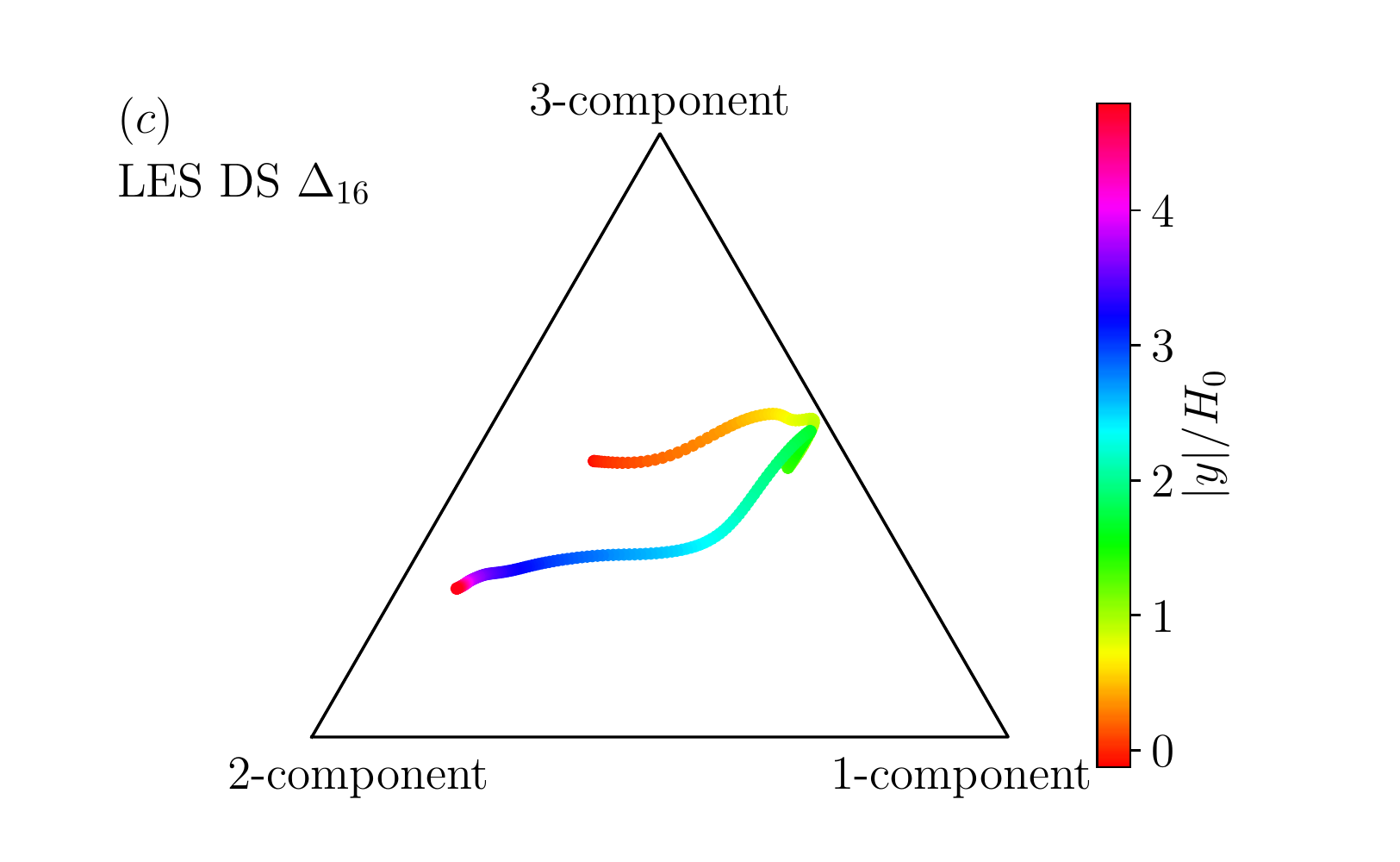}
  \includegraphics[width=0.49\textwidth]{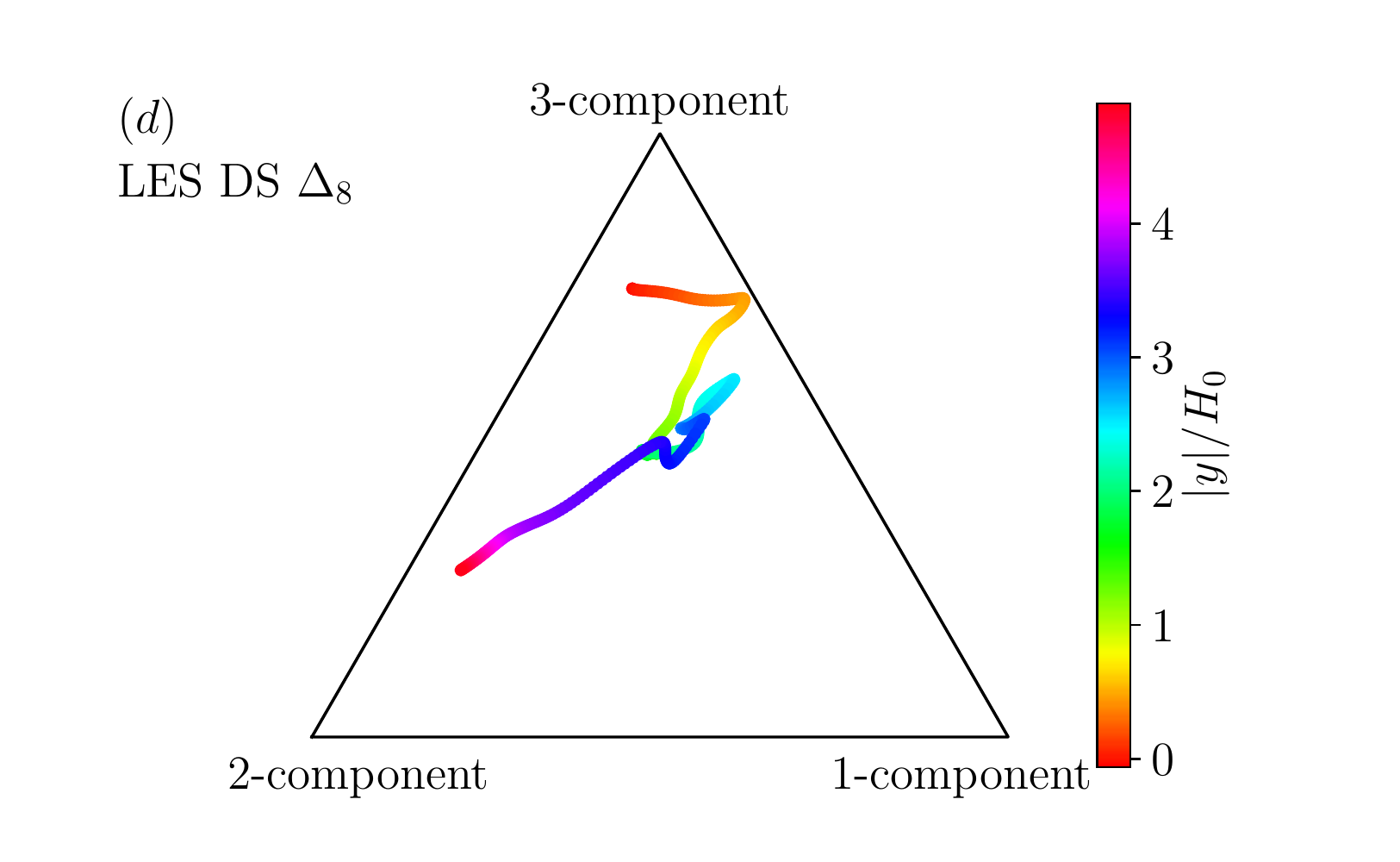} \\
  \includegraphics[width=0.49\textwidth]{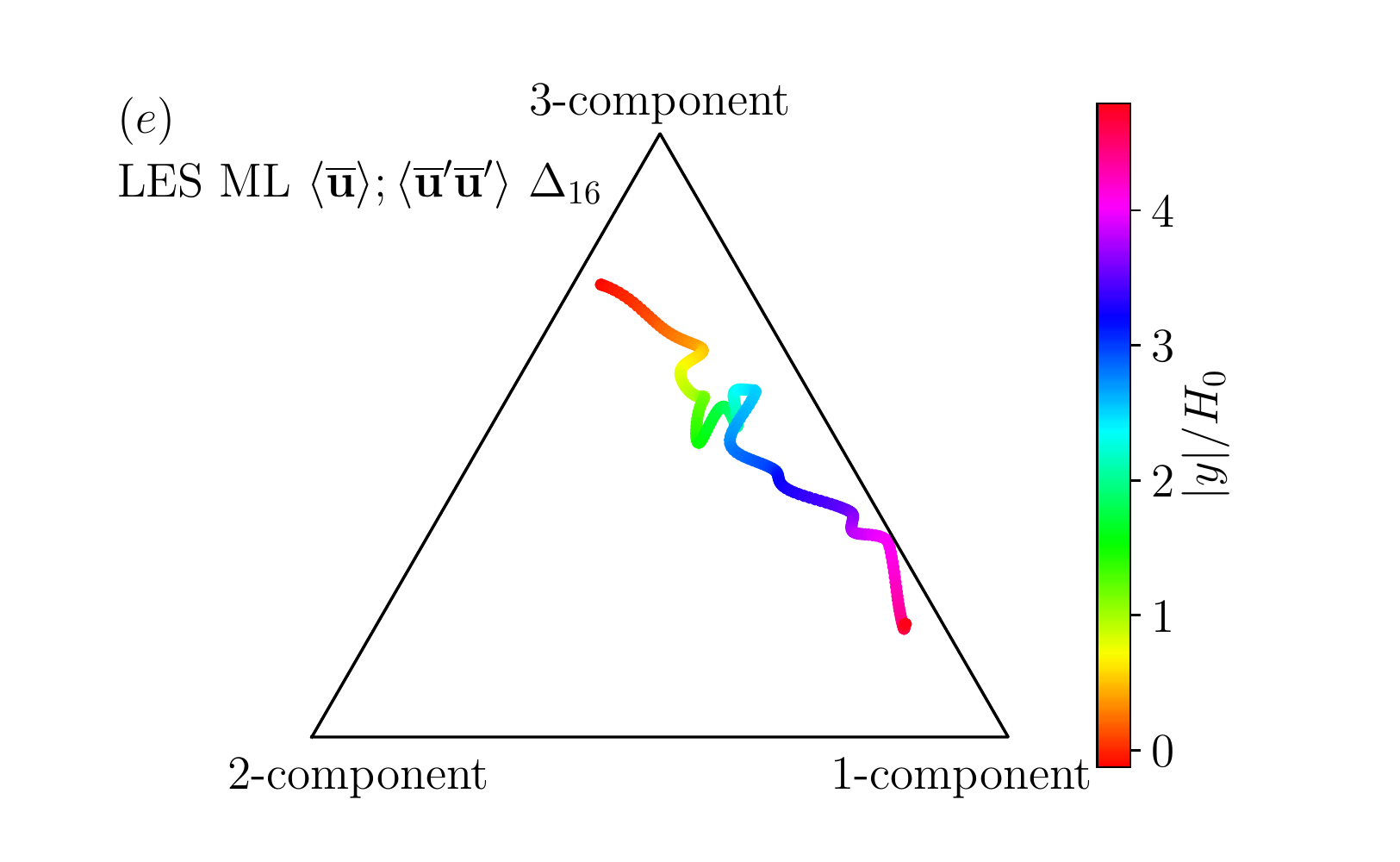} 
  \includegraphics[width=0.49\textwidth]{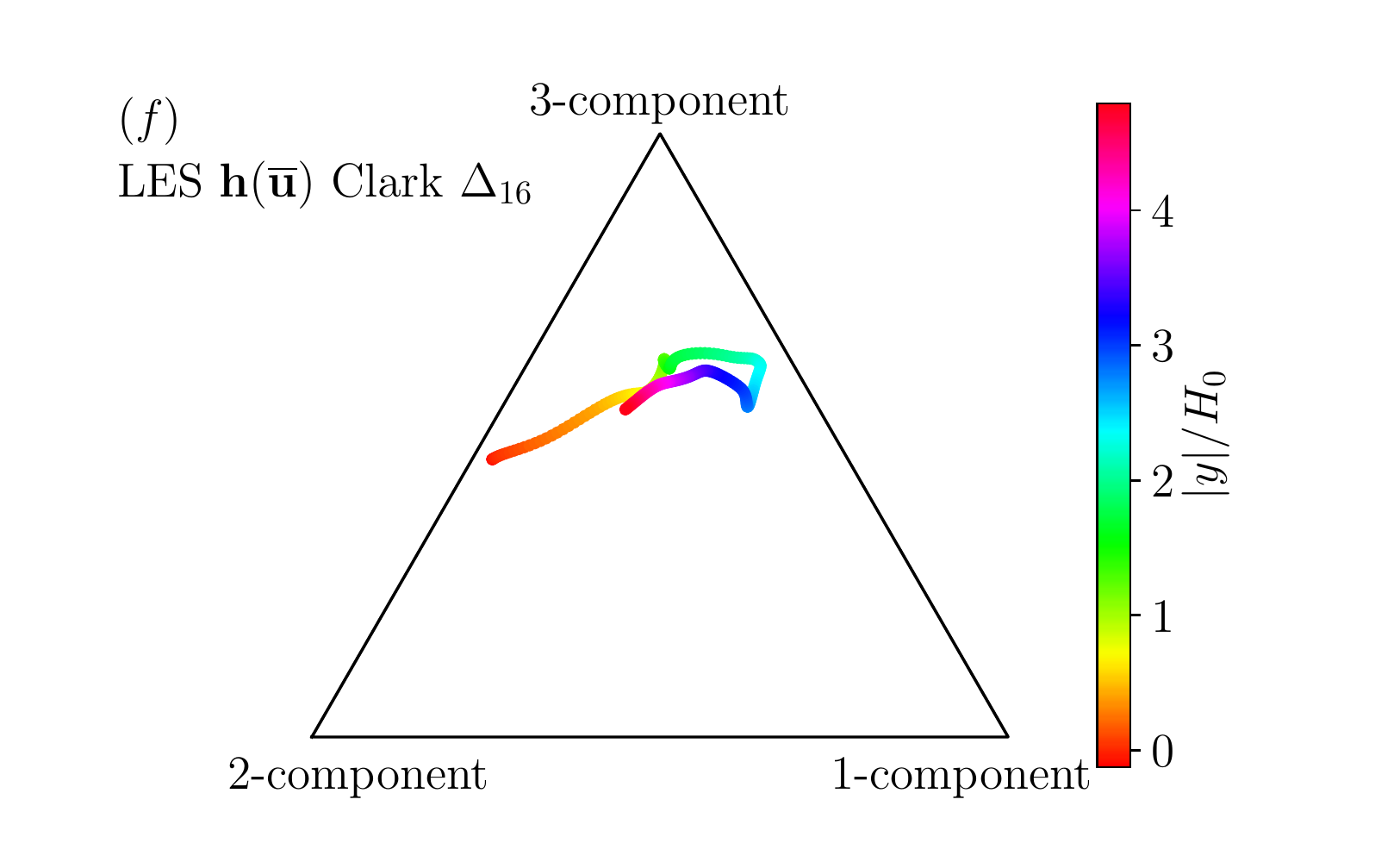}
  \caption{Barycentric maps of normalized resolved Reynolds stress invariants at various cross-stream locations $|y|/H_0$ for case~A at $t=62.5t_{j0}$, with data averaged $\pm y$: 
    (a) direct numerical simulation,
    (b) machine learning model with $\bh(\olbu)$ (see table~\ref{tab:traintest}),
    (c--d) the dynamic Smagorinsky model fo $\Delta_{16,8}$,
    (e) training on averaged statistics $F(\olbu)=(\langle\bu\rangle;\langle\olbu'\olbu'\rangle)$ (see table~\ref{tab:traintest}), and
    (f) the Clark model fitted to $\bh(\olbu)$ (see section~\ref{s:h}).}
  \label{fig:barymaps}
\end{figure}

Barycentric maps for the case~A resolved turbulence and $\Delta_{16}$ machine-learning model are shown in figures~\ref{fig:barymaps}~(a) and~(b). All trajectories remain within the physical bounds, and this is also the case for the dual-jet configurations (not shown).  
The trajectories within the maps show the variation of componentiality
in $y$. Key features of the direct numerical simulation
(figure~\ref{fig:barymaps}~a) include a relative two-component
condition in the outer shear layers ($|y| \approx 4 H_0$),
predominantly plane-strain (the line bridging three- and
two-componentiality) within the shear layers, and three-component
turbulence near the centerline ($|y| \lesssim H_0$).
Also shown are invariant maps of \textit{a posteriori} large-eddy simulation using the dynamic Smagorinsky model, the $F(\olbu)=(\langle\bu\rangle;\langle\olbu'\olbu'\rangle)$ closure, and the $\bh(\olbu)$-fitted Clark model. 
Both the $\Delta_{16}$ $\bh(\olbu)$ states (figure~\ref{fig:barymaps} b) and $\Delta_8$ dynamic Smagorinsky states (figure~\ref{fig:barymaps} d) show the general features of the direct numerical simulation, consistent with the ability of these models to reproduce other turbulence statistics.   However, the coarse $\Delta_{16}$ dynamic Smagorinsky model (figure~\ref{fig:barymaps} c) fails to achieve the same three-componentiality for $|y|\rightarrow 0$.

Trajectories obtained using $\bh$ trained with only mean statistics
($F(\olbu)=(\langle\bu\rangle;\langle\olbu'\olbu'\rangle)$) in
figure~\ref{fig:barymaps}~(e) only recover the turbulence structure
for $|y| \lesssim H_0$. This case fails to recover the two-component limiting conditions in the outer shear layers ($|y|/H_0 \approx 4$), instead trending toward one-component turbulence.  More training data from richer flows might remedy this.\cite{Ling2}

Finally, invariant trajectories of the $\bh(\olbu)$-fitted Clark model from section~\ref{s:h} are shown in figure~\ref{fig:barymaps}~(f). Despite its apparently good mean-flow prediction (figure~\ref{fig:jet_vel_h_fit}~b), the turbulence produced bears little resemblance the direct numerical simulation data.  Thus, this model is not expected to be extrapolative, efforts for which will likely benefit from better faithful representation of the turbulence structure in addition to any particular statistics.  For example, subfilter models for reacting flows will almost certainly benefit from the capability to encode these dynamics.\cite{OBrien2017,MacArt2018a} As a step toward assessing, out-of-sample tests of the $\bh$-fitted Clark model are presented along with the machine-learning model for scalar mixing in the following section.

\section{Mixing of a Passive Scalar}
\label{s:scalar}

In section~\ref{s:apriori}, it was shown that the high-dimensional neural-network model could, not surprisingly, fit the filtered direct numerical simulation sub-grid-scale stresses through a direct mismatch loss function, though this fitted stress failed to provide a viable mode when subsequently used in a simulation.  This exemplifies the risk of fitting without physical constraints, which guard against overfitting and unreliability for extrapolative prediction.  Optimizing the model as embedded in the governing equation (sections~\ref{s:geo_jet} and~\ref{s:geo_doublejet}) provided a model that reproduced turbulence statistics for a class of similar flows, demonstrating some capacity for reliable extrapolation.  Here, we further evaluate the extrapolative capacity of the model by assessing passive scalar $\xi$ mixing predictions.   Filtering \eqref{e:DNSscalar} yields
\begin{equation}
  \pp{\olxi}{t} + \olu_j\pp{\olxi}{x_j} = D\Dparttwo{\olxi}{x_j} - \pp{}{x_j}\left(\ol{u_j\xi} - \olu_j\olxi\right),
  \label{e:LESscalar}
\end{equation}
where $D=0$.  Only numerical diffusion was active in both the direct numerical simulation and the large-eddy simulation, and then only for the scalar advection. The second term on the right-hand side of \eqref{e:LESscalar} is the divergence of the unclosed residual scalar flux $F_{j}^\xi = \overline{u_j\xi} - \olu_j\olxi$. We model $F_{j}^\xi$ using the gradient-diffusion hypothesis, in which the eddy diffusivity is obtained using a scalar analog of the dynamic Smagorinsky model~\cite{Germano1991}.

\begin{figure}
  \centering
  \includegraphics[width=0.48\textwidth]{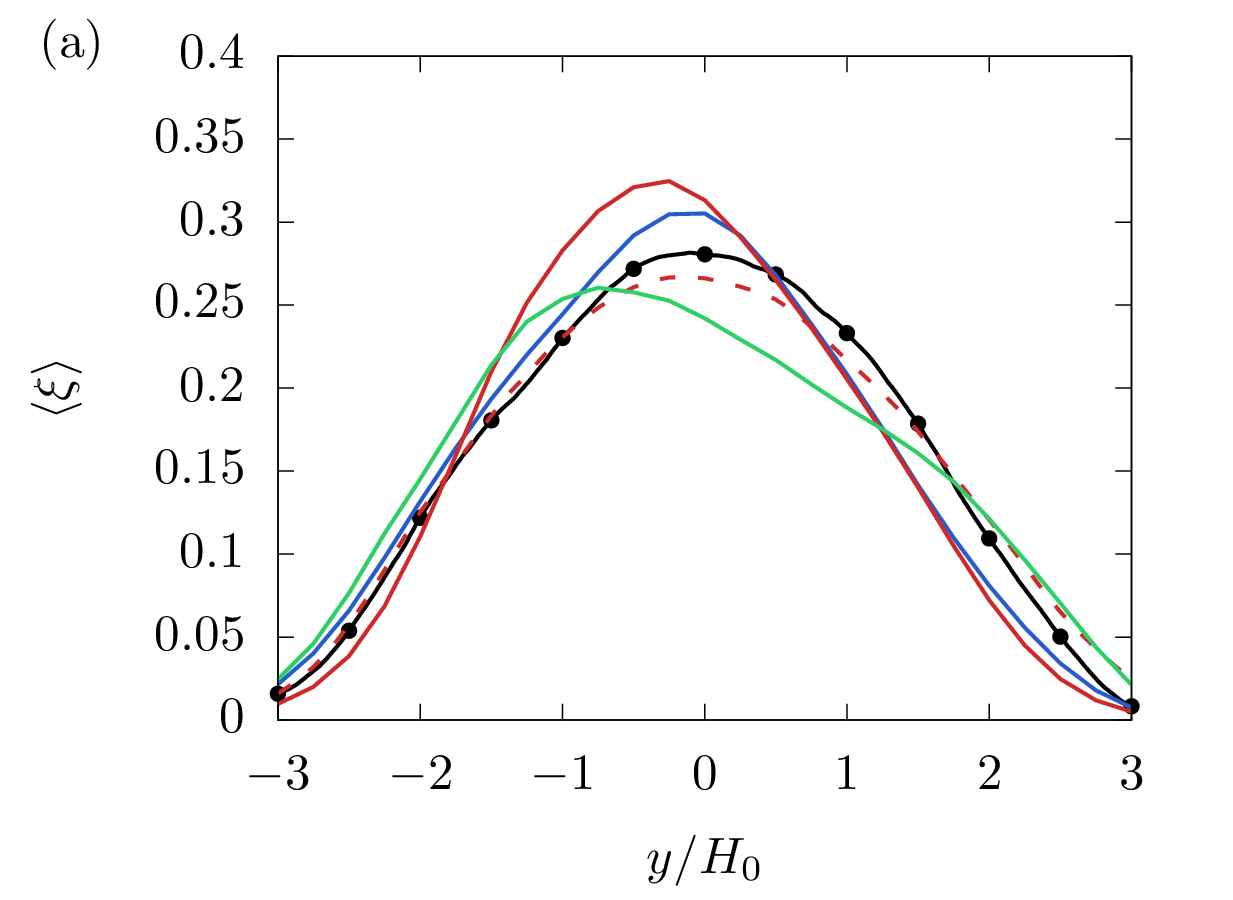}
  \includegraphics[width=0.48\textwidth]{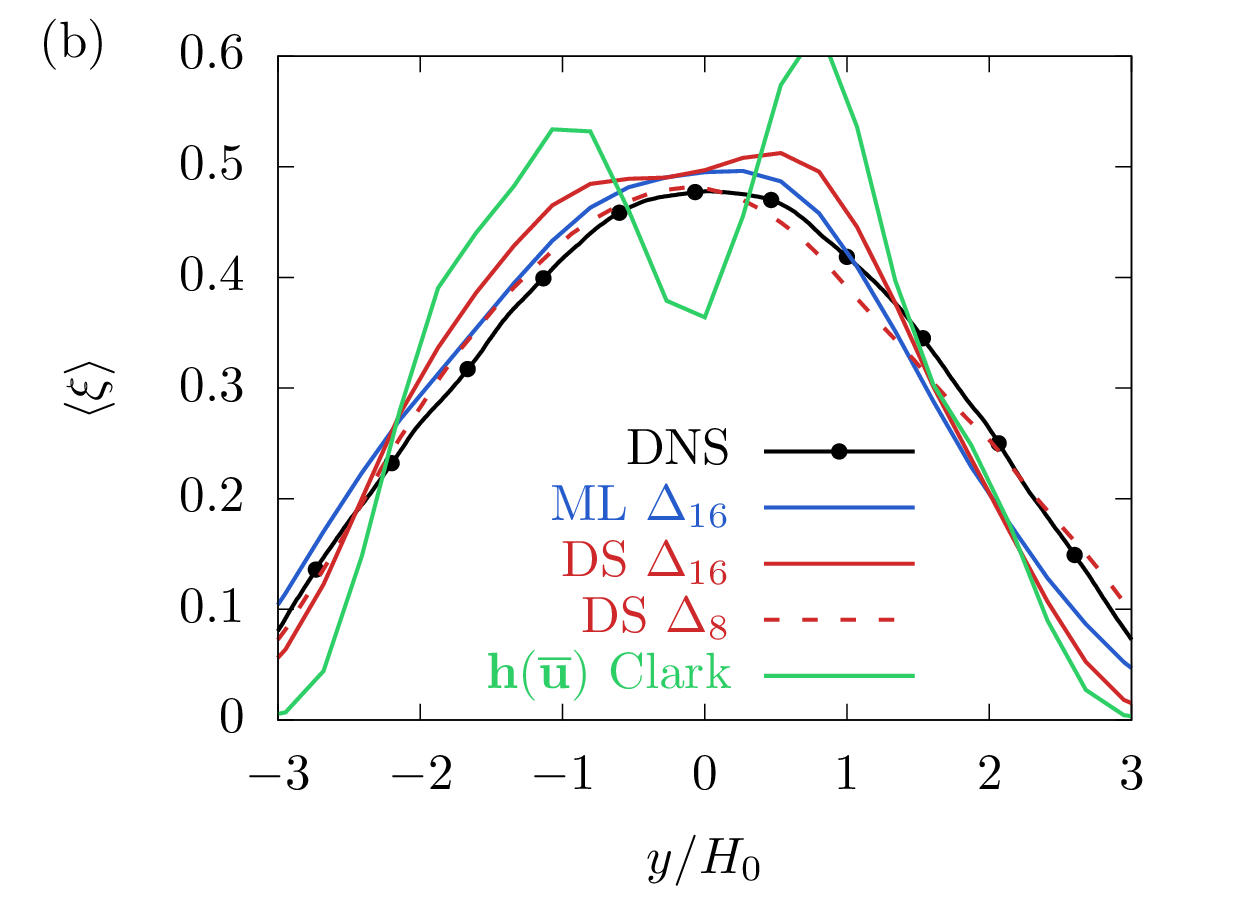} \\
  \includegraphics[width=0.48\textwidth]{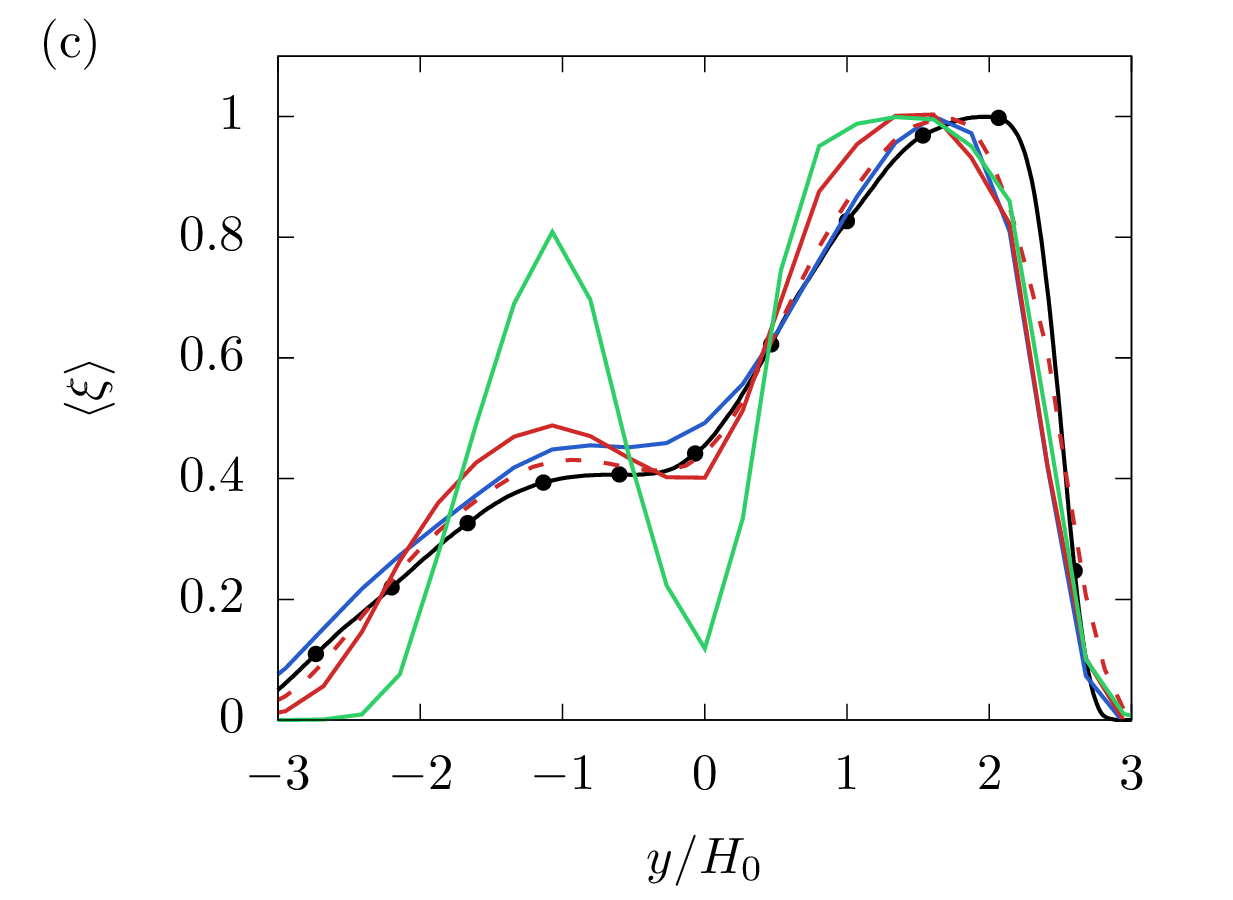}
  \includegraphics[width=0.48\textwidth]{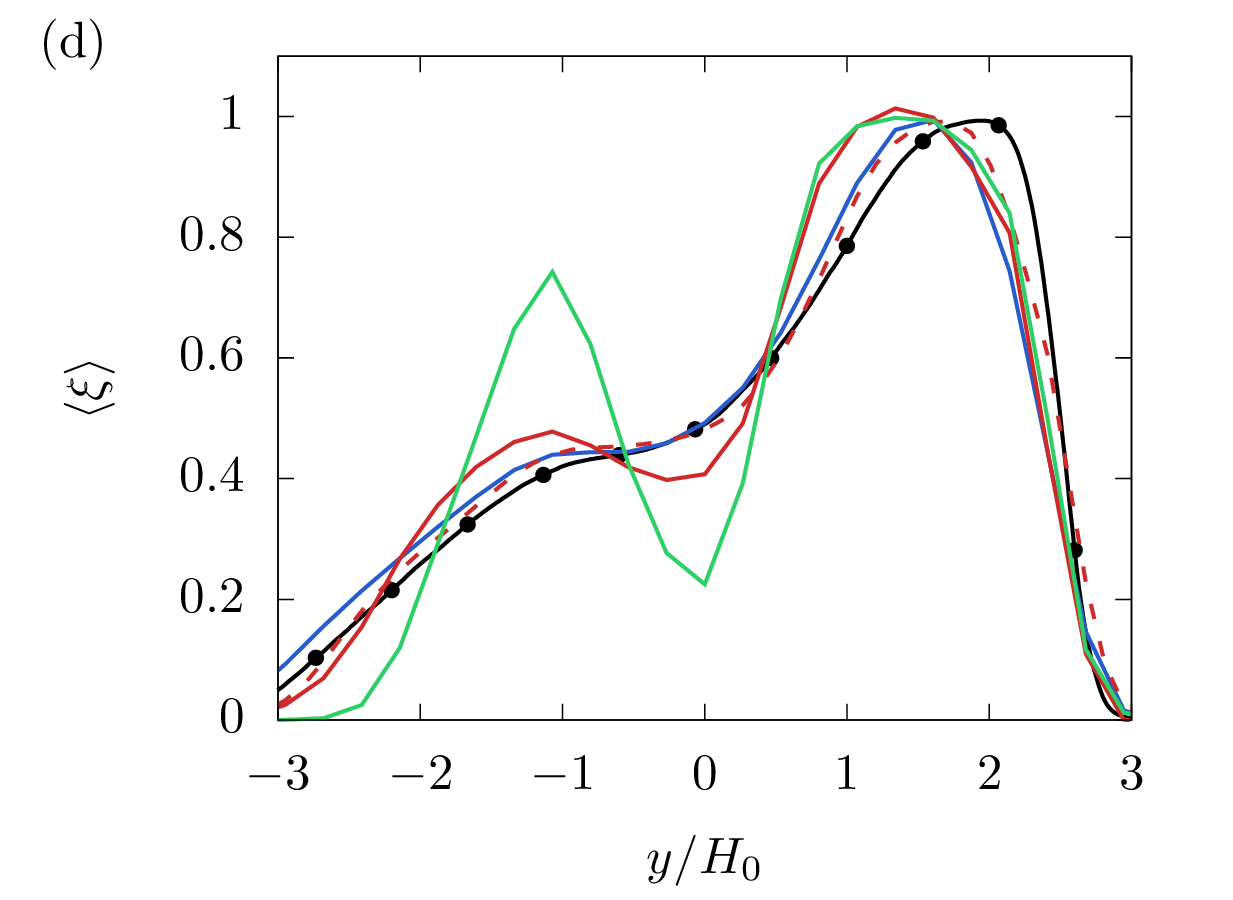}
  \caption{Mean passive scalar for the deep learning (ML), dynamic
    Smagorinsky (DS), and $\bh$-fitted Clark model 
    for $\vtau^r$ for $\Delta_{16}$ for 
    (a) case~A at $t=62.5t_{j0}$, (b)
    case~B at $t=50t_{j0}$, (c) case~C at $t=42.5t_{j0}$, and (d)
    case~D at $t=42.5t_{j0}$. 
    For comparison, $\Delta_8$ DS results are also shown. }

  \label{fig:jet_scalar}
\end{figure}

The predicted mean scalar profiles for the four jet cases, simulated with $\vtau^r$ obtained from either the deep learning model, the dynamic Smagorinsky model, or the $\bh$-fitted Clark model, are evaluated in figure~\ref{fig:jet_scalar}. The deep learning model outperforms the others in all four cases. This is particularly evident in the double-jet cases, for which the same $\Delta_{16}$ using dynamic Smagorinsky underpredicts jet spreading and merging (see figure~\ref{fig:doublejet_vel}).  This result is comparably accurate to  dynamic-Smagorinsky predictions for $\Delta_8$. Conversely, the $\bh$-fitted Clark model, with coefficients obtained for case~A, performs poorly for the out-of-sample dual-jet cases.

The results in this section provide evidence of potentially improved predictive accuracy of coupled-PDE simulations using the deep learning model. However, we note that the more complicated case of two-way-coupled PDEs has yet to be investigated. Variable-density, non-reacting scalar mixing, with a simple equation of state, is an obvious next step in analyzing two-way couplings.

\section{Mesh-dependence}
\label{s:num}

We finally consider the effect of different grid resolutions on the
learned closure $\bh$. The
\textit{a priori} convergence of $\bh$ outputs with grid
refinement is first shown. Then, \textit{a posteriori} performance of
the model on finer grids, for which the model was not explicitly
trained, 
is assessed.

\subsection{Convergence with grid refinement}

The preceding sections center around implicitly filtered large-eddy simulation,
$\Delta_\LESsub/\Delta_\DNSsub = \fDelta/\Delta_\DNSsub$, where
$\Delta_\DNSsub$ and $\Delta_\LESsub$ distinguish the fine direct
numerical simulation mesh and large-eddy simulation meshes. For
constant $\fDelta$, we now consider \textit{a priori} convergence of sub-grid-scale
closure approximations with refinement of $\Delta_\LESsub$. This
enables estimation of discretization errors and the degree to which
$\bh$ deviates from the expected convergence of the differential
approximation of the underlying numerical scheme.

For constant $\fDelta$, standard models for $\vtau^r$ converge as
$O(\Delta_\LESsub^p)$ for order-$p$ spatial discretizations. Here,
$p=2$. The deep learning model is not constrained by this: as
described in section~\ref{ss:model_arch}, although its inputs are the
velocity first and second derivatives at adjacent mesh points, it is 
nonlinear and so not constrained to show second-order convergence,
at least for the resolutions we test. In theory, it can compensate for
coarse-grid truncation error in \textit{a posteriori} large-eddy
simulation, for which numerical errors appear as additional, unclosed
terms when compared to trusted target data\cite{DPM-JCP}.

Velocity gradients are computed from filtered direct numerical
simulation data for $\fDelta = \Delta_{16}$. The finite-difference
approximations are evaluated for
$\Delta_\LESsub/\Delta_\DNSsub=16,8,4,2,1$ using the same staggered variable
placement as the solver (see section~\ref{ss:numerics}).

To quantify the mesh convergence of the different closures, a relative difference\cite{MacArt2016} is computed as
\begin{equation}
  \varepsilon_{\Delta x}\equiv \ell_2\left(\vtau^r_{\Delta x} -
    \vtau^r_{\Delta x/2}\right),
\label{e:epsilon}
\end{equation}
where $\ell_2$ is the 2-norm and $\vtau^r_{\Delta x}$ and
$\vtau^r_{\Delta x/2}$ are closure approximations evaluated for mesh
spacings $\Delta_\LESsub$ and $\Delta_\LESsub/2$.

\begin{figure}
  \centering
  \includegraphics[width=0.5\textwidth]{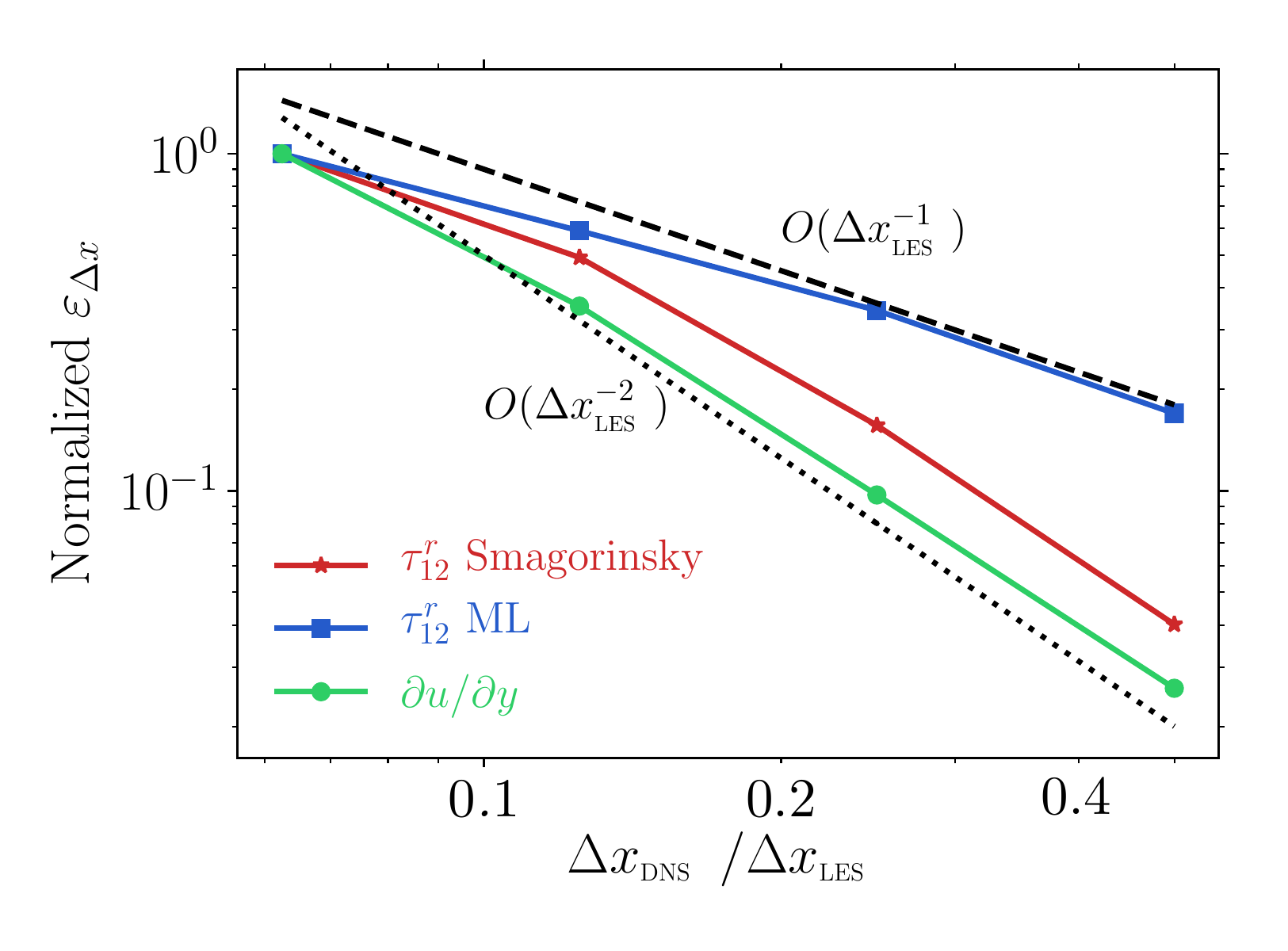}
  \caption{Relative difference $\varepsilon_{\Delta x}$ from
    (\ref{e:epsilon}) showing mesh dependence of $\tau^r_{12}$ for
    case~A with $\fDelta=\Delta_{16}$ training for the indicated models
    (section~\ref{s:geo_jet}).  The corresponding convergence of 
    $\partial u/\partial y$ is shown for reference.  }
  \label{fig:FD_convergence}
\end{figure}

Behavior is shown in figure~\ref{fig:FD_convergence} for both the deep
learning and Smagorinsky models.  Also shown, for reference, is the
cross-stream $u$-velocity gradient evaluated on $\Delta_\LESsub$. Both
the velocity gradients and the Smagorinsky-model $\vtau^r$ appear to
follow $O(\Delta_\LESsub^2)$, as expected. Conversely, the deep learning
model only approaches linear $O(\Delta_\LESsub)$ behavior over the available
range of mesh densities.  This is presumably a consequence of the more
complex and nonlinear
properties that provide its excellent \textit{a posteriori} performance.

In the sense of optimal LES\cite{Langford1999}, discrepancies between
the formal convergence of unclosed LES solutions (or trusted DNS data)
and the ideal closed LES solution are expected. The ideal LES
closure, which achieves the upper limit of accuracy with respect to
trusted data, must approximate Navier--Stokes physics that the filter
operator \eqref{e:filter} maps to its null space. These are, of
course, not representable in an LES solution space even with explicit
filtering\cite{Langford1999}.

Figure~\ref{fig:FD_convergence} may be understood in this context by
considering an optimal closure model
\begin{equation}
  \vtau^r_{\Delta x} = \mathrm{P}\vtau^r_\mathrm{Ideal} +
  \eps_\mathrm{P}(\Delta x) + c\Delta x^p,
\end{equation}
where $\tau^r_\mathrm{Ideal}$ is the ideal LES closure model,
$\mathrm{P}$ is a (non-unique) projection to the truncated $\Delta x$
LES solution space, $\eps_\mathrm{P}(\Delta x)$ is the dynamical error
in the Navier--Stokes solution due to the projection, and $c\Delta
x^p$ is the spatial discretization error of the model inputs. Since
$\mathrm{P}$ is linear and invertible, the relative difference
\eqref{e:epsilon} eliminates $\mathrm{P}\vtau^r_\mathrm{Ideal}$
leaving only the error terms. Traditional LES closures do not model
$\eps_\mathrm{P}(\Delta x)$ and so converge as $O(\Delta
x^p)$. Conversely, the \textit{a posteriori}-trained
\eqref{e:Lindirect} deep learning model does not, even though its
inputs are $O(\Delta x^p)$-accurate, which suggests that it at least
partially accounts for $\eps_\mathrm{P}(\Delta x)$.

\subsection{Performance on out-of-sample meshes}

\begin{figure}
  \centering
  \includegraphics[width=0.48\textwidth]{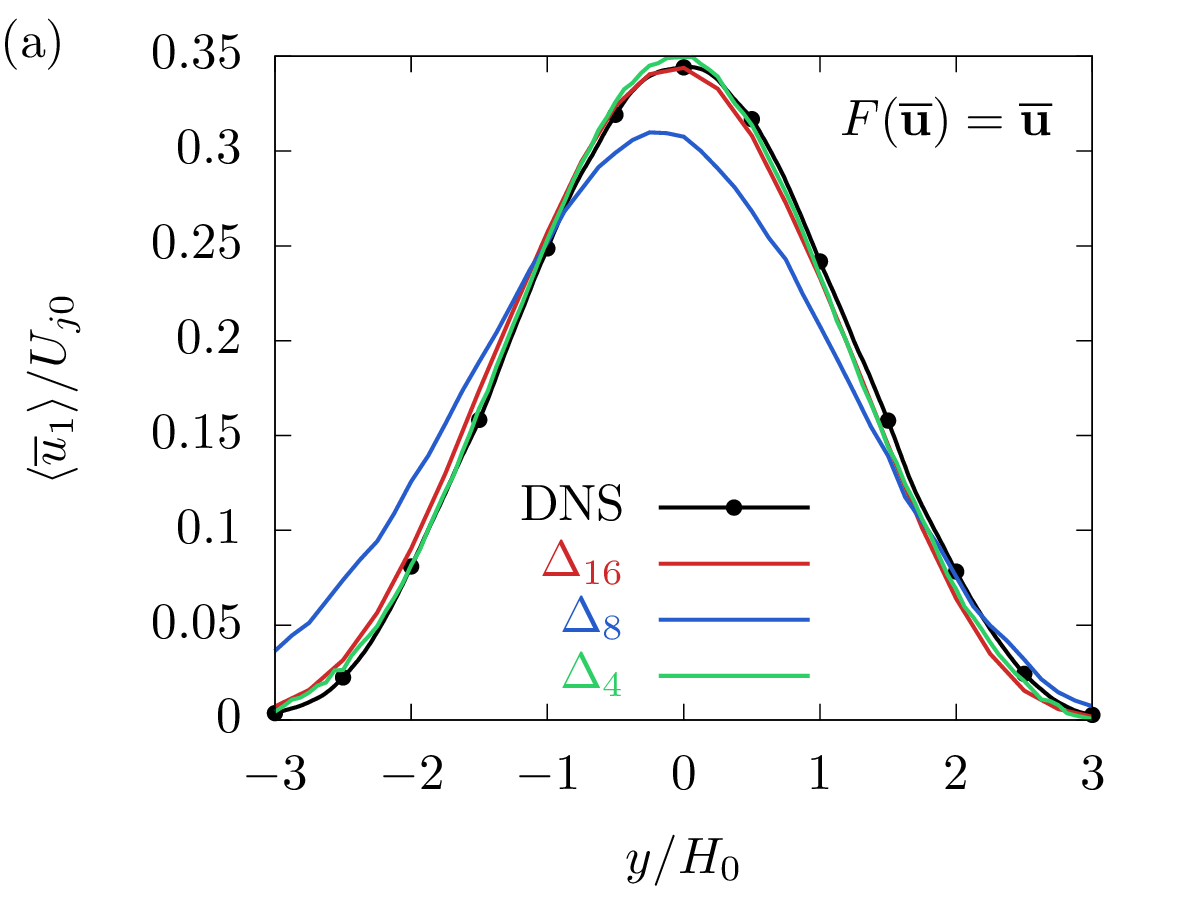}
  \includegraphics[width=0.48\textwidth]{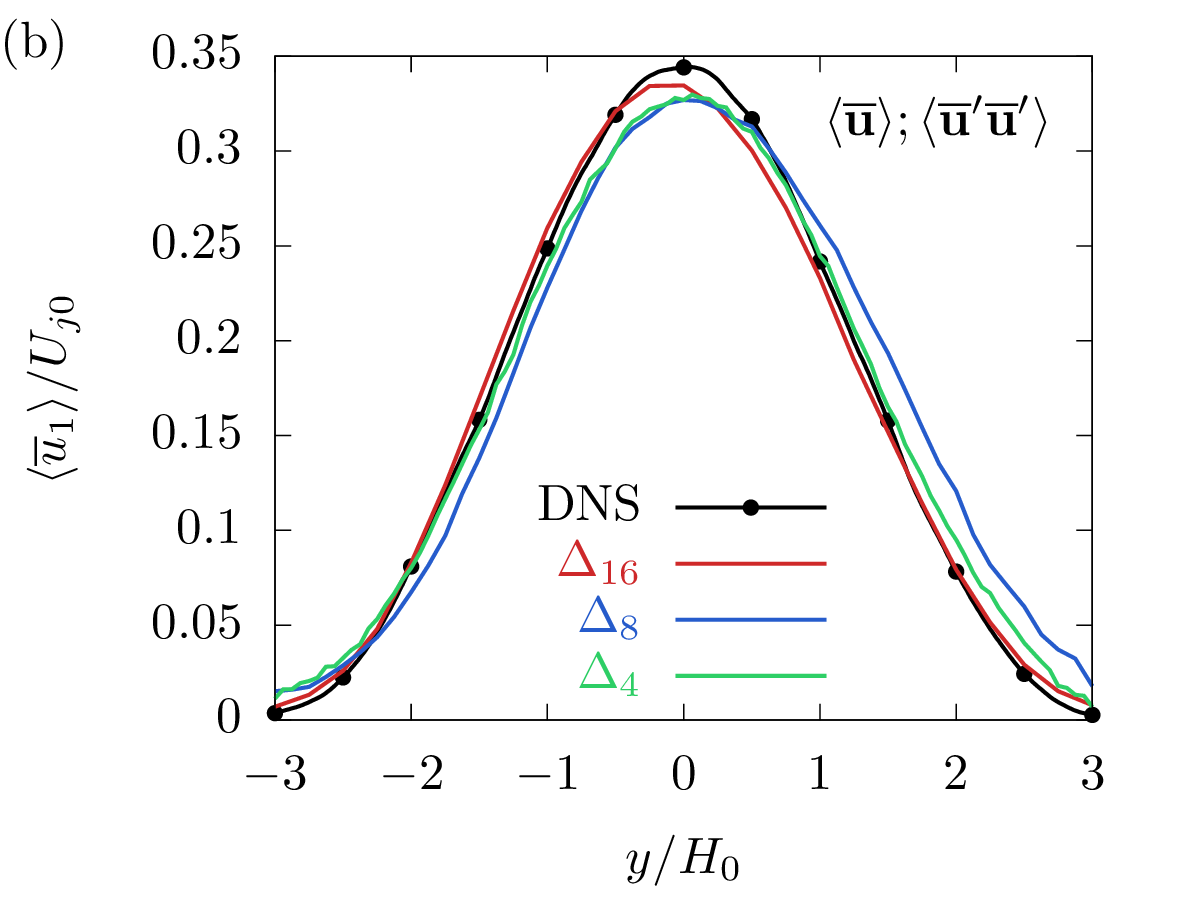}
  \caption{Predicted streamwise mean velocity for the single jet case (A) for grid-filter ratios $\fDelta/\Delta_\DNSsub$. (a) Instantaneous-velocity training; (b) mean-flow and Reynolds stress training. Both models were trained only on $\Delta_{16}$.}
  \label{fig:jet_gridfilter}
\end{figure}

The model is trained for a particular coarse resolution, for which, at
least for established models, discretization errors are known to
couple with model errors. The performance of the model suggests that
it also compensates for discretization errors, which was assessed
previously for isotropic turbulence.\cite{DPM-JCP} Hence, it comes
with no \textit{a priori} assurance that it will be as accurate away
from resolutions for which it was trained. We assess this by applying
the $\bh$ closure to finer grids than the training
$\Delta_{16}$. Figure~\ref{fig:jet_gridfilter} shows the performance
over a range of $\Delta_N$ using both instantaneous-velocity training
and mean-statistics training. The instantaneous-velocity-trained model
in figure~\ref{fig:jet_gridfilter}~(a) reproduces the mean profile on
$\Delta_4$ nearly as well as for the $\Delta_{16}$ trained case, but
not so well on $\Delta_8$.  The mean-flow training is more robust, as
seen in figure~\ref{fig:jet_gridfilter}~(b). However, the implicit
filter definition used for \textit{a posteriori} LES is still a
potential source of error. The mean-flow training
(figure~\ref{fig:jet_gridfilter}b) is presumably less sensitive to
discrepancies between this and the \textit{a priori} explicit filter
definition, although further testing is certainly warranted.

Finally, we note that a deep learning model
trained only for isotropic turbulence\cite{DPM-JCP} performed
reasonably well when tested out-of-sample on the jet case A, although
it was not as accurate as the deep learning models
in figure~\ref{fig:jet_gridfilter}. The model is at least stable and
reasonably accurate when applied to different grids and potentially to
different flows. However, its overall extrapolative capacity, for
example, to different flow phenomena, is likely reduced.

\section{Additional Discussion }
\label{s:summary}

In some sense, it is unsurprising that a
high-degree-of-freedom model can outperform established models with
few if any parameters, at
least for the modest degree of out-of-sample extrapolation we attempted
here.  However, determining viable parameters for this is challenging.
The adjoint-based PDE-coupled training procedures used here seem to
perform well for finding parameters sufficient for 
robust and extrapolative models.  The predictive 
agreement with trusted data for coarse meshes is
striking.  Still, given the number of parameters required and the
challenging optimization procedures needed, this
should not be taken as an unqualified improvement.  Indeed, a fair way to
view results is recognition of just how well dynamic Smagorinsky, as
an example of established models, does
with neither significant parameterization nor a complicated
training regimen.  In closing, we revisit some of the
motivations beyond the basic goal of 
increasing accuracy for fixed mesh resolution.

One particular hope
is that this framework will facilitate
representation of turbulence when it is coupled with additional
physics, especially in cases where this coupling
lacks models of the fidelity available for incompressible turbulence.  The
final demonstration, where the model is shown to also extrapolate to represent
scalar mixing, is an example of this.  Similarly, training
of a robust model based only on mean statistics provides an avenue for
predictions even when part of the physics (e.g., flame dynamics) might
be insufficiently described to craft an explicit model.  
Generalizing high-degree-of-freedom models to
out-of-sample flow conditions or new physics couplings is a main
target for this procedure. 

Even for inert incompressible turbulence, without the complexity of
additional physics, we also anticipate potential 
synergies between the highly parameterized approach we introduce and
established models and modeling approaches.  Our efforts to interpret 
the successful neural-network model by fitting the parameters
of low-degree-of-freedom models are a first step toward
using the successful working representations to improve
conventional models. A
related opportunity is the use of \textit{a posteriori} successful 
neural-network models to aid coefficient
discovery in moderately-parameterized models, based on physical
reasoning, such as those that aim to codify nonlinearly coupled multi-physical
interactions.  We anticipate a fruitful middle ground between fitting accuracy and
generalization where the two modeling approaches can complement and
augment each other.

An outstanding issue, which is shared with all turbulence
sub-grid-scale modeling, is the interplay of numerical errors
associated with the coarse mesh resolution and the model closures.  As
trained through solutions of the discrete flow equations and their
adjoint, the current deep learning models account for errors
introduced by the coarse discretizations, which is important for its
predictive success.  A potential drawback is that it risks linking the
trained model with the specific mesh spacing (and discretization
method) for which it is trained.  In the present case, we demonstrated
that this is not a severe limitation: the model is robustly predictive
for different meshes, though accuracy is indeed reduced.  Looking
forward, it can be anticipated that systematically training for a
range of mesh resolutions might provide improved predictions.  We can
also speculate that the training on the case~A jet, which spreads by
over a factor of two during the training period, might imbue the
learned model with some robustness in this regard since it exposes the
model to different nominal resolutions through the evolution of the
flow.  For example, if the development were exactly self-similar, then
the model requirements for the $\Delta_{16}$ mesh at some later time
would match those for the $\Delta_{8}$ mesh at the earlier time when
the jet has half the later thickness.  Systematic studies of
extrapolative capacity to different resolutions, mesh configurations,
numerical methods, and flows are all warranted.

Finally, it is important to address a general challenge faced when applying
adjoint methods to turbulence.  It is well-understood that the chaos of
turbulence will, for long enough times, lead to divergence of
sensitivity fields, which diminishes the utility of computed
sensitivities.  This is not a challenge for the short-time adjoint
solutions used in the most of the present demonstrations, and we
anticipate that many applications of this method will not suffer from
this because of the character of the problem:  a sub-grid-scale model
to provide a stress correction for the local instantaneous
resolved-scale turbulence.  Since the large turbulence scales are
relatively local in space and time, the sensitivity will not
accumulate significantly.  In general, however, this might limit
applications with limited data availability, especially if it is
sensitive to turbulence that is significantly distant in space and
time.  Further work will be required to understand how the Lyapunov
divergence of turbulence might hinder the adjoint-based training.  We do note
that the adjoint equations remained sufficiently well-behaved when we trained and simulated over
the entire time history of the jet flow when training for mean statistics.

\subsection*{Acknowledgments}

This material is based in part upon work supported by the Department
of Energy, National Nuclear Security Administration, under Award
Number DE-NA0002374.  This research is part of the Blue Waters
sustained-petascale computing project, which is supported by the
National Science Foundation (awards OCI-0725070 and ACI-1238993) and
the State of Illinois. Blue Waters is a joint effort of the University
of Illinois at Urbana--Champaign and its National Center for
Supercomputing Applications.

\end{document}